\documentclass[12pt,letterpaper]{article}
\pdfoutput=1

\usepackage[dvipsnames]{xcolor}
\usepackage{amsmath,amssymb,calc, amsthm,bbm, epsfig,psfrag, mathtools}
\usepackage{graphicx, enumerate}
\usepackage[numbers,sort&compress]{natbib}
\usepackage[font={scriptsize,}]{caption}
\usepackage{tensor}
\usepackage{mathtools}
\usepackage{caption}
\usepackage{subcaption}
\usepackage{tikz-cd}

\usepackage[utf8]{inputenc} 

\newcommand{\Comment}[1]{{}}
\definecolor{darkblue}{rgb}{0.15,0.35,0.55}
\definecolor{reddish}{rgb}{0.65, 0.2, 0.2}

\newcommand{\rto}{\rightarrow}
\newcommand{\lto}{\leftarrow}
\newcommand{\inv}{^{-1}}

\newcommand{\CC}{\ensuremath{\mathbb{C}}}

\usepackage[linktocpage=true]{hyperref}
\hypersetup{
colorlinks=true,
citecolor=darkblue,
linkcolor=reddish,
urlcolor=darkblue,
pdfauthor={},
pdftitle={},
pdfsubject={}
}

\PassOptionsToPackage{linecolor=blue,backgroundcolor=blue!25,bordercolor=blue,textsize=scriptsize}{todonotes}
\usepackage{todonotes}

\makeatletter
\renewcommand\section{\@startsection {section}{1}{\z@}%
                                   {-3.5ex \@plus -1ex \@minus -.2ex}%nn
                                   {2.3ex \@plus.2ex}%
                                   {\normalfont\large\bfseries}}
\renewcommand\subsection{\@startsection{subsection}{2}{\z@}%
                                     {-3.25ex\@plus -1ex \@minus -.2ex}%
                                     {1.5ex \@plus .2ex}%
                                     {\normalfont\bfseries}}
\makeatother

\let\non\nonumber

\newcommand{\del}{\partial}

\def\bea#1\eea{\begin{align}#1\end{align}}

\def\bes #1\ees{\begin{split}#1\end{split}}

\newcommand{\be}{\begin{equation}}
\newcommand{\ee}{\end{equation}}

\newcommand{\bma}{\begin{pmatrix}}
\newcommand{\ema}{\end{pmatrix}}

\newcommand{\R}{{\mathbb R}}

\let\a=\alpha

\def\g{\gamma}

\def\CC{{\mathbb C}}

\newcommand{\C}[1]{$(\ref{#1})$}

\def\IZ{\relax\ifmmode\mathchoice
{\hbox{\cmss Z\kern-.4em Z}}{\hbox{\cmss Z\kern-.4em Z}}
{\lower.9pt\hbox{\cmsss Z\kern-.4em Z}} {\lower1.2pt\hbox{\cmsss
Z\kern-.4em Z}}\else{\cmss Z\kern-.4em Z}\fi}
\def\IR{\relax{\rm I\kern-.18em R}}

\def\one{{\hbox{ 1\kern-.8mm l}}}

\newlength{\bredde}
\def\slash#1{\settowidth{\bredde}{$#1$}\ifmmode\,\raisebox{.15ex}{/}
\hspace*{-\bredde} #1\else$\,\raisebox{.15ex}{/}\hspace*{-\bredde}
#1$\fi}

\newsavebox{\zzzbar}
\sbox{\zzzbar}
  {\setlength{\unitlength}{0.9em}
  \begin{picture}(0.6,0.7)
  \thinlines
  \put(0,0){\line(1,0){0.6}}
  \put(0,0.75){\line(1,0){0.575}}
  \multiput(0,0)(0.0125,0.025){30}{\rule{0.3pt}{0.3pt}}
  \multiput(0.2,0)(0.0125,0.025){30}{\rule{0.3pt}{0.3pt}}
  \put(0,0.75){\line(0,-1){0.15}}
  \put(0.015,0.75){\line(0,-1){0.1}}
  \put(0.03,0.75){\line(0,-1){0.075}}
  \put(0.045,0.75){\line(0,-1){0.05}}
  \put(0.05,0.75){\line(0,-1){0.025}}
  \put(0.6,0){\line(0,1){0.15}}
  \put(0.585,0){\line(0,1){0.1}}
  \put(0.57,0){\line(0,1){0.075}}
  \put(0.555,0){\line(0,1){0.05}}
  \put(0.55,0){\line(0,1){0.025}}
  \end{picture}}

\newfont{\goth}{ygoth.tfm scaled 1200}                   % gothic font (usual)
 \numberwithin{equation}{section}

\def\1{{(1)}}
\def\2{{(2)}}
\def\3{{(3)}}

\newcommand{\ul}{\underline}

\begin{document}
\begin{titlepage}

\begin{center}

\today
\hfill         \phantom{xxx}  EFI-24-6

\vskip 2 cm {\Large \bf Solving the Kinetic Ising Model with Non-Reciprocity} 
\vskip 1.25 cm {\bf Gabriel Artur Weiderpass, Mayur Sharma and Savdeep Sethi}\non\\
\vskip 0.2 cm

{\it Enrico Fermi Institute \& Kadanoff Center for Theoretical Physics \\ University of Chicago, Chicago, IL 60637, USA}
\vskip 0.2 cm

\vskip 0.2 cm

\end{center}
\vskip 1.5 cm

\begin{abstract}
\baselineskip=18pt\textbf{}

Non-reciprocal interactions are a generic feature of non-equilibrium systems. We define a non-reciprocal generalization of the kinetic Ising model in one spatial dimension. We solve the model exactly using two different approaches for infinite, semi-infinite and finite systems with either periodic or open boundary conditions. The exact solution allows us to explore a range of novel phenomena tied to non-reciprocity like non-reciprocity induced frustration and wave phenomena with interesting parity-dependence for finite systems of size $N$. 

We study dynamical questions like the approach to equilibrium with various boundary conditions. We find different regimes, separated by $N^{th}$-order exceptional points, which can be classified as overdamped, underdamped or critically damped phases. Despite these different regimes, long-time order is only present at zero temperature. Additionally, we explore the low-energy behavior of the system in various limits, including the aging and spatiotemporal Porod regimes, demonstrating that non-reciprocity induces unique scaling behavior at zero temperature. Lastly, we present general results for systems where spins interact with no more than two spins, outlining the conditions under which long-time order may exist.

\end{abstract}

\end{titlepage}

\tableofcontents

\section{Introduction} \label{intro}

\subsubsection*{\ul{\it Modeling non-reciprocity}}

Non-reciprocal interactions are generically found in systems that are not in equilibrium~\cite{aguilera_unifying_2021,fruchart_non-reciprocal_2021,ivlev_statistical_2015,you_nonreciprocity_2020,crisanti_dynamics_1987,brauns_nonreciprocal_2024,costanzo_milling-induction_2019,frohoff-hulsmann_nonreciprocal_2023,saha_scalar_2020,liu_non-reciprocal_2023,loos_irreversibility_2020,du_hidden_2024,marchetti_hydrodynamics_2013,hanai_critical_2020,hanai_non-hermitian_2019,avni2023nonreciprocal,rajeev_ising_2024,seara_non-reciprocal_2023,hanai_nonreciprocal_2024,han_coupled_2024,godreche_nonequilibrium_2009,godreche_dynamics_2011,godreche_rates_2013,godreche_dynamics_2014,godreche_dynamics_2015,godreche_generic_2017,godreche_freezing_2018}. By a non-reciprocal interaction between two bodies, we mean the response of body $A$ to body $B$ is different from the response of body $B$ to body $A$. Such interactions, which are possible out of equilibrium, violate Newton's third law of motion; namely, that every action generates an equal and opposite reaction. These interactions cannot, therefore, be easily described in the framework of equilibrium statistical mechanics. Non-reciprocal interactions are seen in a very large variety of systems including synthetic active matter and soft matter systems \cite{brandenbourger_non-reciprocal_2019,veenstra_non-reciprocal_2024,veenstra_non-reciprocal_2023,brandenbourger_non-reciprocal_2024,ghatak_observation_2020,helbig_generalized_2020,kotwal_active_2021,hofmann_chiral_2019,gupta_active_2022,tan_odd_2022,shankar_topological_2022,fruchart_odd_2023,colen_interpreting_2024,scheibner_odd_2020,poncet_when_2022,rosa_dynamics_2020,scheibner_non-hermitian_2020,coulais_topology_2021,dinelli_non-reciprocity_2023}, open quantum systems \cite{metelmann_nonreciprocal_2015,mcdonald_exponentially-enhanced_2020,clerk_introduction_2022,chiacchio_nonreciprocal_2023,bergholtz_exceptional_2021,begg_quantum_2024}, the collective behavior of flocks, herds and social groups \cite{fruchart_non-reciprocal_2021,nagy_hierarchical_2010,morin_collective_2015,ginelli_intermittent_2015,hong_kuramoto_2011}, and neuroscience \cite{sompolinsky_temporal_1986,derrida_exactly_1987,parisi_asymmetric_1986,amir_non-hermitian_2016,montbrio_kuramoto_2018}. 

Given the prominent role played by non-reciprocal interactions in many disciplines, finding exactly solvable many-body systems with non-reciprocity is an endeavor of quite broad interest. Any such model extends the set of integrable theories studied in classical and quantum physics to this new area of high current interest. An exactly solvable model allows us to exhaustively explore the effect of non-reciprocal interactions beyond numerical simulations of finite size systems. The knowledge resulting from an analytic solution can then be used as a building block for a deeper understanding of more complicated systems. 

The most widely studied model in statistical mechanics is the Ising model, which describes the thermodynamics of interacting classical spins. Each spin can take two possible values. Usually one investigates the equilibrium properties of the system at fixed temperature. However, this kind of model cannot describe non-reciprocal interactions as the following example illustrates: suppose we have a system of $N$ classical spins. Non-reciprocity means that the interaction matrix between the $i^{th}$ and $j^{th}$ spins is not symmetric, $J_{ij} \neq J_{ji}$. We can then split the interactions into an even and an odd part,
\begin{align} \label{JeJo}
   J^e_{ij} = \frac{J_{ij}+J_{ji}}{2}\,, \qquad  J^o_{ij} = \frac{J_{ij}-J_{ji}}{2}\,,
\end{align}
which encode the reciprocal and the non-reciprocal interactions, respectively. If we attempt to describe this system within the framework of equilibrium physics, the usual next step is to postulate a partition function $Z= \sum_{s} \exp\{- \beta E(s)\}$ with the energy $E = -\sum_{ij} J_{ij} s_i s_j$. However the energy can be rewritten as
\begin{align} \label{Trivializing-NR}
    & \sum_{i<j} \Big( J_{ij} s_i s_j + J_{ji} s_j s_i \Big) = \sum_{ij} \frac{J_{ij}+J_{ji}}{2} s_i s_j = \sum_{ij} J_{ij}^e s_i s_j\,.
\end{align}
That is, the equilibrium formalism summarily ignores the non-reciprocal interactions.

Since the equilibrium formalism is not useful for studying non-reciprocally coupled spins, we instead start by assigning dynamics to the spin degrees of freedom. There are many choices for this assignment, which depend on the microscopic details of the system under investigation. Our starting point is  the kinetic Ising model pioneered by Glauber \cite{glauber_timedependent_1963}; see also Ref. \cite{godreche_response_2000} for important subsequent work and Refs. \cite{henkel_non-equilibrium_2010,krapivsky_kinetic_2010} for a review. 
This model uses the principle of detailed balance to constrain the spin dynamics. It is among the simplest ways to assign dynamics to spins and the model has  served as a paradigm for non-equilibrium physics. Glauber solved the model in one spatial dimension and it has been extensively studied numerically in higher dimensions. 
 
Let $s(t)$ denote the spin configuration of the system at time $t$. We assume each $s_i$ is a classical spin with value $\pm 1$. The kinetic Ising model assigns a dynamical behavior to classical spins by defining a transition probability from the state $s(t-\delta t)$ to the state $s(t)$ using the local magnetic field that each spin feels at time $t-\delta t$. Glauber first defined the reciprocal kinetic Ising model by assuming detailed balance which fixes the transition rates \cite{glauber_timedependent_1963}. Following Ref. \cite{aguilera_unifying_2021}, we define the kinetic Ising model using the transition probability given by the Markov chain
\begin{align} \label{Markov-C}
    P\big(s(t)\big|s(t-\delta t)\big) = \prod_{j=1}^N\frac{\exp\left({s_j(t) h_j\big[s(t-\delta t)\big]}\right)}{2\cosh h_j\big[s(t-\delta t)\big] } \,,
\end{align}
where %$\alpha$ is a constant with inverse time units and 
$h_j[s(t-\delta t)]$ is the local magnetic field felt by $s_j$ at time $t-\delta t$ and is given by
\begin{align}
    h_j[s] = \sum_i \beta J_{ji} s_i\,.
\end{align}
From the Markov chain (\ref{Markov-C}) we recover Glauber's transition rate when $J_{ij}=J_{ji}$ and we take the limit $\delta t\rightarrow 0$ while allowing at most a single spin to flip per infinitesimal time step.

Expression (\ref{Markov-C}) tells us the probability of finding the system in state $s(t)$ at time $t$ given that it was in state $s(t-\delta t)$ at time $t-\delta t$. Since now we have spins at different times, if we reorganize the summation inside the exponential in (\ref{Markov-C}) as we did in (\ref{Trivializing-NR}), we find
\begin{align}
\begin{aligned}
    \sum_{j} s_j(t) h_j[s(t-\delta t)]  = \frac{1}{2} \sum_{ij} \Big( & J_{ij}^e\left[ s_i(t) s_j(t-\delta t) + s_j(t) s_i(t-\delta t) \right]\\
    & + J_{ij}^o\left[ s_i(t) s_j(t-\delta t) - s_j(t) s_i(t-\delta t) \right] \Big)\,.
\end{aligned}
\end{align}
So the kinetic Ising model defined by (\ref{Markov-C}) is able to encode non-reciprocal interactions and therefore provides a basic theoretical model for non-reciprocal physics.

In this work we formulate a non-reciprocal one-dimensional Ising model; we find an exact solution for this model and study it in detail. There is no unique way to do this but our choice has the virtue of leading to a solvable model in one spatial dimension, analogous to Glauber's reciprocal kinetic Ising model. The interactions between neighboring spins are depicted in Fig. \ref{NR-Ising-Lattice}. In Fig. \ref{NR-Ising-Lattice}, we use $J$ to denote $J^e_{ij}$ and $K$ to denote $J^o_{ij}$ restricted to the case of non-vanishing nearest-neighbor interactions. 

While most discussions of the kinetic Ising model involve reciprocal interactions, there has been other interesting recent work on non-reciprocal Ising models studied via mean field theory, numerical simulations or by performing experiments \cite{avni2023nonreciprocal,rajeev_ising_2024,seara_non-reciprocal_2023,han_coupled_2024}. For example, \cite{avni2023nonreciprocal} builds a non-reciprocal Ising model by starting with two standard reciprocal Ising models and then coupling them to each other in a non-reciprocal way. The construction used in Refs. \cite{seara_non-reciprocal_2023,rajeev_ising_2024} is a possible two-dimensional generalization of our construction. The dynamics of a two spin non-reciprocal kinetic Ising model, which we discuss in detail in Section \ref{Sec-2-Spin}, has been experimentally realized in Ref. \cite{han_coupled_2024} using micro-electromechanical resonators.

\subsubsection*{\ul{\it Overview and summary}}

This paper is organized as follows: in Section \ref{Sec-Markov-Chain} we give a basic overview of the kinetic Ising model and discuss two simple pedagogical examples. The first in Section \ref{Sec-Single-spin} is the classical dynamics of a single spin and the second in Section \ref{Sec-2-Spin} is the classical dynamics of two spins coupled non-reciprocally.  It is worth pointing out that a recent experiment \cite{han_coupled_2024} realizes the two spin model of Section \ref{Sec-2-Spin}, providing a framework for comparing theory with experiment. 

In Section \ref{Sec-N-Spins} we obtain the exact solution of the one-point function of the system which is given by equations (\ref{1-pt-soul-Green}) and (\ref{1-point-propagator}). We show that all the other equal-time $n$-point functions, (\ref{2-point-hom-sol}) and (\ref{n-point-sol-hom}), and therefore all $n$-point functions can be built out of the one-point function; therefore (\ref{1-point-propagator}) is the full solution of the system. We analyze such solutions in detail for infinite spin chains. 

In Section \ref{Sec-Finite-Spin-Chain} we discuss a finite spin chain with both periodic and open boundary conditions. We find phenomena not seen in reciprocal systems which depend on the boundary conditions, even in the limit of large system size. We also discover a mechanism for frustration which is induced by non-reciprocity as well as seeing different wave-like phenomena that allow movement through the lattice. We then analyze the two-point function $C_{x}(t,\tau) = \langle s_{x+i}(t+\tau) s_{i}(\tau) \rangle$ of the system in Section \ref{Sec-Two-Point}. In the low-energy limit, we show that the effect of non-reciprocity, at least for small enough times $t\ll \tau_{eq}$, is to deform the two-point function of the reciprocal kinetic Ising model, see Ref.~\cite{godreche_response_2000}, by the transformation:
\begin{align}
    x \rightarrow x + \gamma_o t\,,
\end{align}
where $\gamma_o$ encodes the non-reciprocity. Furthermore in some cases, discussed around equations (\ref{2-pt-lim2-eq}) and (\ref{2-pt-Porod-Equilibrium}), this result is actually valid for all times. We also show that the two point function of the system is symmetric under the scaling transformation:
\begin{align} \label{Scale-Transf-Age-3}
    x \rightarrow \lambda^{-1}\left(x+ (1-\lambda^{-1}) \gamma_o t \right) \, , \qquad t \rightarrow \lambda^{-2} t\,, \qquad \tau \rightarrow \lambda^{-2} \tau\, .
\end{align}
Furthermore in the aging (\ref{2-pt-ageing}) and the spatiotemporal Porod regimes (\ref{2-pt-Porod}), there is an extra scale invariance given by the transformation:
\begin{align} \label{Scaling-Non-Trivial}
    x \rightarrow \lambda^{-1} x\, , \qquad t \rightarrow \lambda^{-1} t\, , \qquad \tau \rightarrow \lambda^{-2} \tau \, .
\end{align}
These non-trivial scaling behaviors are a direct result of non-reciprocity.
 
\begin{figure}
    \centering
    \includegraphics[width=0.8\linewidth]{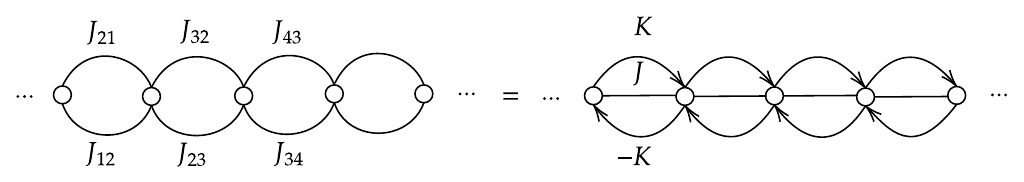}
    \caption{Non-reciprocal Ising model with non-reciprocity between neighboring spins.}
    \label{NR-Ising-Lattice}
\end{figure}

In Section \ref{Sec-Random-Couplings} we generalize the construction of Figure \ref{NR-Ising-Lattice} and allow each spin to couple to any other spin with any possible coupling strength. The only restriction we impose is that each spin couples to at most two spins. In this general setup, we show that there is no long-time order for systems of finite size at finite temperature, regardless of non-reciprocity.

For systems of infinite size the situation is more subtle. We show that if the system has an infinite order exceptional point with a single eigenvalue, it cannot exhibit long-time order at finite-temperature. This leaves the possibility of finite temperature long-time order in systems with more than one eigenvalue, where at least one eigenvalue has infinite algebraic multiplicity. Further investigation is needed to address this possibility. There are also a myriad of fascinating related questions mirroring the development of equilibrium statistical physics involving, for example, systems with higher dimension or more complicated interactions where the method we used is not applicable. Lastly we summarize our conclusions in Section~\ref{conclusions}.

\vskip 0.2in
\noindent {\bf Note Added:} After we completed this work, we were informed of very interesting prior work \cite{godreche_dynamics_2011}, which has both complimentary results and some overlap with the results of Section \ref{Sec-N-Spins}, Section \ref{genfunc_infchain} and equations (\ref{Low-E-Gx}) and (\ref{2-pt-ageing}). We are grateful to Claude Godrèche for bringing these results to our attention.

\section{Markov Chains and Master Equations} \label{Sec-Markov-Chain}

Even though (\ref{Markov-C}) is a good conceptual starting point for our discussion, this expression is not very well-suited for explicit calculations. Therefore we use the equivalent framework of the master equation which is a continuous time description of (\ref{Markov-C}). Since we are making the time steps infinitesimally small, we assume that at most a single spin flips at each time step. There are other interesting possible dynamical update rules that one might consider~\cite{krapivsky_kinetic_2010}. We can write a time evolution equation for the probability of finding the system at state $s$ at time $t$ which is the master equation,
\begin{align} \label{Master-Eq}
    \partial_t P(s,t) = \sum_j \Big[ & w(s_j \leftarrow -s_j)P(s_1,\ldots,-s_j,\ldots,s_N;t)  \cr & - w(-s_j \leftarrow s_j) P(s_1,\ldots,s_j,\ldots,s_N;t) \Big]\, ,
\end{align}
where $w(-s_j \leftarrow s_j)$ is the transition rate of the spin flipping from $s_j$ to $-s_j$. Notice that this equation is more general than the specific form describing a Markov chain in equation (\ref{Markov-C}) because the transition rates do not need to be compatible with (\ref{Markov-C}). Indeed all that is needed for (\ref{Master-Eq}) to define a good probability theory is that
\begin{align}
    \sum_{s_j} w(-s_j \leftarrow s_j) = \alpha\,,
\end{align}
where $\alpha$ is the characteristic time in which a single spin flips. Here we are interested in transition rates that are compatible with (\ref{Markov-C}), which we are taking as the definition of the kinetic Ising model. The transition rate for flipping the spin $s_j$ given that the whole system is in the state $s$ at time $t-\delta t$ is therefore, 
\begin{align}
    \begin{aligned}
    w(s_j\leftarrow -s_j) & 
    = \frac{\alpha}{2} \left( 1 + s_j \tanh h_j\big[-s_j\big] \right) \,  ,\\[10pt]
    w(-s_j\leftarrow s_j) & 
    = \frac{\alpha}{2} \left( 1 - s_j \tanh h_j\big[s_j\big] \right)\,, \nonumber
    \end{aligned}
\end{align}
where we are using the shorthand $h_j[-s_j] = h_j[s_1,\ldots,-s_j,\ldots,s_N]$. So the master equation compatible with (\ref{Markov-C}) is just
\begin{align}
    \partial_t P(s,t) = \frac{\alpha}{2} \sum_j \Big[ \left( 1 + s_j \tanh h_j\big[-s_j\big] \right) P(-s_j,t) - \left( 1 - s_j \tanh h_j\big[s_j\big] \right) P(s_j,t) \Big] \,,
\end{align}
where we use similar shorthand $P(-s_j) = P(s_1,\ldots,-s_j,\ldots,s_N)$.
The physical observables of interest in this system are the equal-time $n$-point functions and the time-ordered $n$-point functions. Let us begin our discussion with the equal-time $n$-point functions which are defined as follows:
\begin{align} \label{Equal-t-n-pt}
    r_{i_n \ldots i_1 }(t) = \langle s_{i_n} \ldots s_{i_1} \rangle (t) = \sum_s s_{i_n} \ldots s_{i_1} P(s,t) \,.
\end{align}
If we take the derivative of (\ref{Equal-t-n-pt}) and use the master equation (\ref{Master-Eq}) inside the summation, it is easy to show that 
\begin{align} \label{Eq-t-n-pt DE}
    \frac{d}{dt} r_{i_n \ldots i_1 }(t)
    & = -2 \left \langle s_{i_n} \ldots s_{i_1} \left[ \sum_{j=1}^n w(-s_{i_j}\leftarrow s_{i_j}) \right]\right \rangle\,. 
\end{align}
For example, for the one and two-point functions we see that
\begin{align}
    \label{1-point-Homex}
    \frac{d}{dt} q_i & = -2 \langle s_i w(-s_i\leftarrow s_i) \rangle\,, \\ \label{2-point-Hom}
    \frac{d}{dt} r_{ij} & = -2 \langle s_i s_j \big[ w(-s_{i}\leftarrow s_{i}) + w(-s_{j}\leftarrow s_{j}) \big] \rangle\,,
\end{align}
where we use $q_j(t) = \langle s_j(t) \rangle$ to denote the one-point function because of its importance. If these transition rates come from a joint probability distribution of exponential type (\ref{Markov-C}), we get
\begin{align} \label{1-point-eom-h}
    \frac{d}{dt} q_j & = -\alpha q_j + \alpha \langle s_j \tanh h_j [s_j] \rangle\,, \\ \label{2-point-eom-general}
    \frac{d}{dt} r_{ij} & = -2\alpha r_{ij} + \alpha \langle s_j \tanh h_i [s_i] \rangle + \alpha \langle s_i \tanh h_j [s_j] \rangle\,.
\end{align}
It is important to note that the second equation is actually an inhomogeneous differential equation because $r_{ii}(t) = 1$. The complete solution to the system must reflect this fact. In Sections \ref{Sec-N-Spins} and \ref{Sec-Two-Point} we show how to deal with the fact that they are inhomogeneous differential equations. Similarly notice that to solve any equal-time higher $n$-point function, we must again deal with this inhomogeneity.

Let us now look at the two-point function at different times
\begin{align}
    \langle s_i(t) s_j(t') \rangle = \sum_{s s'} s_i P(s,t|s',t') s'_j P(s',t')\,.
\end{align}
Notice that $\sum_{s} s_i P(s,t|s',t')$ is just the one-point function $r_i(t-t')$ with the initial condition $r_i(0) = s'_i$. If we define $G_{ii'}(t)$ to be the one-point function propagator such that
\begin{align} \label{1-pt-Prop-def}
    q_i(t) = \sum_{i'} G_{ii'}(t) q_{i'}(0) \,,
\end{align}
then we can write the two-point function as follows:
\begin{align} \label{2-point-General}
\begin{aligned}
    \langle s_i(t) s_j(t') \rangle & = \sum_{ s'} \sum_{i'} G_{ii'}(t-t') s'_{i} s'_j P(s',t')\,,  \\
    & = \sum_{i'} G_{ii'}(t-t') r_{i'j}(t') \,. 
\end{aligned}
\end{align}
If the system is translationally invariant then correlations depends only on the distance between the two points, therefore the propagator is $G_{ii'}(t) = G_{i-i'}(t)$. We are not making this assumption a priori though. Now we define the propagator of the equal-time two-point function, denoted $G^{(2)}_{ii',jj'}(t)$, so that $r_{ij}(t)$ can be written as
\begin{align}
    r_{ij}(t) = \sum_{i'j'} G^{(2)}_{ii',jj'}(t) r_{i'j'}(0)\,.
\end{align}
Note that for the condition $r_{ii}(t)=1$ to be satisfied, we must have
\begin{align}
    G^{(2)}_{ii',ij'}(t) = \delta_{i'i}\delta_{ij'} \frac{r^{part}_{i'j'}}{r_{i'j'}(0)}\,,
\end{align}
where $r^{part}_{ij}$ is a particular solution of the inhomogeneous differential equation satisfying $r_{ii}^{part}=1$. For an example, see equation (\ref{2-point-propagator-sol}) where we built the explicit two-point propagator for the non-reciprocal model corresponding to the Markov chain (\ref{N-spin-NR-Markov}). The two-point function at different times can then be written as
\begin{align}
    \langle s_i(t) s_j(t') \rangle= \sum_{i'j'i''} G_{ii'}(t-t') G^{(2)}_{i'i'',jj'}(t') r_{i'' j'}(0)\,.
\end{align}
The interpretation of this equation is quite straightforward: the two-point function propagator $G_{ii',jj'}^{(2)}(t')$ takes the initial condition $r_{ij}(0)$ and propagates two spins in time until time $t'$. We then use the single spin propagator $G_{ii'}(t-t')$ to propagate the spin at site $i$ from time $t'$ to time $t$. This also highlights an important fact about the one-point function propagator. If we set $t'=0$, have uncorrelated initial conditions $r_{ij}(0)=\delta_{ij}$, and the system is translationally invariant then
\begin{align} \label{2-pt-tau=0-uncor-second-sec}
    \langle s_{x+i}(t) s_i(0) \rangle = G_{x}(t)\,.
\end{align}
So the one-point function is crucial for understanding the behavior of the two-point function and, as we see in our specific system, the one-point propagator is what really contains data about the non-reciprocity.

\subsection{Dynamics of a single classical spin}
\label{Sec-Single-spin}

As a pedagogical example to develop some intuition, let us discuss the classical dynamics of a single spin which is defined by the Markov chain
\begin{align}
    \begin{aligned}\includegraphics{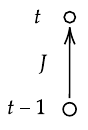}\end{aligned} = P\big(s(t)\big|s(t-\delta t)\big) = \frac{\exp\left[\beta J s(t) s(t-\delta t)\right]}{2\cosh\left[\beta J s(t-\delta t)\right]}\,.
\end{align}
The local magnetic field in this system is $h=\beta J s$. This means that the spin feels its own magnetic field and takes it into account when it updates its configuration; namely, that its configuration at $t-\delta t$ affects its next configuration at time $t$. If $J>0$ the spin will favor alignment with its former configuration, while if $J<0$ it will favor anti-alignment with its former configuration. Notice that this is a minor modification of Glauber's original single spin model \cite{glauber_timedependent_1963}, and in this paper vertical arrows in the Markov chain diagrams will mean self-interactions of this kind. Since $s\in \{\pm1\}$, $\tanh \beta J s = s \tanh \beta J$. Therefore from this Markov chain we  obtain the transition rates
\begin{align}
    w(s \leftarrow - s) = w(- s \leftarrow s) =\frac{\alpha}{2} \big( 1 - \tanh \beta J \big)\,,
\end{align}
and thus the master equation is
\begin{align}
    \partial_t P(s) & = \frac{\alpha}{2} \big( 1 - \tanh \beta J \big) \Big(P(-s) - P(s) \Big).
\end{align}
The only correlator in this system is the one-point function 
\begin{align}
\begin{aligned}
    q(t) & = \sum_s s P(s,t) = P(1,t)-P(-1,t)\,.
\end{aligned}
\end{align}
From the master equation it is clear that the equation of motion for the average value of the spin is
\begin{align}
    \dot q = -\alpha (1 - \tanh \beta J) q\,.
\end{align}
If the system is then initialized in a state $P_0(s)$ such that $q_0 = \sum_s s P_0(s)$, we see that
\begin{align}
    q(t) = q_0 e^{-\alpha (1 - \tanh \beta J)t}\,, \qquad P(s,t) = \frac{1}{2}(1+s q(t))\,.
\end{align}
The self-interaction coefficient $J$ helps to set the relaxation timescale of the spin,
\begin{align}
    \tau_{eq} = \frac{1}{\alpha(1-\tanh \beta J)}\,.
\end{align}
Consider first the ferromagnetic case when $J \geq 0$. At infinite temperature corresponding to $\beta J=0$, the system relaxes to equilibrium at the fastest rate possible $\tau_{eq}=1/\alpha$. As we lower the temperature, the system relaxes slower and slower until at zero temperature, $\beta J\rightarrow \infty$, the system freezes completely and it takes an infinite time to relax to equilibrium that is,  $\tau_{eq} \rightarrow\infty$. In the antiferromagnetic case, $J\leq 0$, the system relaxes to equilibrium slower at infinite temperature and faster at zero temperature: $\beta J\rightarrow - \infty$ sends $\tau_{eq}=1/2\alpha$.

In the next section we  solve a simple system which already has non-reciprocal interactions. This will help us build intuition about the effects of non-reciprocity in this simpler setup.

\subsection{A two spin non-reciprocal system}

\label{Sec-2-Spin}

As a nice example of non-reciprocity, in this section we study the dynamics of two classical spins that are coupled to each other. This toy model is useful for generating intuition about non-reciprocity. We define this model by the Markov chain
\begin{align}
    \begin{aligned}
        \includegraphics{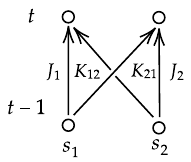}
    \end{aligned} =
    P\big(s(t)\big|s(t-\delta t)\big) = \frac{\exp s_1(t)h_1(t-\delta t)}{2\cosh h_1(t-\delta t)} \times \frac{\exp s_2(t)h_2(t-\delta t)}{2\cosh h_2(t-\delta t)}\,,
\end{align}
where the local magnetic fields felt by spins $s_1$ and $s_2$ are respectively,
\begin{align}
    h_1 = \beta J_1 s_1 + \beta K_{12} s_2\,, \qquad
    h_2 = \beta J_2 s_2 + \beta K_{21} s_1\,.
\end{align}
In this section from now on, we absorb $\beta$ into the definition of the coupling constants to simplify notation. Again we are considering self-interactions given by the couplings $J_1$ and $J_2$ which diagrammatically are represented by the vertical arrows. The coupling between the two spins $K_{12}$ and $K_{21}$ can be split into even and odd parts like (\ref{JeJo}),
\begin{align}
    K_e = \frac{K_{12}+K_{21}}{2}\,, \qquad K_o = \frac{K_{12}-K_{21}}{2} \,,
\end{align}
which respectively encode the reciprocal and non-reciprocal interactions between the spins. The transition rates are
\begin{align}
\begin{aligned}
   & w( s_1 \leftarrow- s_1)  = \frac{\alpha}{2} \big(1+ s_1 \tanh (-J_1 s_1 + K_{12} s_2) \big)\,, \cr & w( - s_1 \leftarrow s_1) = \frac{\alpha}{2} \big(1 - s_1 \tanh (J_1 s_1 + K_{12} s_2) \big) \,, \cr 
   & w( s_2 \leftarrow- s_2)  = \frac{\alpha}{2} \big(1+ s_2 \tanh (-J_2 s_2 + K_{21} s_1) \big)\,, \cr & w( - s_2 \leftarrow s_2) = \frac{\alpha}{2} \big(1 - s_2 \tanh (J_2 s_2 + K_{21} s_1) \big) \,. \nonumber
\end{aligned}
\end{align}
Since $(s_1,s_2)$ can only take the values $\pm 1$, the hyperbolic tangents above simplify to
\begin{align}
\begin{aligned}
    \tanh (J_1 s_1 + K_{12} s_2) & = \frac{s_1 + s_2}{2} \tanh(J_1 + K_{12}) + \frac{(s_1-s_2)}{2} \tanh (J_1 - K_{12})\,, \\
    \tanh (-J_1 s_1 + K_{12} s_2) & = - \frac{s_1 + s_2}{2} \tanh(J_1 - K_{12}) - \frac{(s_1-s_2)}{2} \tanh (J_1 + K_{12})\,.
\end{aligned}
\end{align}
There are similar expressions for the hyperbolic tangent with $h_2$. Therefore we can rewrite the transition rates as
\begin{align}
\begin{aligned}
    w(s_1\leftarrow-s_1) & = \frac{\alpha}{2}\big(1- A_{1}+ s_{1} s_{2} B_{12}\big) \,, \\
    w(- s_{1} \leftarrow s_{1}) & =\frac{\alpha}{2}\big(1-A_{1}- s_{1} s_{2} B_{12} \big)\, , \\
    w(s_2\leftarrow-s_2) & =\frac{\alpha}{2}\big(1-A_{2}+s_{1} s_{2} B_{21}\big) \,, \\
    w(- s_{2} \leftarrow s_{2}) & =\frac{\alpha}{2}\big(1- A_{2} - s_{1} s_{2} B_{21}\big)\,,
\end{aligned}
\end{align}
where
\begin{align}
\begin{aligned}
    A_1 & = \frac{1}{2} \big[\tanh(J_1+ K_{12}) + \tanh(J_1- K_{12})  \big]
    \, , \\
    B_{12} & = \frac{1}{2} \big[\tanh(J_1+ K_{12}) - \tanh(J_1- K_{12})  \big] \, , \\
    A_2 & = \frac{1}{2} \big[\tanh(J_2+ K_{21}) + \tanh(J_2- K_{21})  \big] \, , \\
    B_{21} & = \frac{1}{2} \big[\tanh(J_2+ K_{21}) - \tanh(J_2- K_{21})  \big] \, .
\end{aligned}
\end{align}
Then the master equation is
\begin{align}
\begin{aligned}
    \partial_t P = \frac{\alpha}{2} \Big( &\big[ 1- A_1 +s_1 s_2 B_{12} \big] P(-s_1,s_2)  + \big[ 1 - A_2+s_1 s_2 B_{21} \big] P(s_1,-s_2) \\
    & - \big[ 2 - A_1 - A_2 - s_1 s_2 (B_{12}+B_{21}) \big] P(s_1,s_2) \Big)\,.
\end{aligned}
\end{align}
It is convenient to study three correlators for this system: the $1$-point functions $q_i = \langle s_i(t)\rangle$, the equal-time $2$-point function $r_{ij}(t) = \langle s_i s_j \rangle(t)$ and the unequal time $2$-point function $\langle s_i(t) s_j(t') \rangle$. We start with the $1$-point function. The equations of motion for the $1$-point function are
\begin{align}
\left\{\begin{aligned}
    \dot q_1 & = -\alpha (1-A_1) q_1 + \alpha B_{12} q_2\,, \\[8pt]
    \dot q_2 & = -\alpha (1-A_2) q_2 + \alpha B_{21} q_1\,.
\end{aligned}\right.
\end{align}
After substituting one equation into the other, we can rewrite this system as
\begin{align}
    &\ddot q_1 + \alpha(2 - A_1 - A_2) \dot q_1 + \alpha^2 [(1-A_1)(1-A_2) - B_{12}B_{21}] q_1 = 0\, ,
\end{align}
which are the equations of motion for a damped harmonic oscillator.
The solution to the equations of motion for $q_1(t)$ and $q_2(t)$ given initial conditions $(q_1(0),q_2(0))$ is
\begin{align} \label{2-spin-1-pt-sol}
    &\left\{\begin{aligned}
    & q_1(t) = \left( \cosh \Omega t + \alpha(A_1-A_2)\frac{\sinh \Omega t}{\Omega} \right) e^{-\eta t} q_1(0) + \alpha B_{12} \frac{\sinh \Omega t }{\Omega} e^{-\eta t} \, q_2(0) , \\[8pt]
    & q_2(t) = \left( \cosh \Omega t - \alpha(A_1-A_2)\frac{\sinh \Omega t}{\Omega} \right) e^{-\eta t} q_2(0) + \alpha B_{21} \frac{\sinh \Omega t}{\Omega} e^{-\eta t} \, q_1(0), 
\end{aligned}\right. \\
\end{align}
where
\begin{align} \label{eta-Omega}
    \eta & = \frac{\alpha}{2}(2-A_1 -A_2), \qquad
    \Omega = \alpha \sqrt{\left(\frac{A_1 - A_2}{2} \right)^2 + B_{12} B_{21}}\,.
\end{align}
From these solutions we can read the one-point function propagators
\begin{align}
\begin{aligned}
    G_{11}(t) & = \left( \cosh \Omega t + \alpha(A_1-A_2)\frac{\sinh \Omega t}{\Omega} \right) e^{-\eta t}\,, \\
    G_{12}(t) & = \alpha B_{12} \frac{\sinh \Omega t}{\Omega} e^{-\eta t} \,, \\
    G_{21}(t) & = \alpha B_{21} \frac{\sinh \Omega t}{\Omega}e^{-\eta t} \,, \\
    G_{22}(t) & = \left( \cosh \Omega t - \alpha(A_1-A_2)\frac{\sinh \Omega t}{\Omega} \right) e^{-\eta t}\,.
\end{aligned}
\end{align}
The parameter space of this system has three distinct regimes:
\begin{align} \label{2-spin-3Regimes}
\begin{aligned}
    \left(\frac{A_1 - A_2}{2} \right)^2 + B_{12} B_{21} > 0\,, && &\left\{\begin{aligned}
        &\text{\textbf{overdamped}: Reciprocal interactions} \\ &\text{are larger than non-reciprocal.}
    \end{aligned} \right.\\ \\
    \left(\frac{A_1 - A_2}{2} \right)^2 + B_{12} B_{21} < 0\,, && &\left\{\begin{aligned}
        &\text{\textbf{underdamped}: non-reciprocal interactions} \\ &\text{are larger than reciprocal.}
    \end{aligned}\right. \\ \\
    \left(\frac{A_1 - A_2}{2} \right)^2 + B_{12} B_{21} = 0\,, && &\left\{\begin{aligned}
        &\text{\textbf{critical damping}: non-reciprocal interactions} \\ &\text{cancel reciprocal.}
    \end{aligned}\right.
\end{aligned}
\end{align}
Notice that there are many choices of parameters that can satisfy these conditions and we discuss them later in this section. The first two cases have the same solution (\ref{2-spin-1-pt-sol}). We only need to take an analytic continuation that turns hyperbolic functions into trigonometric ones. The critical damping case, however, has a different solution:
\begin{align}
\left\{\begin{aligned}
    q_1(t) & = \left(1+\alpha\frac{A_1-A_2}{2}\, t\right) e^{-\eta t} q_1(0) + \alpha B_{12} t\, e^{-\eta t}\, q_2(0) , \\[8pt]
    q_2(t) & = \left(1- \alpha\frac{A_1-A_2}{2}\, t\right) e^{-\eta t} q_2(0) + \alpha B_{21} t\, e^{-\eta t}\, q_1(0).
\end{aligned}\right.
\end{align}
Now let us look at the equal-time two-point function. Using (\ref{2-point-eom-general}) we can get the equations of motion of the two-point function
\begin{align} \label{2-spin-2-point-eom}
    \dot r_{12} & = -\alpha(2-A_1-A_2) r_{12} + \alpha(B_{12} + B_{21})\,.
\end{align}
The equation of motion for $r_{21}$ is exactly the same so $r_{12}(t)=r_{21}(t)$, and by definition $r_{11}(t)=r_{22}(t) =1$. The solution to this equation, given an initial condition $r_{12}(0)$, is
\begin{align}
    r_{12}(t) = r_{12}(0) e^{-\alpha (2-A_1 - A_2) t} +\frac{B_{12}+B_{21}}{2-A_1-A_2} \big(1- e^{-\alpha (2-A_1 - A_2 ) t} \big)\,.
\end{align}
Therefore this system has a stationary long-time distribution
\begin{align}
    r_{12}^{ss} = \frac{B_{12}+B_{21}}{2-A_1-A_2}\,.
\end{align}
Finally using equation (\ref{2-point-General}), the unequal time two-point functions are
\begin{align}
\begin{aligned}
    &\langle s_1(t) s_2(t') \rangle = G_{11}(t-t') r_{12}(t') + G_{12}(t-t')\, ,\\
    &\langle s_2(t) s_1(t') \rangle = G_{21}(t-t') + G_{22}(t-t') r_{21}(t')\,, \\
    &\langle s_1(t) s_1(t') \rangle = G_{11}(t-t') + G_{12}(t-t') r_{21}(t')\, ,\\
    &\langle s_2(t) s_2(t') \rangle = G_{21}(t-t') r_{12}(t') + G_{22}(t-t')\,.
\end{aligned}
\end{align}
The last two of these two-point functions are the self-correlation functions, which show us how the spin configuration at time $t$ correlates with the configuration at time $t'$. Notice that when $t=t'$ both functions are a constant which is consistent with 
\begin{align}
    \langle s_i(t) s_i(t) \rangle = \sum_s s_i s_i P(s,t) = \sum_s P(s,t) =1 \,.
\end{align}

Now let us examine the behavior of this system in more detail for some of the different cases. To do so we first notice that our solution (\ref{2-spin-1-pt-sol}) contains terms like
\begin{align}
    e^{-\eta t}\, (e^{\Omega t} \pm e^{-\Omega t})\, ,
\end{align}
where $(\eta, \Omega)$ appear in \C{eta-Omega}. The relaxation rate, $\tau_{eq}$, and oscillation period, $\tau_{osc}$, of the system are therefore
\begin{align}
\begin{aligned}
    \tau_{eq} & = \frac{1}{\eta-\Omega}\,, && \tau_{osc}= \infty\,, && \text{overdamped}\,, \\
    \tau_{eq} & = \frac{1}{\eta}\,, && \tau_{osc}= \infty\,, && \text{critical damping}\,, \\
    \tau_{eq} & = \frac{1}{\eta}\,, && \tau_{osc}= \frac{1}{|\Omega|}\,, && \text{underdamped}\,.
\end{aligned}
\end{align}
So we can understand the divergence of the oscillation time as a feature of the transition from the underdamped to overdamped regimes. We also notice that in the overdamped regime, $\Omega$ can balance out $\eta$ to increase the relaxation time. Since non-reciprocity decreases the value of $\Omega$ until it reaches zero and then becomes imaginary, we conclude that non-reciprocity tends to spoil long-time order. This widely known feature of damped harmonic motion might be the central reason why non-reciprocity seems to spoil order. We also see this effect later in Section \ref{Sec-N-Spins} for a spin chain; non-reciprocity spoiling long-time order in many-body systems was already reported in Ref. \cite{avni2023nonreciprocal}. 

Now we restrict our analysis to some specific cases. We set the self-interaction of both spins to have equal magnitude $J_1=J_2=J$ and split the coupling strength into even and odd parts $K_{12}=K_e+K_o$ and $K_{21}=K_e-K_o$. In this notation if $K_o > 0$, we have
\begin{align}
    & K_o < K_e\,, \qquad \text{$s_1$ wants to strongly align with $s_2$, $s_2$ weakly aligns with $s_1$}\,, \\
    & K_o > K_e\,, \qquad \text{$s_1$ wants to align with $s_2$, $s_2$ anti-aligns with $s_1$}\,.
\end{align}
If $K_o < 0$, the inequality conditions are reversed and we swap $s_1$ and $ s_2$.

Now we turn off the self-interaction $J=0$ which leads to $A_1=A_2=0$. Then we have
\begin{align}
    \eta= 1 \,, \qquad \Omega = \sqrt{\tanh(K_e+K_o)\tanh(K_e-K_o))}\,,
\end{align}
and thus the underdamped, critical damping, and overdamped regimes are given by:
\begin{align}
    \begin{aligned}
        |K_e|> |K_o|\, && &\text{overdamped} \,,\\
        |K_e|=|K_o|\, && &\text{critical damping} \,, \\
        |K_e|<|K_o|\, && &\text{underdamped} \, .
    \end{aligned}
\end{align}
Notice that $0\leq|\Omega|\leq 1$ therefore in the underdamped regime it is not possible to have oscillations that are faster than the relaxation rate of the system. 

In the next case we keep the self interaction, but we turn off the even part of the coupling between the spins $K_e=0$ and we set $J>0$. Then we have
\begin{align} \label{eta-omega-case2}
    &\eta = 1 - \frac{1}{2} \big[\tanh(J+K_o) + \tanh(J-K_o) \big]\,, \cr  &\Omega = \frac{i}{2}\big|\tanh(J+K_o)-\tanh(J-K_o)\big|\,. 
\end{align}
So the system is always underdamped unless we set $K_o=0$ which decouples the spins completely and the problem reduces to two copies of the system discussed in Section \ref{Sec-Single-spin}. Both $\eta$ and $|\Omega|$ are bounded between zero and one, and there is no combination of $(J,K_o)$ that can make $|\Omega| > \eta$. Therefore the system is not able to complete many oscillation cycles before reaching equilibrium. To see this more clearly we restrict ourselves to $K_o=J$ for which
\begin{align}
    \eta = 1 - \frac{1}{2} \tanh 2J \,, \qquad \Omega = \frac{i}{2}\tanh2J\,. 
\end{align}
Therefore at $J=0$ we have $\eta=1$ and $\Omega=0$ and as we increase the interaction strength $\eta$ decreases and $\Omega$ increases until at $J=\infty$ they are given by $\eta=1/2$, and $\Omega=i/2$. Therefore we see clearly that the oscillation scale $\tau_{osc}$ is never larger than the relaxation scale $\tau_{eq}$.

Finally we allow $J\neq0$ and $K_e\neq0$. In this case there is no simplification of $\eta$ and $\Omega$, and thus there is no easy way to determine where in the $(J,K_e,K_o)$ parameter space the transition between regimes happens. The central difference between this and the preceding two cases is that we can now use $J$ and $K_e$ to increase $\tau_{eq}$ and allow for longer lived non-equilibrium states. Indeed depending on how we tune the parameters, the system might evolve towards equilibrium so slowly that it looks as if the system stabilized in a state with nonzero $q_1(t)$ and $q_2(t)$. However again as we increase non-reciprocity, $\Omega$ goes to zero and the system relaxes quickly to equilibrium.

\begin{figure}
    \centering
    \includegraphics[scale=0.87]{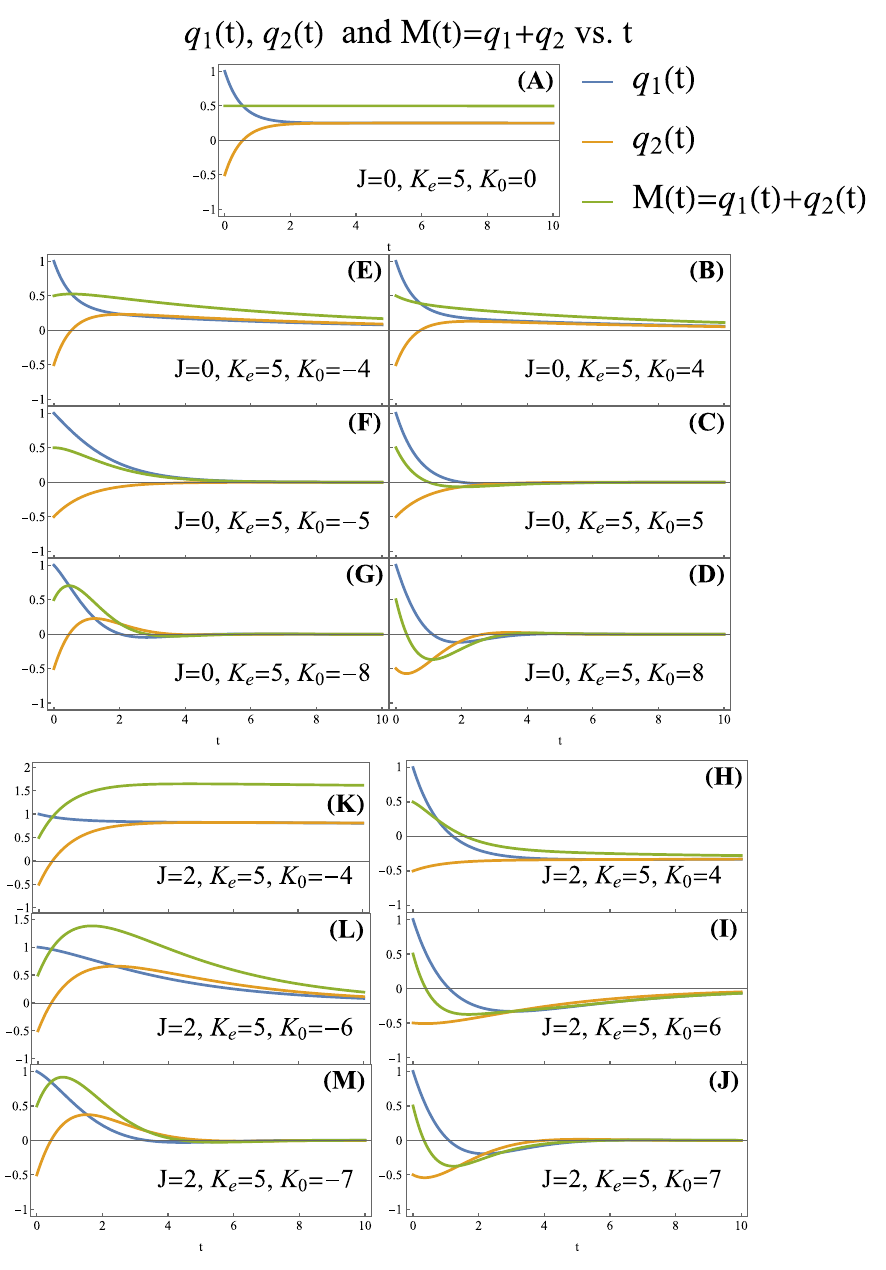}
    \caption{We plot $q_1(t)$, $q_2(t)$ and  $M(t)=q_1(t)+q_2(t)$ as a function of $t$ with initial conditions $q_1(0) = 1$ and $q_2(0)=-0.5$. In plot $\textbf{(A)}$ we study the system for $J=K_o=0$. Because of the initial condition the system rapidly evolves to an intermediate non-equilibrium state with $q_1=q_2=0.25$ and then slowly decays to equilibrium. In $\textbf{(B-D)}$ and $\textbf{(E-G)}$ we turn on non-reciprocity with $K_o>0$ and $K_o<0$, respectively. In $\textbf{(H-J)}$ and $\textbf{(K-M)}$ we turn on both self-interaction and non-reciprocity with $K_o>0$ and $K_o<0$, respectively. As we see, non-reciprocity always tends to spoil non-equilibrium states which are evolving slowly to equilibrium. Another interesting feature seen from $\textbf{(K)}$ and $\textbf{(H)}$ is that by turning on self-interaction and a bit of non-reciprocity, we can shift the value of $M$ for the long-lived intermediate non-equilibrium state from what we see in $\textbf{(A)}$.}
    \label{Fig-2-Spin-Plot}
\end{figure}

The discussion in the last few paragraphs is made explicit in Figure \ref{Fig-2-Spin-Plot}. There we plot $q_1(t)$, $q_2(t)$ and $M(t)=q_1(t)+q_2(t)$ with the initial condition $q_1(0)=1$ and $q_2(0)=-0.5$ for different values of $(J,K_e,K_o)$. In Figure  \hyperref[Fig-2-Spin-Plot]{\ref*{Fig-2-Spin-Plot}(A)} we study the system with $J=K_o=0$. Because of the initial conditions, we see that the system rapidly evolves to a non-equilibrium state with $q_1=q_2=0.25$ and then very slowly evolves to equilibrium. Then in Figures \hyperref[Fig-2-Spin-Plot]{\ref*{Fig-2-Spin-Plot}(B-D)} and \hyperref[Fig-2-Spin-Plot]{\ref*{Fig-2-Spin-Plot}(E-G)} we turn on non-reciprocity with $K_o>0$ and $K_o<0$, respectively. As we see, non-reciprocity spoils the slowly decaying non-equilibrium state. In Figures \hyperref[Fig-2-Spin-Plot]{\ref*{Fig-2-Spin-Plot}(H-M)} we turn on the self-interaction $J$ and non-reciprocity with either $K_o>0$ or $K_o<0$ for \hyperref[Fig-2-Spin-Plot]{\ref*{Fig-2-Spin-Plot}(H-J)} or \hyperref[Fig-2-Spin-Plot]{\ref*{Fig-2-Spin-Plot}(K-M)}, respectively. The self-interaction together with the non-reciprocity allows the initial state of one of the spins to dominate the dynamics. In Figure \hyperref[Fig-2-Spin-Plot]{\ref*{Fig-2-Spin-Plot}(H)} we see that $K_e>0$ makes the initial value of $q_2(t)$ dominate the intermediate time dynamics, while in Figure \hyperref[Fig-2-Spin-Plot]{\ref*{Fig-2-Spin-Plot}(K)} we see that $K_e<0$ makes the initial value of $q_1(t)$ dominate the dynamics.

\section{Non-Reciprocal $N$-Spin System}
\label{Sec-N-Spins}

\begin{figure}
    \centering
    \includegraphics[width=1\linewidth]{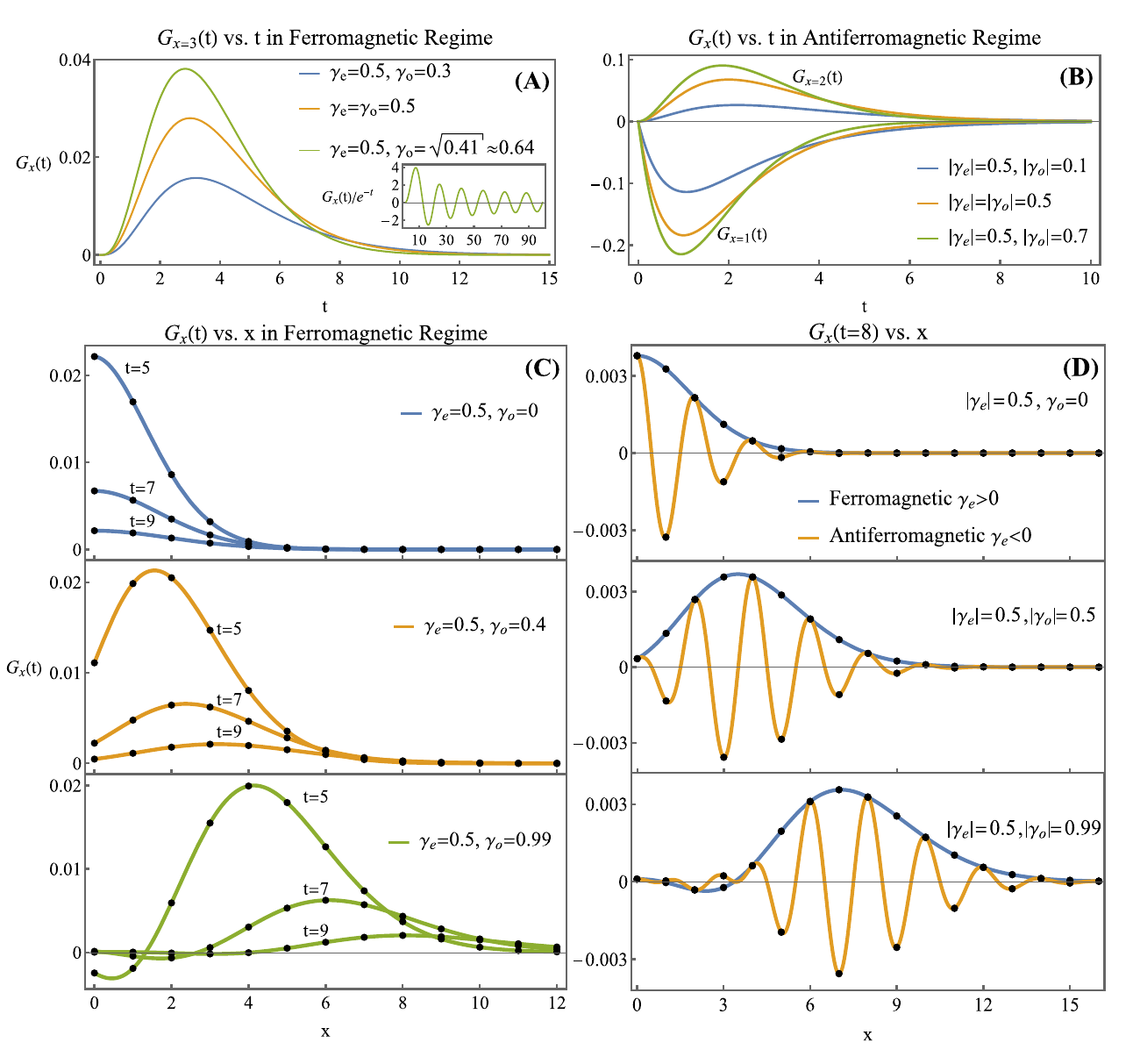}
    \caption{In \textbf{(A)} and \textbf{(B)} we plot $G_x(t) = \langle s_{x+i}(t) s_i(0)\rangle$ of equations (\ref{1-point-propagator}) and (\ref{2-pt-tau=0-uncor-second-sec}) as a function of time for fixed $x$ both in the ferromagnetic regime $\gamma_e>0$ in \textbf{(A)} and the antiferromagnetic regime $\gamma_e<0$ in \textbf{(B)}. In both cases we fix the reciprocal interaction to $|\gamma_e|=0.5$ and change $\gamma_o$ to show the overdamped $|\gamma_e|>|\gamma_o|$, the critical damping $|\gamma_e|=|\gamma_o|$, and the underdamped regime $|\gamma_e|<|\gamma_o|$. The inset in \textbf{(A)} shows that in the underdamped regime the correlation function oscillates because of the $J_x(t)$ Bessel function of (\ref{1-point-propagator}). In the underdamped regime the correlation has a higher peak but also falls faster. In panel \textbf{(B)} we see the same behavior for the antiferromagnetic system as we saw in \textbf{(A)} but when $x=\text{even}$ spins correlate and anticorrelate from $x=\text{odd}$. In \textbf{(C)} and \textbf{(D)}, we plot $G_x(t)$ as a function of $x$ for some fixed values of $t$. We start at zero non-reciprocity on the top and increase the non-reciprocal coupling in the subsequent plots. Even though the system is discrete in the variable $x$, we choose to analytically continue the plots for better visualization. As we increase the non-reciprocity, the correlation forms a wave packet that travels with velocity $\gamma_o$. This feature, a bit unclear from expression (\ref{1-point-propagator}), is made evident by the low-energy behavior of the system analyzed in Section~\ref{Sec-Two-Point}.
    There we see that the non-reciprocity changes the equilibrium correlations by adding a wavepacket motion with the transformation $x \rightarrow x+\gamma_o t$. These results are captured in the plots \textbf{(C)} and \textbf{(D)}. Also notice that in \textbf{(D)}, the $G_x(t)$ for the antiferromagnetic spin chain is exactly enveloped by the negative of the $G_x(t)$ function for the ferromagnetic spin chain.}
    \label{Gx-Infinite}
\end{figure}

Having introduced the formalism and discussed simple examples in Section \ref{Sec-Markov-Chain}, we proceed to the central model of our paper. We consider a system of $N$ spins in a one-dimensional lattice with lattice spacing $a$. We define the dynamical behavior of this system by the Markov chain,
\begin{align} \label{N-spin-NR-Markov}
    \begin{aligned}
        \includegraphics{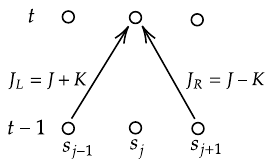}\end{aligned} = P\big(s(t)\big|s(t-\delta t)\big) = \prod_{j=1}^N \frac{\exp\left[ s_j(t) h_j(t-\delta t)\right]}{2\cosh\left[ h_j(t-\delta t)\right]} \, ,
\end{align}
where the local magnetic field is
\begin{align}
    h_j = \beta J\big(s_{j-1} + s_{j+1}\big) + \beta K\big(s_{j-1} - s_{j+1}\big)\,.
\end{align}
The hyperbolic tangent of the local magnetic field can be written in the form
\begin{align} \label{tanh-in-NR}
    \tanh h_j = \frac{\gamma_e}{2} \big(s_{j-1}+s_{j+1}\big) + \frac{\gamma_o}{2} \big( s_{j-1}-s_{j+1} \big)\,,
\end{align}
where $\gamma_e = \tanh 2\beta J$ and $\gamma_o = \tanh 2\beta K$. The master equation is then
\begin{align}
\begin{aligned}
    \partial_t P(s) = \frac{\alpha}{2} \sum_{j}\Big[&\left(1+\frac{\gamma_{e}}{2} s_{j}\left(s_{j-1}+s_{j+1}\right)+\frac{\gamma_{o}}{2} s_{j}\left(s_{j-1}-s_{j+1}\right)\right) P(-s_j) \\
    &- \left(1-\frac{\gamma_{e}}{2} s_{j}\left(s_{j-1}+s_{j+1}\right)-\frac{\gamma_{o}}{2} s_{j}\left(s_{j-1}-s_{j+1}\right) \right) P(s_j)\Big]\, ,
\end{aligned}
\end{align}
where $\alpha$ is again the characteristic time for a single spin to flip. Using (\ref{1-point-eom-h}) we can easily obtain the equation of motion for the $1$-point function:
\begin{align}  \label{1-point-eom-h-NR}
    \del_t q_j = -\alpha q_j + \frac{\alpha \gamma_e}{2}(q_{j-1} + q_{j+1}) + \frac{\alpha \gamma_o}{2}(q_{j-1} - q_{j+1})\,.
\end{align}
We want to solve this equation for three different situations: an infinitely long chain, and a finite chain with both periodic and open boundary conditions. In this section we first solve for the periodic chain and then take $N\rightarrow \infty$ to get the infinite chain. In Section \ref{Sec-Open-Spin} we use the method of images to get the solution for the open chain. We first decompose the system into its Fourier modes,
\begin{align}
    q_j(t) & = \sum_k b_k(t) e^{i ka j }\,, \qquad b_k(t) = \frac{1}{N}\sum_{j=1}^{ N} q_j(t) e^{-i kaj }\,,
\end{align}
where the periodic boundary conditions require $k = \frac{2\pi m}{N a}$ with $m=0,1,\ldots,N-1$ and $a$ is the lattice spacing. The equations of motion for the Fourier modes are given by,
\begin{align}
    \dot b_k = - \alpha\left( 1 -  \gamma_e \cos ka + i\gamma_o \sin ka \right) b_k\, .
\end{align}
These equations are trivially solved and we can invert the Fourier transformation to get
\begin{align} \label{1-pt-sol-Fourier}
    q_j(t) & = e^{-\alpha t }\sum_{l=1}^{N} q_l(0) \sum_k \frac{1}{N} \exp\Big[ika(j-l) + \alpha t \big( \gamma_e \cos ka - i \gamma_e \sin ka\big) \Big]\,.
\end{align}
It is possible to explicitly evaluate the sum over Fourier modes above and rewrite it in terms of Bessel functions
\begin{align}
    \sum_{m=0}^{N} \frac{1}{N} \exp\left[i\frac{2\pi}{N} m p + x \cos \frac{2\pi}{N} m - i y \sin \frac{2\pi}{N} m \right] = \sum_{n,s=-\infty}^\infty (-1)^s I_{p+s+n N}(x) J_s(y) \,.
\end{align}
This identity, which is an extension of an earlier result by Sung and Hovden \cite{sung_infinite_2022}, is proven in the beginning of Appendix \ref{Appendix-Bessel} using the Jacobi-Anger identities \cite{Watson-Tretesie-1922}. Therefore the exact solution for the $1$-point function is
\begin{align}
    \label{exact-1pt}
    q_j(t) & = e^{-\alpha t} \sum_{l=1}^{ N } q_l(0) \sum_{n,s=-\infty}^{\infty} (-1)^s I_{j-l+s+n N}\big(\alpha \gamma_e t\big) J_s\big(\alpha \gamma_o t\big)\,.
\end{align}
The product of the exponential and the Bessel function in the summation on $s$ and $n$ completely describes the dynamical behavior of our system. This product defines the one-point function propagator of the theory discussed in \C{1-pt-Prop-def}. As we see, this same function is used to build all the other equal-time $n$-point functions. We can further simplify the summation using Graff's generalization of Neumann's addition theorem \cite{Watson-Tretesie-1922},
\begin{align} \label{Graff-Neumann}
    &\sum_{s=-\infty}^\infty J_{s+p}(Z) J_s(z) e^{is \phi} = J_p(\omega)  \left(\frac{Z-z e^{-i\phi}}{Z-z e^{i\phi}} \right)^{\frac{p}{2}}\,,
\end{align}
where $\omega = \left(Z^2+z^2 - 2Z z \cos \phi\right)^{1/2}$. It is important to notice that this result only holds if $|Z| > |ze^{\pm i\phi}|$. Therefore this summation must be treated separately for the $|\gamma_e|>|\gamma_o|$ and $|\gamma_e|<|\gamma_o|$. We can get the case $|\gamma_e|=|\gamma_o|$ via analytic continuation from the other two cases. So in the infinite spin chain case, the one-point function can be written in the compact form
\begin{align} \label{1-pt-soul-Green}
    q_j(t) = \sum_{l} q_l(0) G_{j-l}(t) \,,
\end{align}
where the one-point function propagator is given by
\begin{align} \label{1-point-propagator}
    G_x(t) = \left\{ \begin{aligned}
        &e^{-\alpha t} I_{x}\big( \sqrt{\gamma_e^2-\gamma_o^2} \alpha t\big) \left( \frac{\gamma_e+\gamma_o}{\gamma_e-\gamma_o}\right)^{\frac{x}{2}} && |\gamma_e| > |\gamma_o| \\ &\left\{\begin{aligned}
            &e^{-\alpha t} \frac{(\gamma \alpha t)^{|x|}}{|x|!} && x\geq 0 \\
            &0 && x<0
        \end{aligned}\right. && \gamma_e = \gamma_o = \gamma \\
        &\left\{\begin{aligned}
            &0 && x>0 \\
            &e^{-\alpha t} \frac{(\gamma \alpha t)^{|x|}}{|x|!} && x\leq0
        \end{aligned}\right. && \gamma_e = - \gamma_o = \gamma \\
        &e^{-\alpha t} J_{x}\big( \sqrt{\gamma_o^2-\gamma_e^2} \alpha t\big) \left( \frac{\gamma_e+\gamma_o}{\gamma_o-\gamma_e}\right)^{\frac{x}{2}} && |\gamma_e| < |\gamma_o|
    \end{aligned} \right..
\end{align}
Notice that these three expressions are connected via analytic continuation. For periodic boundary conditions we just use (\ref{1-pt-soul-Green}) with\footnote{One might be concerned about whether the sum in \C{b.c.-propagator} converges. To see convergence for the critical damping case, we can use Stirling's approximation for the gamma function: $x! \sim \sqrt{2\pi x} \left( \frac{x}{e} \right)^x$ . The gamma function dominates over the numerator which grows like $t^{|x|}$ for large $x$ guaranteeing convergence. This can be checked explicitly by numerically integrating $G_x$ from some large $x$ to infinity. A very similar argument works for $J_x$ which is also gamma function suppressed at large $|x|$. The most interesting case involves $I_x$. Convergence at fixed $t$ follows from $I_x(z) = i^{-x} J_x(iz)$. The difference is that $I_x(z)$ grows like $e^z/\sqrt{z}$ at fixed $x$. For this Green's function, note that $\left(\g_e^2 - \g_o^2\right)^{1/2} \leq 1$ so the growth of $I_x$ at large $t$ is never larger than the overall $e^{-\alpha t}$ prefactor.}
\begin{align} \label{b.c.-propagator}
    G^{N,per}_x (t) & = \sum_{n=-\infty}^\infty G_{x+n N} (t)\, .
\end{align}
Notice that the three cases of (\ref{1-point-propagator}) are analogous to the overdamped, critical damped and underdamped cases that we saw in (\ref{2-spin-3Regimes}). That is, 
\begin{align} \label{N-Spin-OCU}
    \begin{aligned}
        |\gamma_e|>|\gamma_o|, \qquad &\text{overdamped} \,,\\
        |\gamma_e|=|\gamma_o|, \qquad &\text{critical damped} \,, \\
        |\gamma_e|<|\gamma_o|, \qquad &\text{underdamped} \, .
    \end{aligned}
\end{align}
We want to highlight that the critical damping regime is in an $N^{th}$-order exceptional point~\cite{ashida_non-hermitian_2020}. To see this, notice that equation (\ref{1-point-eom-h-NR}) at critical damping is
\begin{align}
\begin{aligned}
    \partial_t q_j & = - \alpha q_j + \alpha \gamma q_{j-1}\,, \qquad \gamma_e = \gamma_o = \gamma\,, \\
    \partial_t q_j & = - \alpha q_j + \alpha \gamma q_{j+1}\,, \qquad \gamma_e = - \gamma_o = \gamma\,,
\end{aligned}
\end{align}
which is already in Jordan normal form with a single eigenvector. We discuss exceptional points further in Section \ref{Sec-Random-Couplings}. In Figure \ref{Gx-Infinite} we explore the function $G_x(t)$ of (\ref{1-point-propagator}) for the underdamped, critical damped and overdamped limits both for ferromagnetic chains, $\gamma_e>0$, and for antiferromagnetic chains, $\gamma_e<0$. We plot $G_x(t)$ as a function of time for fixed $x$ and also as a function of $x$ for fixed $t$. In this latter case $x$ is discrete but we choose to analytically continue $G_x(t)$ for clarity.

We highlight some important features: Figure \hyperref[Gx-Infinite]{\ref*{Gx-Infinite}(A)} and \hyperref[Gx-Infinite]{\ref*{Gx-Infinite}(B)} show that increasing the non-reciprocity, $\gamma_o$, does not lead to an increase in the timescale over which the system is correlated. Instead these plots suggest that as we increase the non-reciprocity, the correlations decay a little faster at fixed $x$ even though the peak of the correlation is higher. So Figures \hyperref[Gx-Infinite]{\ref*{Gx-Infinite}(A)} and \hyperref[Gx-Infinite]{\ref*{Gx-Infinite}(B)} provide evidence that non-reciprocity tends to equilibrate the system a little bit faster and thus spoils hopes of long-time order. This is a feature observed in other works \cite{avni2023nonreciprocal,ghimenti_sampling_2023}. In Section \ref{Sec-Two-Point}, we analyze the low-energy behavior of this system and we see that the relaxation time $\tau_{eq}$, the correlation length $\xi_{eq}$ and the gap $\mu^2$ only depend on the reciprocal coupling $\gamma_e$ as shown in~(\ref{Gap-etal}); there is only long-range order at $T=0$. The low-energy analysis turns out to be insensitive to the non-reciprocity and does not teach us about how the correlation time changes which we can see explicitly in the plots.  

Figures \hyperref[Gx-Infinite]{\ref*{Gx-Infinite}(C)} and \hyperref[Gx-Infinite]{\ref*{Gx-Infinite}(D)} indicate that when we turn on non-reciprocity the peak correlation length tends to propagate in time as a wave packet. This behavior is confirmed and made explicit by the low-energy analysis. As we will see in Section \ref{Sec-Two-Point}, the low-energy behavior seen in the purely reciprocal kinetic Ising model is changed by the transformation,
\begin{align}
    x \rightarrow x+\gamma_o t\, ,
\end{align}
when we turn on non-reciprocity. This simple modification is valid for short times, $t\ll \tau_{eq}$. A more involved modification valid for all times can be found in equations (\ref{2-pt-lim2-eq}) and (\ref{2-pt-Porod-Equilibrium}).

Now let us turn to the equal-time two-point function and, subsequently, the equal time $n$-point functions. The main feature we are going to see is that $G_x(t)$ given by (\ref{1-point-propagator}) is the central building block of all equal-time $n$-point functions, and therefore of the whole system. Plugging (\ref{tanh-in-NR}) into (\ref{2-point-eom-general}) gives
\begin{align} \label{2-point-eom}
    \frac{d}{dt}{r}_{i j}= &-2 \alpha r_{i j}+\frac{\alpha \gamma_e}{2}\left(r_{i-1 j}+r_{i +1 j}+r_{i j-1}+r_{i j+1}\right) \cr & +\frac{\alpha \gamma_o}{2}\left(r_{i-1 j}-r_{i+1 j}+r_{i j-1}-r_{i j+1}\right)\,.
\end{align}
As we discussed before, this is an inhomogeneous differential equation because $r_{ii} = 1$. Therefore the equations of motion for $r_{ii}$, $r_{i-1i}$, $r_{i+1i}$, $r_{ii-1}$ and $r_{ii+1}$ have constant terms and are inhomogeneous, making the whole system inhomogeneous. To solve the system, we obtain a particular solution satisfying $r_{ii}^{part}=1$ and then add to it a solution of the homogeneous equation with the condition $r_{ii}^{hom}=0$. To get the particular solution for this system, we look for the equilibrium distribution of the system satisfying
\begin{align} \label{2-pt-eom-Stationary}
    2 \alpha r_{i j} = \frac{\alpha \gamma_e}{2}\left(r_{i-1 j}+r_{i +1 j}+r_{i j-1}+r_{i j+1}\right)+\frac{\alpha \gamma_o}{2}\left(r_{i-1 j}-r_{i+1 j}+r_{i j-1}-r_{i j+1}\right)\,.
\end{align}
To solve this, we can use the ansatz $r_{ij}^{eq} = \zeta^{|i-j|}$ which assumes translational invariance in the equilibrium state. This assumption gets rid of the non-reciprocal term and we get the equation
\begin{align}
    2 = \gamma_e (\zeta^{-1}+\zeta)\,,
\end{align}
which gives $\zeta = \gamma_e\inv \left[1-(1-\gamma_e)^{1/2}\right]$. The equilibrium two-point function is therefore
\begin{align} \label{equilibrium-sol}
    r_{ij}^{eq} = \left(\tanh \beta J \right)^{|i-j|}.
\end{align}
To solve the homogeneous equation we again decompose the system into its Fourier modes, but now with two indices:
\begin{align}
    & r_{i j}(t)=\sum_{k q} b_{k q}(t) e^{i\left(k a i+q a j \right)}\,.
\end{align}
The equation of motion for each Fourier mode is given by,
\begin{align}
    \dot{b}_{k q}= -\alpha \Big(2 - \gamma_e(\cos k a+\cos q a)+ \gamma_o i(\sin k a+\sin q a)\Big) b_{k q}\,.
\end{align}
Much of the analysis proceeds exactly as before and the solution to the homogeneous equation is then
\begin{align}
\begin{aligned}
     r_{ij}(t) = e^{-2\alpha t} \sum_{m,n=1}^{N} r_{mn}(0)& \sum_{s,\nu=-\infty}^\infty (-1)^s I_{i-m+s+\nu N}(\alpha \gamma_e t) J_s(\alpha \gamma_o t) \\
     &\times \sum_{r,\mu=-\infty}^\infty (-1)^r I_{j-n+r+\mu N}(\alpha \gamma_e t) J_r(\alpha \gamma_o t)\,.
\end{aligned}
\end{align}
The product of Bessel functions under the summation signs are exactly the one-point propagator (\ref{1-point-propagator}) that we obtained earlier. Therefore this expression is just
\begin{align} \label{2-point-hom-sol}
    r_{ij}(t) = \sum_{m,n=-\infty}^{\infty} r_{mn}(0) G_{i-m}(t) G_{j-n}(t)\,,
\end{align}
where we dropped the summations on $\mu$ and $\nu$ which made the system periodic and take the number of sites $N$ to infinity; thus for the remainder of this section we be look at the infinite chain. This is a homogeneous solution of the problem. To get the full solution of (\ref{2-point-eom}) that satisfies $r_{ii}=1$ we must combine the particular solution (\ref{equilibrium-sol}) which satisfies $r_{ii}^{eq}=1$ with a homogeneous solution satisfying $r_{ii}^{hom}=0$. To do this we antisymmetrize (\ref{2-point-hom-sol}) and get
\begin{align} \label{equal-two-point-sol}
    r_{ij}(t) = \zeta^{|i-j|} + \sum_{m,n=-\infty}^{ \infty} \Big(r_{mn}(0) - \zeta^{|m-n|} \Big)\Big( G_{i-m}(t) G_{j-n}(t) - G_{i-n}(t) G_{j-m}(t) \Big)\,.
\end{align}
We can also write this as
\begin{align}
    r_{ij}(t) = \sum_{m,n=-\infty}^{\infty} r_{mn}(0) G^{(2)}_{i-m,j-n}\,,
\end{align}
where the two-point propagator is
\begin{align} \label{2-point-propagator-sol}
    G^{(2)}_{i-m,j-n}(t) = G^{(2)hom}_{i-m,j-n}(t) + \big(\delta_{i-m}\delta_{j-n} - G^{(2)hom}_{i-m,j-n}(t) \big) \frac{\zeta^{|m-n|}}{r_{mn}(0)}\,,
\end{align}
and $G^{(2)hom}_{i-m,j-n}(t) = G_{i-m}(t) G_{j-n}(t) - G_{i-n}(t) G_{j-m}(t)$ is the homogeneous two-point propagator which satisfies the condition $G^{(2)hom}_{i-m,i-n}(t)=0$.
The fact that the two-point propagator can be expressed in terms of one-point propagators is a feature that is also true for higher point propagators. To check this notice that the equations of motion for all the other equal-time $n$-point functions follow the same structure as we saw for these two cases:
\begin{align} \label{n-point-eom-h}
    \dot r_{j_n\ldots j_1}
     = \alpha \sum_{l=1}^n \Big[ -r_{j_1\ldots j_l\ldots j_n}& + \frac{\gamma_e}{2} (r_{j_n\ldots j_{l}-1\ldots j_1}+r_{j_n\ldots j_{l}+1 \ldots j_1}) \cr &+ \frac{\gamma_o}{2} (r_{j_n \ldots j_{l}-1 \ldots j_1}-r_{j_n \ldots j_{l}+1 \ldots j_1}) \Big]\,.
\end{align}
Using the Fourier decomposition,
\begin{align}
    r_{j_n \ldots j_1}(t) = \sum_{k_n \ldots k_1} b_{k_n \ldots k_1}(t) e^{i\sum_{l=1}^n(k_l a j_l)}\,,
\end{align}
we find the equation of motion for the Fourier modes:
\begin{align}
    \dot b_{k_n \ldots k_1} = - \alpha \left[ \sum_{j=1}^n \big( 1 - \gamma_e \cos k_l a + i \gamma_o \sin k_l a \big) \right] b_{k_n \ldots k_1} \,.
\end{align}
Following the same steps as before we arrive at a close expression for the solution of the homogeneous part of the $n$-point function,
\begin{align} \label{n-point-sol-hom}
    r_{j_n \ldots j_1}(t) & = \sum_{j_n' \ldots j_1'} r_{j_n' \ldots j_1'}(0) \prod_{l=1}^n G_{j_l-j_l'}(t)\,.
\end{align}
To get the full solution we now must account for the inhomogeneous part of this system. The inhomogeneity arises when some of the $n$ points coincide. This process is involved and we will do not perform it for general equal-time $n$-point function. Nonetheless, the structure of the homogeneous solution (\ref{n-point-sol-hom}) makes it clear that $G_x (t)$ is the central building block of all equal-time $n$-point functions and therefore of all $n$-point functions and of the system as a whole.

\section{Open and Periodic Boundary Conditions}

\label{Sec-Finite-Spin-Chain}

We can also compute the $n$-point functions for our system through the method of generating functionals as motivated by Glauber \cite{glauber_timedependent_1963}. This method is relatively simple, allowing us to solve for the $n$-point function values using some function identities and providing geometric intuition for how the system behaves and how to apply different boundary conditions to the system. As we saw in equations (\ref{n-point-eom-h}) and (\ref{n-point-sol-hom}), the building block of all $n$-point functions are the one-point functions, and so we focus on solving (\ref{1-point-eom-h-NR}).

We take a few cases with different topology encoded in the choice of boundary conditions as examples to understand this method and its applicability to various lattice problems. To plot these functions, we take the Graf-Neumann simplified function (\ref{1-point-propagator}) for the infinite correlator case. For the case of a finite chain, we truncate the infinite sums of (\ref{exact-1pt}) as follows, 
\begin{align}
    \label{truncated-1pt}
    q_j(t) & = e^{-\alpha t} \sum_{l=1}^{ N } q_l(0) \sum_{n=-L_n}^{L_n}\sum_{s=-L_s}^{L_s} (-1)^s I_{j-l+s+n N}\big(\alpha \gamma_e t\big) J_s\big(\alpha \gamma_o t\big)\,,
\end{align}
with $L_s$ and $L_n$ at least 25, so that the truncation error is small for the specific choices of system size $N$ that we examine. 

\subsection{The infinite chain} \label{genfunc_infchain}
We first start with a 1D lattice of infinite size with general $\gamma_e,\gamma_o$ coupling strengths, so \C{1-point-eom-h-NR} gives the equation of motion for the one-point function,
\begin{align}
    \label{1pt-DE}
    \del_t q_j(t) = -\alpha q_j + \frac{\alpha \gamma_e}{2}(q_{j-1}+q_{j+1}) + \frac{\alpha \gamma_o}{2}(q_{j-1}-q_{j+1})\, .
\end{align}
We now define the generating functional for one-point functions to simplify this infinite set of coupled equations,
\begin{align}
    F(\lambda,t) = \sum_{n=-\infty}^\infty \lambda^n q_n(t) \, .
\end{align}
If we now sum equation \C{1pt-DE} over all indices with different powers of $\lambda$, it gives us a single equation for the generating functional which we can trivially solve,
\begin{align}
    \del_t F(\lambda,t) &= -\alpha F(\lambda,t) + \frac{\alpha \gamma_e}{2}(\lambda + \lambda\inv)F(\lambda,t) + \frac{\alpha \gamma_o}{2}(\lambda - \lambda\inv)F(\lambda,t) \,,
\end{align}
by integration to get:
\begin{align}
    F(\lambda,t) &= F(\lambda,0)\exp\left(-\alpha t + \frac{\alpha \gamma_e}{2}(\lambda + \lambda\inv)t + \frac{\alpha \gamma_o}{2}(\lambda - \lambda\inv)t\right)\,.
\end{align}
Now applying Bessel function identities, we can simplify this exponential where the $\gamma_e$ term gives an $I_n$ summation and the $\gamma_o$ term gives a $J_n$ summation from \C{besselI} and \C{besselJ}, 
\begin{align}
    \label{gen_func_eom}
    F(\lambda,t) = F(\lambda,0)\exp(-\alpha t) \left[\sum_{n=-\infty}^\infty \lambda^n I_n(\gamma_e \alpha t) \right]\left[\sum_{m=-\infty}^\infty 
    \lambda^m J_m(\gamma_o \alpha t) \right]\,,
\end{align}
which gives the $\lambda^k$ term,
\begin{align}
    q_k(t) &= e^{-\alpha t}\sum_{l=-\infty}^\infty q_l(0)\sum_{n=-\infty}^\infty I_{k-l+n}(\gamma_e \alpha t)(-1)^{n}J_{n}(\gamma_o \alpha t)\,.
\end{align}
This gives the same 1-point function value as we found using Fourier transforms and will serve as a foundation to understand other interesting boundary cases.
\begin{figure}
    \centering
    \includegraphics[scale=0.5]{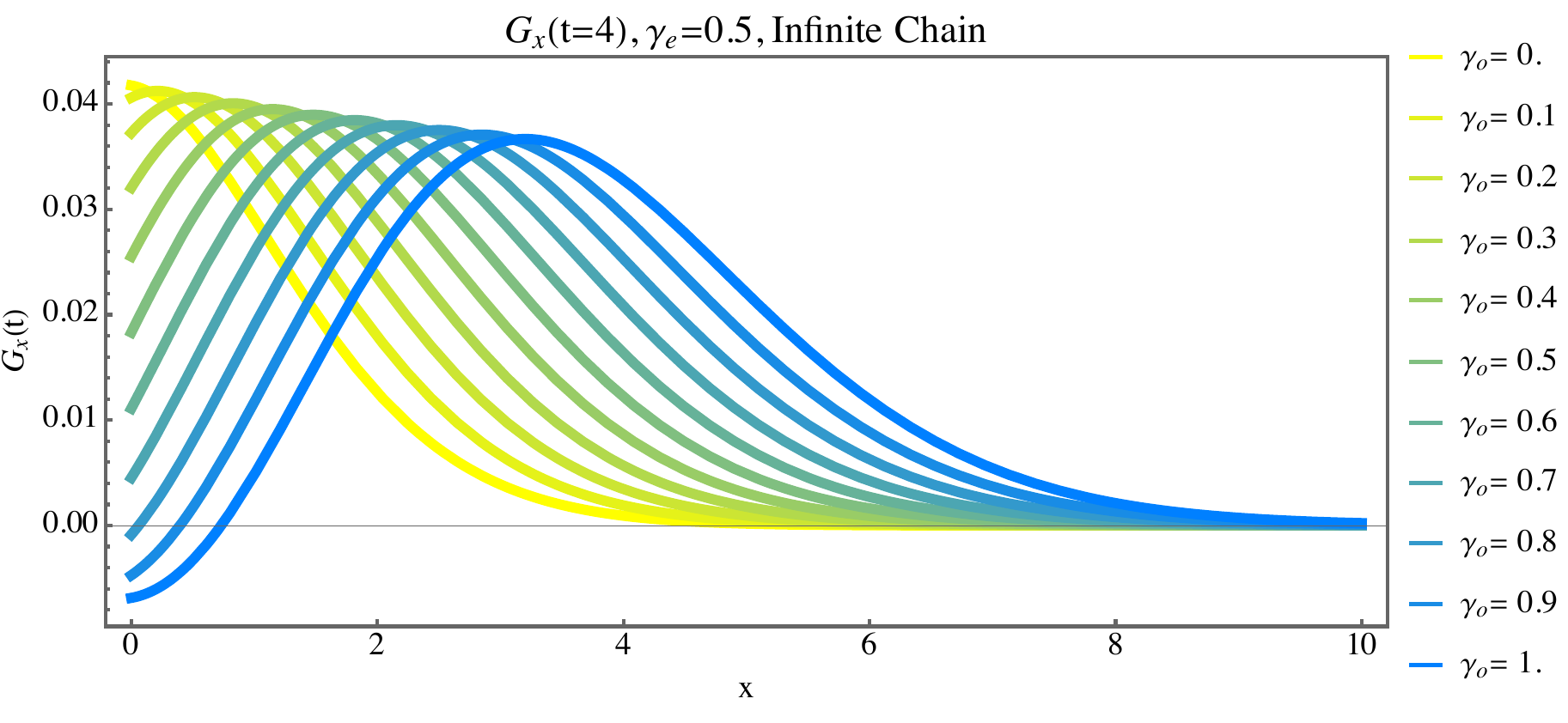}
    \caption{We plot different $G_x(t=4) = \langle s_{x+i}(4) s_i(0)\rangle$ as functions of $x$ for fixed $t=4$ while varying $\gamma_o \in [0,1]$ at fixed $\gamma_e=0.5$ given initial condition $s_0(0) =1$. Smaller $\gamma_o$ values are plotted in yellow, with larger $\gamma_o$ colors shown in green and then blue. Here we note that $\gamma_o\neq 0$ results in the excitation of the system moving in time, similar to the plots in Figure \ref{Gx-Infinite}. Larger $\gamma_o$ values result in this excitation traveling further along the chain by time $t=4$.}
    \label{Inf_ge-go}
\end{figure}
\par  Figure \ref{Inf_ge-go} contains a plot of this function. We see how larger $\gamma_o$ values allow for the excitation to move faster within the system. For $\gamma_o=0$ (a purely reciprocal system), exciting the spin at site $k=0$ results in an excitation that stays localized around zero and thermalizes over time. However, when $\gamma_o \neq 0$ we instead see that the peak of our excitation moves within the chain as a result of non-reciprocity while thermalizing. 

We further note that for larger $\gamma_o$ values, $q_0(t)$ is negative at later times. To explain this let us rewrite \C{1pt-DE} as
\begin{align}
    \del_t q_j(t) = -\alpha q_j + \frac{\alpha (\gamma_e +\gamma_o)}{2}q_{j-1} + \frac{\alpha (\gamma_e - \gamma_o)}{2}q_{j+1}\, .
\end{align}
As we increase $\gamma_o$ with fixed $\gamma_e$ we see that the coupling to the spins on the right decreases until we reach $\gamma_e-\gamma_o=0$ where information only travels from the left to the right. Beyond that point when $\gamma_e-\gamma_o<0$, the spins want to anti-align with the spin on their right. This means that while spin $s_j$ interacts ferromagnetically with $s_{j-1}$, it interacts antiferromagnetically with $s_{j+1}$ and thus we get negative correlations traveling to the left.

Imposing different boundary conditions on this behavior then creates interesting phenomena because the motion of the moving excitations is now constrained. 

\subsection{The periodic spin chain}

We take a periodic spin chain of size $N$ with the same local magnetic field as above but with periodicity in our lattice indexing. First note that we can compute this in the same way as we did for $G^{N,\text{per}}_x(t)$ in \C{b.c.-propagator}, namely, by taking our infinite chain result and summing over translates to impose periodicity. However, we can also do this by defining our generating functional as follows:
\begin{align}
    F(\lambda,t) = \sum_{n=1}^{N} \lambda^n q_n(t)\,,
\end{align}
where we apply periodicity by requiring $\lambda^N =1$. From the different indexing, we have the following 3 equations for $k= 1$, $1 <k < N$, and $k=N$, respectively:
\begin{align}
\left\{\begin{aligned}
    \frac{1}{\alpha}\frac{d}{dt}q_1(t) &= -q_1(t) + \frac{\gamma_e}{2}(q_{2}(t) + q_{N}(t)) + \frac{\gamma_o}{2}(q_{2}(t) - q_{N}(t))\,,\\
    \frac{1}{\a}\frac{d}{ dt}q_k(t) &= -q_k(t) + \frac{\gamma_e}{2}(q_{k+1}(t) + q_{k-1}(t)) + \frac{\gamma_o}{2}(q_{k+1}(t) - q_{k-1}(t))\,, \\
    \frac{1}{\alpha}\frac{d}{dt}q_{N}(t) &= -q_{N}(t) + \frac{\gamma_e}{2}(q_{1}(t) + q_{N-1}(t)) + \frac{\gamma_o}{2}(q_{1}(t) - q_{N-1}(t))\,.
\end{aligned}\right.
\end{align}
To solve this system of equations, we follow the same steps as before using the generating functional method with this specific $\lambda$. However, once we reach \C{gen_func_eom} we must enforce $\lambda^N=1$ and thus we get
\begin{align}
    F(\lambda, t) = F(\lambda,0)e^{-\alpha t} &\left[\sum_{n=1}^N\sum_{p=-\infty}^\infty \lambda^n I_{n+pN}(\gamma_e \alpha t) \right]\left[\sum_{m=1}^N\sum_{r=-\infty}^\infty 
    \lambda^{m} J_{m+rN}(\gamma_o \alpha t) \right]\,.
\end{align}
From this expression, we can again choose the $k^{th}$ degree term to get $q_k(t)$,
\begin{align}
    q_k(t) = \exp(-\alpha t) \sum_{m,l=1}^N q_l(0) \sum_{r=-\infty}^\infty \left( \sum_{p=-\infty}^\infty I_{k-l-m+pN}(\gamma_e\alpha t)J_{m+rN}(\gamma_o\alpha t) \right)\,.
\end{align}
While this summation looks different from  \C{exact-1pt}, we note that the $m$ and $r$-dependence can be absorbed into a single sum $s=-m-rN$ to give
\begin{align}
    q_k(t) = \exp(-\alpha t) \sum_{l=1}^N q_l(0) \sum_{s,p=-\infty}^\infty (-1)^s I_{k-l+s+pN}(\gamma_e\alpha t)J_{s}(\gamma_o\alpha t) \label{G_per} \, .
\end{align}
We again see that our expectation values agree with directly applying periodicity to our infinite correlators. 

\subsubsection*{\ul{\it Probability redistribution and parity dependence}}

As a result of periodicity, the excitations in our system do not simply move to the left or to the right, as in the case of the infinite chain, but instead must cycle back to the same location after some fixed amount of time. 
\begin{figure}
    \centering
    \includegraphics[scale=0.55]{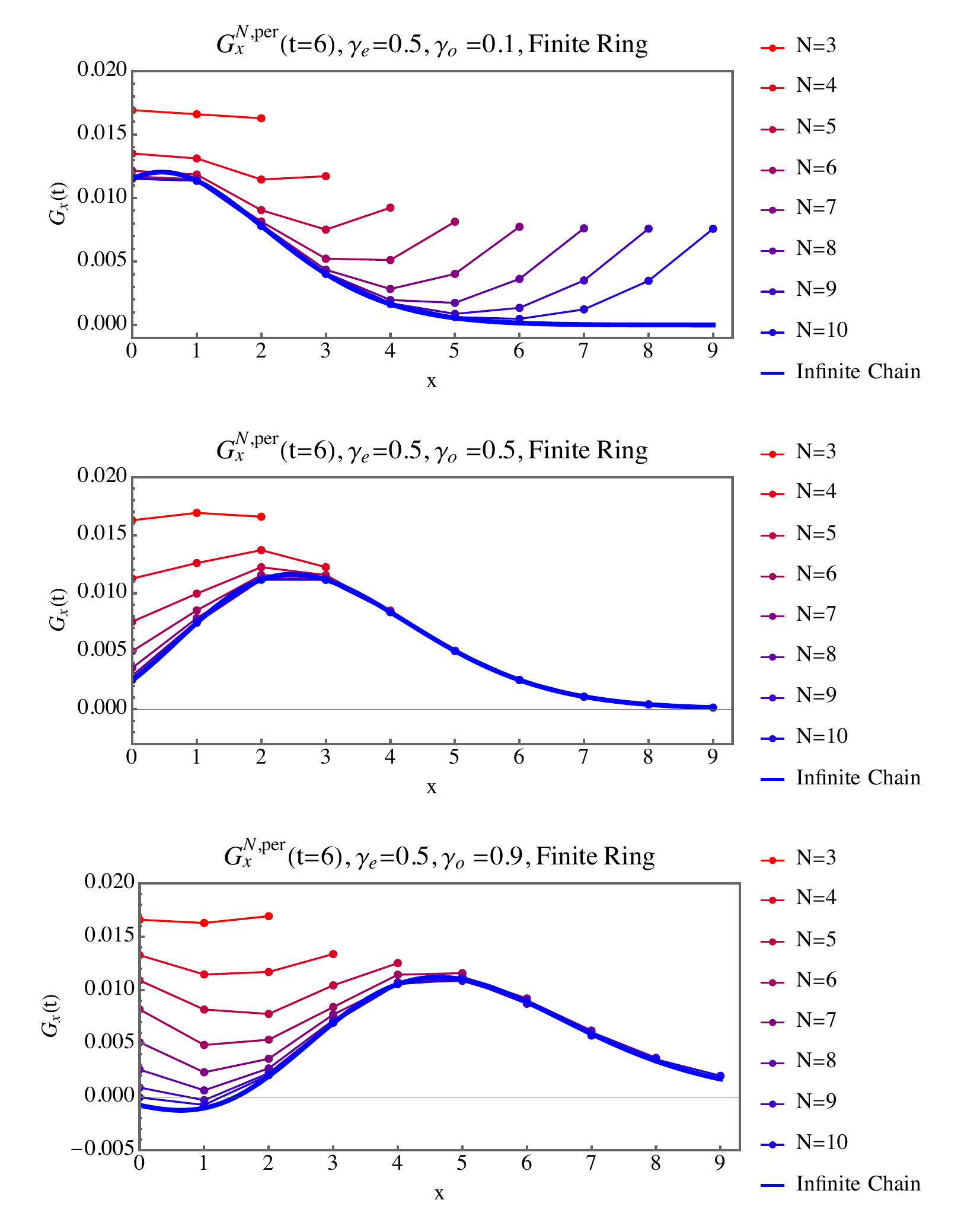}
    \caption{Here we plot $G_x(t=6)$  for periodic systems with varying $N$ values. This gives the correlator for the entire ring at a definite fixed time, where we have set $\gamma_e=0.5$ in all cases and  $\gamma_o=0.1$ \textbf{(top)}, $\gamma_o=0.5$ \textbf{(middle)}, and $\gamma_o=0.9$ \textbf{(bottom)}. We use a color gradient with red for small $N$ and blue for large $N$ with the infinite limit given in solid blue to compare how the correlator changes over time for different $N$ values. We see that smaller $N$ values deviate from the infinite case more and are far flatter than the wave we see in the infinite case. In particular, across all plots we note that the $N=3,4$ look flatter than would be expected across all $(\gamma_e, \gamma_o)$ choices.}
    \label{per_dev}
\end{figure}
\par After some equilibration time, this results in systems with  periodic boundary conditions reaching a uniform nonzero expectation value across $x$-values as in Figure \ref{per_dev}.  We can qualitatively explain this behavior by noting that the $G_x(t)$ values of the wave in the infinite chain case need to be summed over translates to give our $G^{N,\text{per}}_x(t)$ value,
\begin{align}
    G^{N,\text{per}}_x(t) = \sum_{n=-\infty}^\infty G_{x+Nn}(t) \,.
\end{align}
Thus, as the wave widens, summing over $N$-sized slices of our system gives flatter expectation values in our periodic systems. This gives the systems immediate $N$-dependence because larger $N$ will take longer to flatten than small $N$ values. As seen in Figure \ref{per_dev}, $N=3$ has already mostly flattened at $t=6$ whereas $N=9,10$ still agree closely with the wave solution. 

From Figure \ref{Inf_ge-go}, we further observe that larger $|\gamma_o|$ values result in the excitation moving faster and widening at earlier times. As such, the equilibration time has explicit $\gamma_o$ dependence, with larger $\gamma_o$ values resulting in the wave widening and cycling back onto itself earlier, leading to earlier flattening within the system.

\begin{figure}
    \centering
    \includegraphics[scale=0.45]{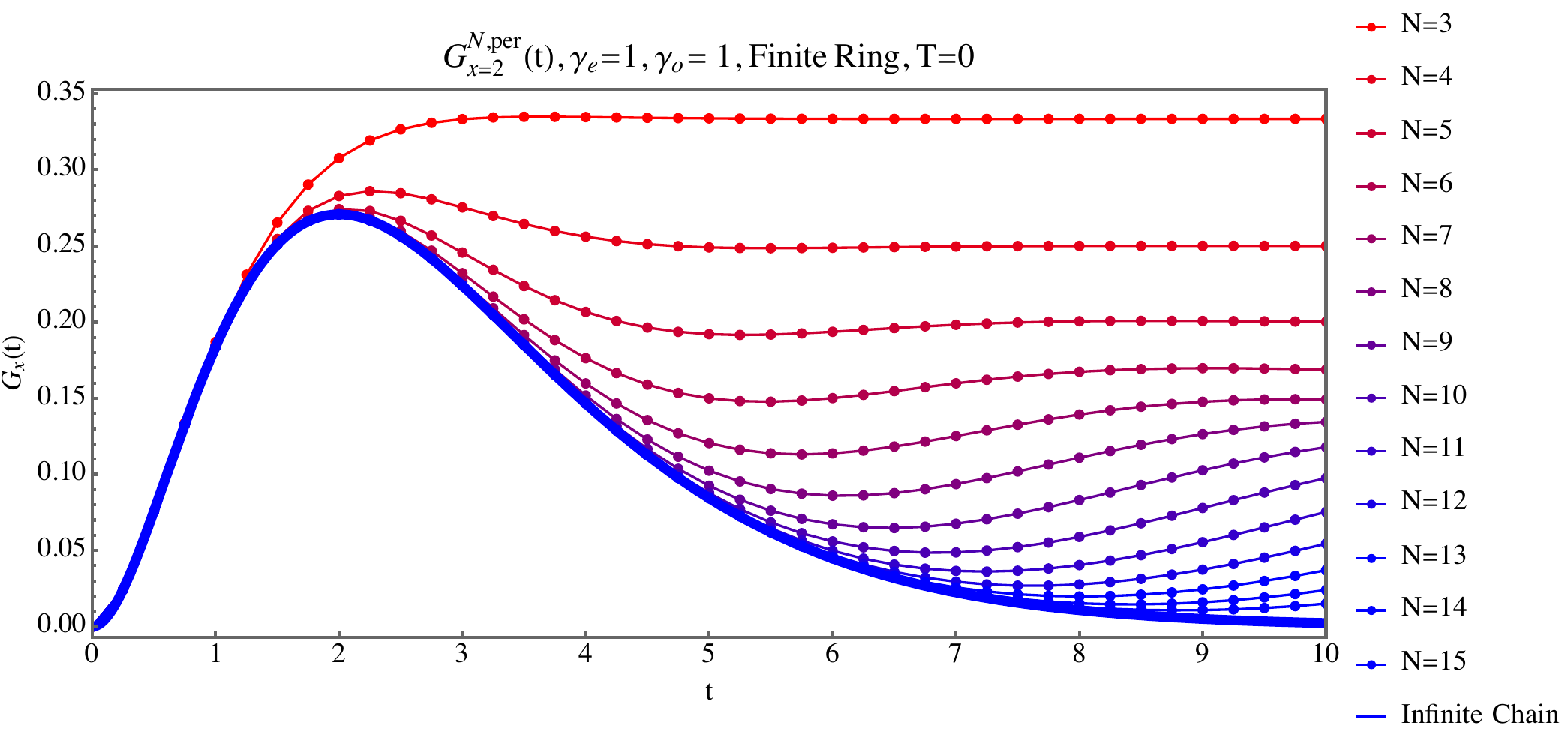}
    \caption{We plot $G_2(t)$ for finite rings of size $N$ ranging from 3 to 15 (seen in dotted lines) with $G_2(t)$ for the infinite chain shown in the solid blue line, all computed with $\gamma_e=\gamma_o=1$. The finite ring correlator is plotted in colors going from red to blue with  bluer lines indicating a larger $N$ value. From these plots, the flattening phenomenon can be seen more clearly with the probabilities in each finite ring eventually stabilizing to a fixed value. Here we specifically see how the expected values all stabilize to roughly $1/N$, where $N$ is the size of the finite ring. These limits are approached earlier by rings with smaller $N$, because they deviate from the infinite limit earlier. This late time limit is independent of $\gamma_o$ as seen in \C{magnetization_decay}.}
    \label{per_N}
\end{figure}

The flattening seen in Figure \ref{per_dev} can also be interpreted physically as a result of probability being redistributed through the system. In particular, once the thermal dissipation given by $\exp( -\alpha t)$ is removed by taking $T=0$ or equivalently $\gamma_e = 1$ because $\gamma_e \propto \tanh(T\inv)$, this gives a system where the $G_x(t)$ values become fixed. Figure \ref{per_N} shows this phenomenon more clearly, where for rings of size $N$ we see that
\begin{align}
    G_2(t\rto \infty) \rightarrow N\inv \,,
\end{align}
and this limiting value is reached for all $x$ values.
\par We see this convergence to nonzero $q_k(t)$ analytically in the case of $\gamma_e=1$ by using periodicity of our correlator. In particular, we note that the sum of the total spins in this system is given by setting $\lambda\rightarrow 1$ in \C{gen_func_eom}
\begin{align}
\label{total_magnet}
    \sum_{x=1}^{N} q_x(t) = \left(\sum_{x=1}^{N} q_x(0) \right)\exp(-\alpha t) \left[\sum_{n=-\infty}^\infty I_n(\gamma_e \alpha t) \right]\left[\sum_{m=-\infty}^\infty 
    J_m(\gamma_o \alpha t) \right] \,,
\end{align}
where the sum of $q_x(0)$ on the right-hand side is our initial total magnetization $m_0$ multiplied by the number of spins $N$. Using \C{besselI} and \C{besselJ}, we then simplify both of these summations giving 
\begin{align}
    \label{magnetization_decay}
    \sum_{x=1}^{N} q_x(t) = m_0 N e^{-\alpha(1-\gamma_e) t} \,.
\end{align}
This follows as a special case of the Jacobi-Anger identity \cite{Watson-Tretesie-1922} where we use 
\begin{align}
    e^{\frac{x}{2}(t+t\inv)}&=\sum_{n=-\infty}^\infty I_n(x)t^n\, , \qquad e^{\frac{x}{2}(t-t\inv)}=\sum_{n=-\infty}^\infty J_n(x)t^n\,,
\end{align}
and set $t=1$ in each case. Then when $\gamma_e = 1$, this gives 
\begin{align}
    \sum_{x=1}^{N} q_x(t) = m_0 N\, ,
\end{align}
as a constant value in the system. As a result, if we initialize our system with $q_0(0)=1$, $m_0 N =1$, we see that as the $N$-sized ferromagnetic system reaches equilibrium, every particle will have $q_k(t)\rto N\inv$. If instead we had an infinite 1D system at $T=0,\gamma_e=1$ then as $t\rto \infty$ we would again expect the probability to be distributed evenly over infinitely many points, which gives $q_n(t)\propto t^{-1/2} \rto0$ \cite{glauber_timedependent_1963}. However, now with finitely many lattice points, the total excitation must be completely redistributed over $N$, and so we reach the limiting value of $N\inv$ at each point. This result also explains the flattening of expectation values in Figure \ref{per_dev}; we can view the flattening as resulting from the probability in the system redistributing and then dissipating thermally. 

\par In the original reciprocal systems studied by Glauber, this behavior in the periodic and infinite case is not described in detail but is still present  \cite{glauber_timedependent_1963}. It is worth noting, as seen in Figure \ref{per_N}, that even with $\gamma_o=1$ periodicity of the system forces the probability to flatten and be maintained at these fixed $N\inv$ values.\footnote{For the cases with $|\gamma_o|=\gamma_e=1$, we cannot use \C{total_magnet}. Instead we must use the polynomial case of \C{1-point-propagator}. However in this case, after applying the same summation method, we find the total polynomial sum again gives $e^{\gamma \alpha t}$ and we again have $m_0N$ conserved.}
\begin{figure}
    \centering
    \includegraphics[scale=0.62]{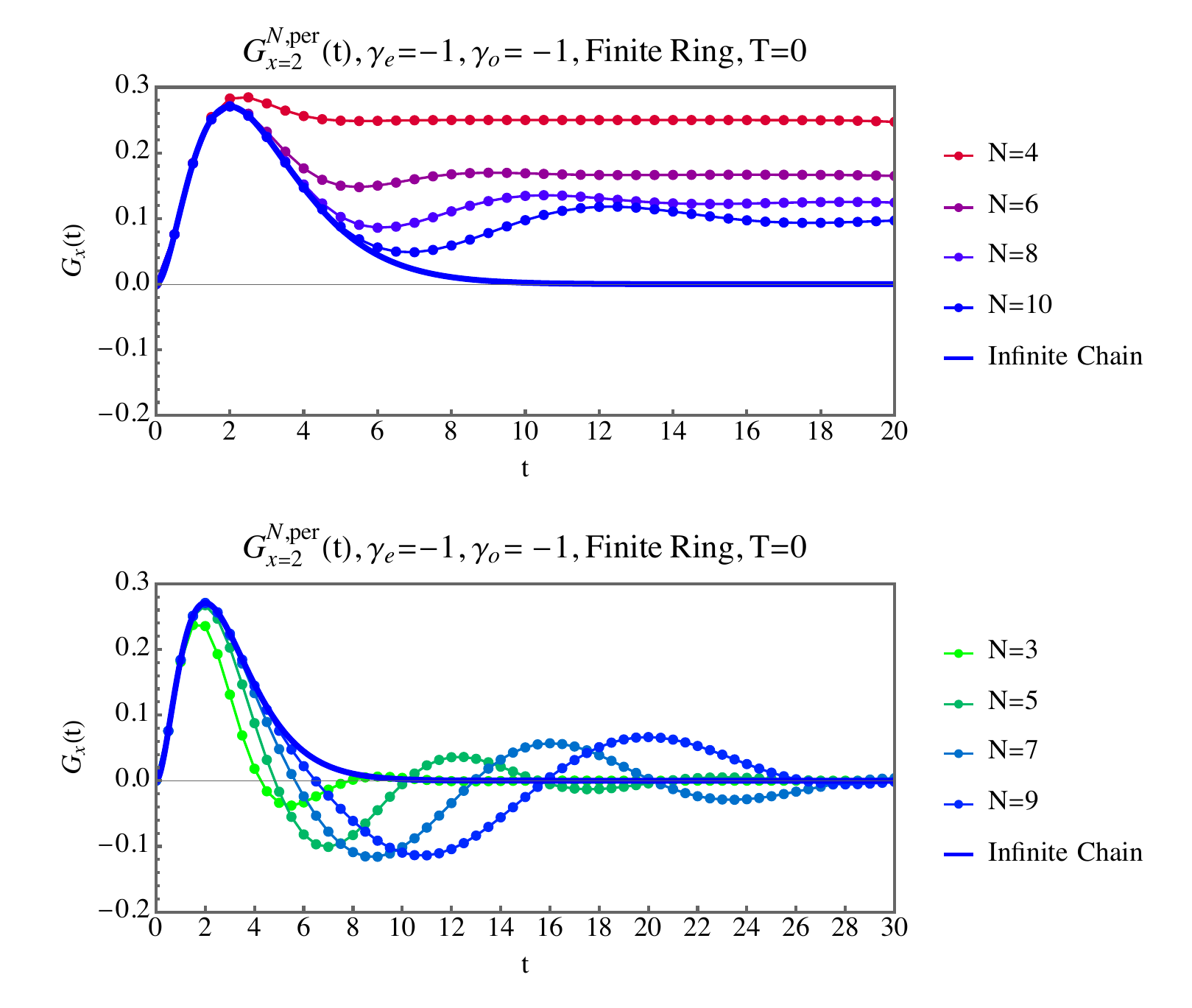}
    \caption{Here we again plot $G^{N,\text{per}}_{x=2}(t)$ for different $N$-sized finite rings, but now in the antiferromagnetic case. \textbf{Top:} we take the case of $N$ even. Red to blue corresponds to increasing $N$ with the infinite chain plotted in blue for comparison. \textbf{Bottom:} we take the case of $N$ odd from green to blue. We observe that in odd $N$ cases, the antiferromagnetic correlator no longer stabilizes to a fixed nonzero value and instead oscillates to zero. This is unlike the even $N$ case where the behavior is similar to the ferromagnetic ($\gamma_e > 0 $) case.}
    \label{per_parity}
\end{figure}
\begin{figure}[hbt!]
    \centering
    \includegraphics[width=0.5\linewidth]{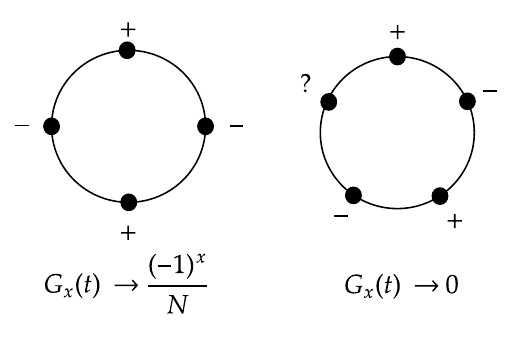}
    \caption{Here we show a sketch of the $N=4$ and $N=5$ lattices with antiferromagnetic conditions at equilibrium to illustrate whether or not alternating expectation values can be imposed as a final equilibrium configuration.}
    \label{Even-Odd}
\end{figure}
\par This behavior does not persist when the system becomes antiferromagnetic, as we observe in Figure \ref{per_parity}. In this case, the even $N$ rings still seem to have $G^{N,\text{per}}_x\rto (-1)^x N\inv$ but the odd cases now have $G^{N,\text{per}}_x(t)$ oscillate until they limit towards zero for all $x$ and $N$ values. We further observe how in Figure \ref{per_parity}, the odd cases deviate from the infinite correlator value by being smaller, whereas the even $N$ cases always deviate by being greater than the infinite case. 
\begin{figure}
    \centering
    \includegraphics[scale=0.55]{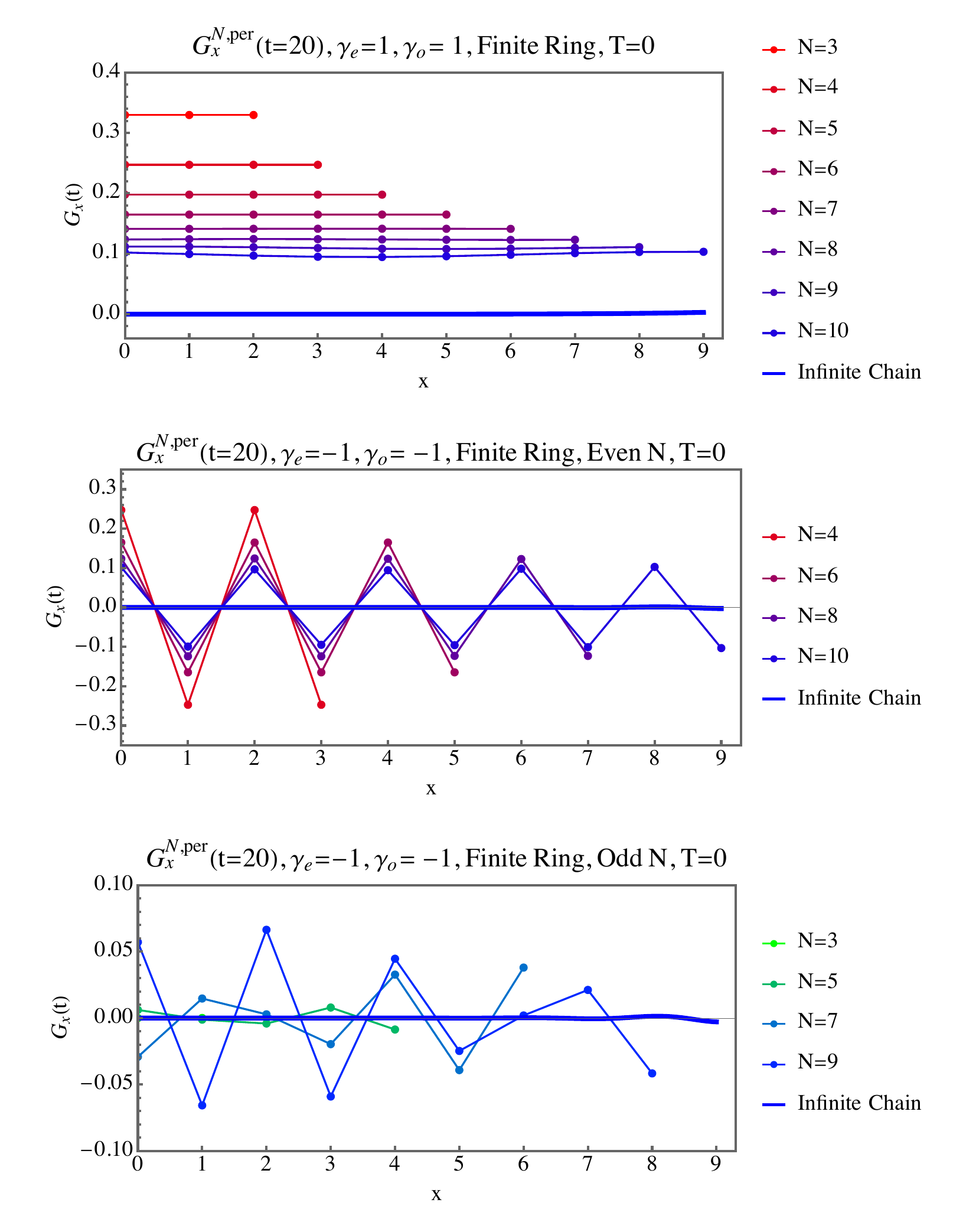}
    \caption{\textbf{Top:} we show $G_x(t=20)$ under ferromagnetic conditions with $\gamma_e=\gamma_o=1$. This is long enough to give the late time stable result for the system. Each ring of size $N$ stabilizes to $1/N$. \textbf{Middle and Bottom:} here we see $G_x(t=20)$ under antiferromagnetic conditions with $\gamma_e=\gamma_o=-1$ for even $N$ and odd $N$. For even $N$, the result is as expected with positions having alternating sign and magnitude $1/N$. However odd $N$ cases are far below this value and instead exhibit more complex behavior as $G^{N,\text{per}}_x(t)\rightarrow 0$.} 
    \label{per_fvaf}
\end{figure}
\par Considering the full system at long times as in Figure \ref{per_fvaf} makes this phenomenon clearer. For the ferromagnetic case, at long times ($t=20$), $G^{N,\text{per}}_x(t) \rto N\inv$ as in the top plot. However, when the system is antiferromagnetic, the even $N$ systems (seen in red and purple) go to $\pm N\inv $ whereas the odd $N$ systems approach $0$ in a more complicated way. We summarize the behavior of the $1$-point function for these different cases below,
\begin{align} \label{FrustrationGammae<0}
    \lim_{t\rightarrow \infty}G_x^{N,\text{per}}(t) = \left\{\begin{aligned}
        &\frac{1}{N} && \gamma_e=1 \\
        &\left\{\begin{aligned}
            &\frac{(-1)^x}{N} && N=\text{even}\\
            &0 && N=\text{odd}
        \end{aligned}\right. && \gamma_e=-1
    \end{aligned}\right. \, .
\end{align}
To explain the difference between the ferro- and antiferromagnetic cases, we use a heuristic sketch of the equilibrium system at long times. It will be helpful to look at Figure \ref{Even-Odd}. For $t\rto\infty$, a periodic system with $N$ even can have $q_k(t)$ with alternating signs, allowing for an equilibrium limit of $\pm N\inv$. In systems with $N$ odd, however, we cannot have $q_k(t)$ with alternating signs and we instead observe oscillating behavior until the expectation values of each spin decays to zero. 

We can explain this more clearly by expanding on our preceding analysis for $\gamma_e=1$ by now summing the absolute value of $q_k(t)$: 
\begin{align}
    \sum_{x=0}^{N-1} |q_x(t)| = \sum_{x=0}^{N-1} \Bigg|\exp(-\alpha t) \sum_{l=1}^N q_l(0) \sum_{s,p=-\infty}^\infty (-1)^s I_{x-l+s+pN}(\gamma_e\alpha t)J_{s}(\gamma_o\alpha t) \Bigg|\,.
\end{align} 
For $N$ even and $\gamma_e < 0$, heuristically, every term in the summation of $p$ will have the same sign and the absolute value sum will be roughly identical to the $\gamma_e > 0$ case. However if $N$ is odd and $\gamma_e < 0$, $I_n$ will alternate sign for different values of $p$ because $I_n(-x)=(-1)^nI_n(x)$, seen by applying \C{besselIandJreln}. As a result, the sum of absolute values for odd $N$ will be the absolute value of an alternating sum, which is smaller than the other two cases and heuristically will decay. Antiferromagnetism seems, therefore, to lead to an interesting dependence on the parity of $N$.
\begin{figure}
    \centering
    \includegraphics[scale=0.55]{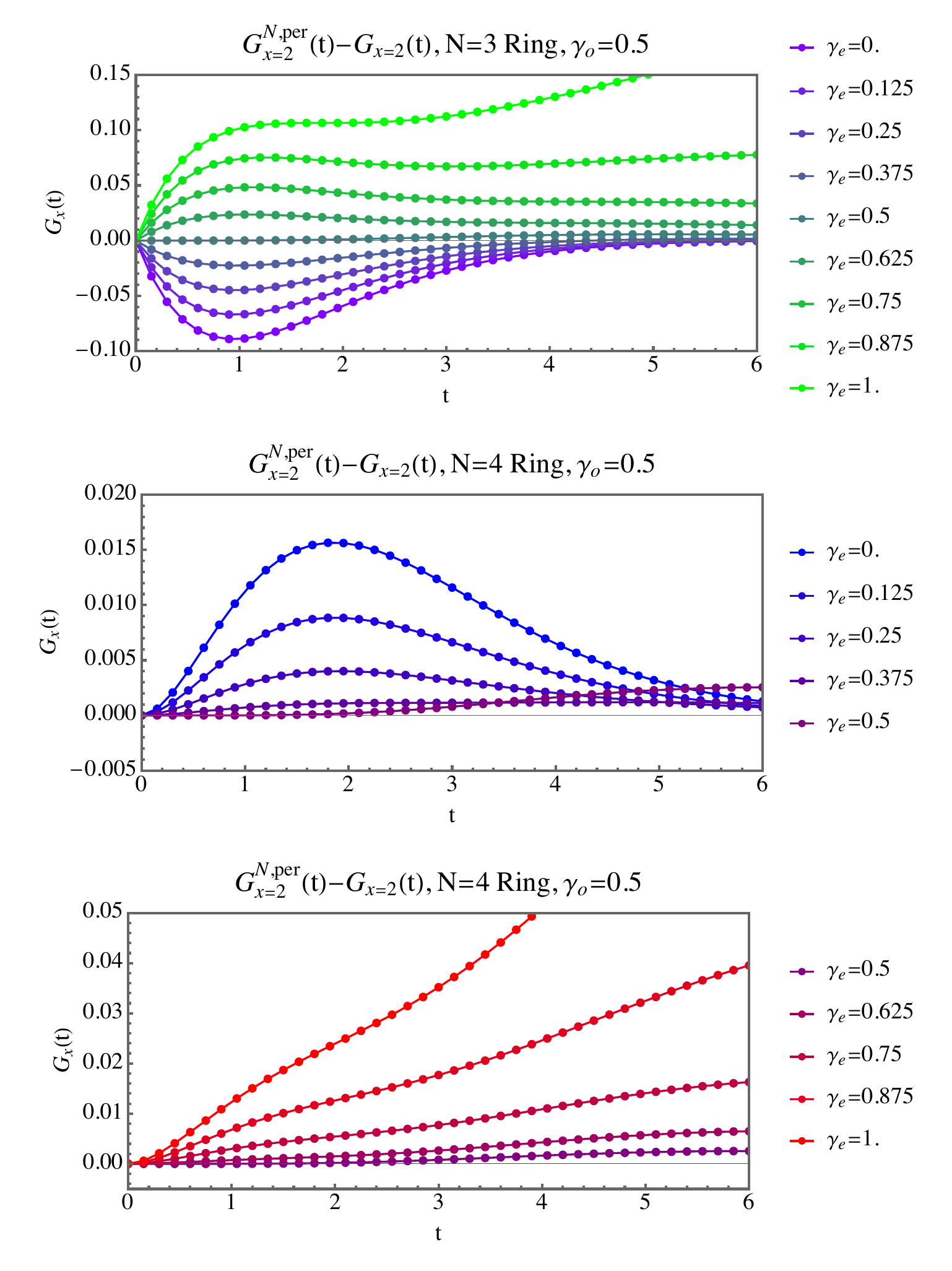}
    \caption{We plot $G^{N,\text{per}}_{x=2}(t)-G_{x=2}(t)$ to see the difference between the infinite and periodic correlator  keeping $\gamma_o=0.5$ fixed and varying $\gamma_e$. \textbf{Top:} the $N=3$ periodic ring with $\gamma_e$ going from 0 to 1 with larger $\gamma_e$ plotted in green and smaller $\gamma_e$ plotted in purple. \textbf{Middle:} the $N=4$ periodic ring with $\gamma_e$ going from 0 to 0.5 with smaller $\gamma_e$ blue and larger $\gamma_e$ purple. \textbf{Bottom:} the $N=4$ periodic ring with $\gamma_e$ going from 0.5 to 1 with  smaller $\gamma_e$ purple and larger $\gamma_e$ red. Fixing $\gamma_o$ results in the excitation moving through the system at the same speed, and makes the transition between different $(\gamma_e,\gamma_o)$ regimes most clear. We have not plotted the case of $\gamma_e\in [-1,0]$ because there is no significant qualitative difference in behavior from the displayed cases.}
    \label{per_ge}
\end{figure}

The one-point function for the periodic spin chain approaches the one-point function for the infinite case in the large-$N$ limit: $\lim_{N\rightarrow\infty}G_{x}^{N,per}(t) = G_x(t)$. However how the system approaches this limit depends on whether the interactions are ferromagnetic or antiferromagnetic, and will be further modified by non-reciprocity. We can get a feel for the different behaviors by looking at $G_2(t)^{N,\text{per}} - G_2(t)$ which is plotted in Figure \ref{per_ge}. Before a more involved discussion, we can summarize the primary conclusion of that plot for the various cases:
\begin{align} \label{FrustrationGammae>0}
    G_2(t)^{N,\text{per}} - G_2(t) = \left\{\begin{aligned}
        &>0 && \gamma_e-|\gamma_o| \gtrapprox 0 \\
        &\left\{\begin{aligned}
            >0 && N=\text{even} \\
            <0 && N=\text{odd}
        \end{aligned}\right. && \gamma_e-|\gamma_o| \lessapprox 0
    \end{aligned}\right. \, .
\end{align}
That is, if the system is roughly ferromagnetic $\gamma_e-|\gamma_o|\gtrapprox0$ the large-$N$ limit is independent of whether the lattice has an even or odd number of spins. If, however, the system is roughly antiferromagnetic $\gamma_e-|\gamma_o|\lessapprox0$ then this limit depends on whether the spin chain has an even or odd number of sites.

In more detail, Figure \ref{per_ge} plots $G_2(t)^{N,\text{per}} - G_2(t)$ with $\gamma_o=0.5$ fixed and $\gamma_e \in [0,1]$ varied for the $N=3$ and $N=4$ rings. Comparing the top and bottom plots of Figure \ref{per_ge}, we observe that for both $N=3,4$, the $\gamma_e\gtrapprox \gamma_o$ regimes appear roughly similar, with $G^{N,\text{per}}_2(t)-G_{2}(t)$ increasing with increasing $\gamma_e$. However, in the top plot as $\gamma_e$ goes from 0 to 1, we observe that the difference goes from negative to positive, whereas the difference is always positive in the middle and bottom plot, showing the parity dependent way the ring correlators differ from the infinite correlators. 

\par We can connect this to Figures \ref{per_N} and \ref{per_parity} by an argument similar to the one used above in the critical case. Note that this parity-dependent to parity-independent transition occurs when we have a negative neighbor interaction. By this we mean that $q_i(t)$ is correlated with its neighbors $q_{i+1}(t)$ and $q_{i-1}(t)$ with coupling  $\gamma_e+\gamma_o$ and $\gamma_e-\gamma_o$, respectively. For $\gamma_e \gtrapprox |\gamma_o|=0.5$, these values are both positive, and we have positive correlation for every interaction along our ring and there is no parity dependent issue. However, when a negative interaction exists, by similar reasoning to our preceding discussion, the system again shows parity dependence. As a result, the antiferromagnetic case ($\gamma_e<0$, where at least one of the neighbor interactions are negative) will always have this parity dependent behavior. From this observation, the results of Figures \ref{per_N} and \ref{per_parity} fit better into the larger $(\gamma_e,\gamma_o)$-dependent phenomena of finite rings shown in Figure \ref{per_ge}.\footnote{One could try to extend the plots in Figure \ref{per_ge} to higher values of $N$ but the $\gamma_e$ transition value is harder to determine because the sums are both harder to approximate accurately and the actual numerical values are very small.}

\par We further note that at even $N$, there is still a difference in behavior between $\gamma_e\gtrapprox |\gamma_o|$ and $\gamma_e \lessapprox |\gamma_o|$. Rather than simply flatten as $\gamma_e$ is decreased, the initial peak instead increases with decreasing $\gamma_e$ then subsequently decreases. Apart from the sign difference, this growth also generally matches the $\gamma_e \lessapprox \gamma_o$ plots in the $N=3$ case and is reminiscent of the non-reciprocal behavior in Figure \ref{Gx-Infinite}. Given this peak is for $G_2(t)-G_{2,\text{inf}}(t)$, this seems to indicate that for $\gamma_e \lessapprox |\gamma_o|$ where we have a negative interaction, periodicity magnifies the effect of non-reciprocity on the system. 

\subsection{Semi-infinite chain and one spin fixed}

To solve the case of the semi-infinite chain, we can directly apply the method of image charges outlined by Glauber \cite{glauber_timedependent_1963}. To do so, we define image charges $q_{-n}(t) = -q_n(t)$ at negative indices. This will make the spin expectation value an odd function of $n$, forcing $q_0(t) = 0$. Glauber specifically studies this system in the purely reciprocal case where we have
\begin{align}
\begin{aligned}
    q_k(t) &= \exp(-\alpha t) \sum_{l=1}^\infty q_l(0) \left(I_{k-l}(\gamma_e \alpha t) - I_{-k-l}(\gamma_e \alpha t) \right)\,, \\
    &= \exp(-\alpha t) \sum_{l=1}^\infty q_l(0) \left(I_{k-l}(\gamma_e \alpha t) - I_{k+l}(\gamma_e \alpha t) \right) \,,
\end{aligned}
\end{align}
and the $I_{k-l}(\gamma_e \alpha t) - I_{-k-l}(\gamma_e \alpha t)$ term includes the negative image charges and we have used Bessel function properties to flip the $I_{-k-l}$ index; see, Appendix~\ref{Appendix-Bessel}. This will ensure that $q_n$ is odd with respect to $n$, and gives a solution to the desired differential equations. To generalize this approach to our non-reciprocal case, we simply replace these $I_n(\gamma_e \alpha t)$ functions with our more general one-point functions,
\begin{align}
\label{semi_fin_equil}
\begin{aligned}
    q_k(t) = \exp(-\alpha t) \sum_{l=1}^\infty q_l(0) \Bigg[&\sum_{n=-\infty}^\infty I_{k-l+n}(\gamma_e \alpha t)(-1)^{n}J_{n}(\gamma_o \alpha t)\\ & - \sum_{n=-\infty}^\infty I_{-k-l+n}(\gamma_e \alpha t)(-1)^{n}J_{n}(\gamma_o \alpha t) \Bigg]\,,
\end{aligned}
\end{align}
to give our full solution. We will use this solution to describe the behavior of a system tending to equilibrium with one spin fixed, which will lead to a polarization cloud as outlined by Glauber~\cite{glauber_timedependent_1963}.

\par In the case where we have one spin fixed at all times, we choose to enforce the condition $q_0(t) = 1$. We now observe that the system will not reach the equilibrium value $q_k(t \rto \infty)=0$ but will instead form a polarization cloud around $k=0$. We model this behavior by keeping all differential equations for the system unchanged, except we now lose the equation for index $k=0$ where the behavior is known and fixed. This amounts to cutting the chain from infinite to semi-infinite in either the positive or negative $k$ directions. Equilibrium now corresponds to the case where we set $\del_t q_k(t) = 0$ in  (\ref{1-point-eom-h-NR}) which gives the following equation for the equilibrium system:
\begin{align}
\label{eq_1pt_func}
    q_k = \frac{\gamma_e}{2}(q_{k-1}+q_{k+1}) + \frac{\gamma_o}{2}(q_{k-1}-q_{k+1}) \qquad k \neq 0 \,.
\end{align}
This equation can be solved by taking 
\begin{align}
    q_k = \eta^k\,,
\end{align}
for some $\eta$. This gives an equation to solve for the equilibrium configuration,
\begin{align}
    \label{eq_1pt_quad}
    \frac{\gamma_e-\gamma_o}{2}\eta^2 - \eta + \frac{\gamma_e+\gamma_o}{2} = 0 \,.
\end{align}
We first consider the case where $\gamma_e\neq |\gamma_o|$. Here our equation will be quadratic and we get two roots,
\begin{align}
    \eta_{\pm} = \frac{1\pm \sqrt{1-(\gamma_e^2-\gamma_o^2)}}{(\gamma_e-\gamma_o)}\,.
\end{align}
The general solution for $q_k(t_{eq})$ is then as follows,
\begin{align}
\label{gen_1pt_eq_fixed0}
    q_k(t_{eq}) = p\eta_+^k + (1-p)\eta_-^k\,,
\end{align}
with $0\leq p\leq1$ given that $q_0(t_{eq}) = 1$. First we will show that $|\eta_+| \geq 1$ and $|\eta_-| \leq 1$. To prove this, we observe that $\eta_+>0$ only when $ \gamma_e>\gamma_o$. We now assume $\eta_+\leq 1$ for the sake of constructing a contradiction. We want to determine the conditions this assumption imposes on $(\gamma_e, \gamma_o)$,
\begin{align}
    \frac{1+\sqrt{1-\gamma_e^2+\gamma_o^2}}{\gamma_e-\gamma_o} &\leq 1\,,\\
    \label{inequalitybounds}
    \sqrt{1-\gamma_e^2+\gamma_o^2} &\leq \gamma_e-\gamma_o - 1 \,.
\end{align}
The second line assumes $\gamma_e \geq \gamma_o$. Since the left-hand side is positive, this inequality holds when both sides are squared:
\begin{align}
    1-\gamma_e^2+\gamma_o^2 &\leq \gamma_e^2 + \gamma_o^2 + 1 -2\gamma_e\gamma_o - 2\gamma_e + 2\gamma_o\,, \\
     (\gamma_e-1)(\gamma_e-\gamma_o) &\geq 0 \,, \label{middleline} \\
    \gamma_o &\geq \gamma_e \, ,
\end{align}
where we note that the first factor, $ (\gamma_e-1)$ in \C{middleline}, is either zero for $\gamma_e=1$ or negative. If $\g_e \neq 1$, this is the contradiction we seek and we conclude that $\gamma_e > \gamma_o$ implies $ \eta_+ > 1$. If $\g_e=1$ then $\eta_+ = \frac{1+|\g_o|}{1-\g_o} \geq 1$. Similarly for the opposite case of $\gamma_e < \gamma_o$, we note that $(\gamma_e,\gamma_o) \rto (-\gamma_e,-\gamma_o)$ sends $\eta\rto -\eta$ and we conclude that if $\gamma_e < \gamma_o$ then $\eta_+ \leq -1$. 

We similarly verify $\eta_-$ bounds by instead taking the case of $\gamma_e > \gamma_o$ and assuming that $\eta_- \geq 1$. This then gives 
\begin{align}
    \frac{1-\sqrt{1-\gamma_e^2+\gamma_o^2}}{\gamma_e-\gamma_o} &\geq 1 \,,\\
     1-\gamma_e+\gamma_o &\geq \sqrt{1-\gamma_e^2+\gamma_o^2} \,.
\end{align}
Squaring as in \C{inequalitybounds} we obtain $\gamma_e \leq \gamma_o$ which is a contradiction similar to the preceding case. We again take $(\gamma_e,\gamma_o) \rto (-\gamma_e,-\gamma_o)$ to get the other bound, and we conclude that $-1\leq \eta_-\leq 1$.

As a result, \C{gen_1pt_eq_fixed0} must be modified in the following sense: for $k>0$ we will use one solution of the form \C{gen_1pt_eq_fixed0} while for $k<0$ we use another solution. Note that $|\eta_+^k| > 1 $ for $k > 0$ and $|\eta_-^k| > 1 $ for $k < 0$ and so we take
\begin{align}
\label{1pt_eq_fixedNR}
    q_k(t_{eq}) &= \left\{\begin{aligned}
        \eta_+^k && k < 0 \\
        1 && k = 0\\
        \eta_-^k && k > 0
    \end{aligned} \right. \,.
\end{align} 
We can also directly solve the critical case $\gamma_e = \pm \gamma_o$ and interpret the solution in the spirit of \C{1pt_eq_fixedNR}. If we have $\gamma_e = \pm \gamma_o$, we transform \C{eq_1pt_func} to give 
\begin{align}
    q_k = \gamma_e q_{k-1}\,, \qquad q_k = \gamma_e q_{k+1}\,,
\end{align}
respectively. Using the same ansatz as before, we get $\eta=\gamma_e$ for $\gamma_e = \gamma_o$ and $\eta=\frac{1}{\gamma_e}$\footnote{Notice that \C{eq_1pt_quad} naively seems to have an extra root $\eta=0$ for $\gamma_e=-\gamma_o$. But the actual equation that we have to solve is \C{eq_1pt_func},
\begin{align}
    1 = \frac{\gamma_e+\gamma_o}{2} \eta^{-1} + \frac{\gamma_e-\gamma_o}{2} \eta \,, \non
\end{align} 
which has solutions $\eta= \gamma_e$ and $\eta=\frac{1}{\gamma_e}$ for $\gamma_e = \pm \gamma_o$.} for $\gamma_e = -\gamma_o$. Summarizing these cases, 
\begin{align}
    q_k(t_{eq}) =\left\{\begin{aligned}
        &\left\{\begin{aligned}
        &0 && k<0 \\
        &\gamma_e^k && k\geq0
        \end{aligned}\right. && \gamma_e = \gamma_o \,, \\
        &\left\{\begin{aligned}
        & \gamma_e^{-k} 
        && k\leq 0 \\
        &0 && k>0
        \end{aligned}\right. && \gamma_e = - \gamma_o \,. 
    \end{aligned}\right.
\end{align}
However, note that taking $\gamma_o \rto -\gamma_o$ is equivalent to swapping the lattice points $k$ and $-k$ so these two solutions must be related by this symmetry.

We further note that because we are in the critical case, our system only couples in one direction. On one side of the spin at zero, $q_k(t_{eq})$ falls as $\eta^k$ while in the other direction the spin expectation value will be zero. This picture is reversed when we change the sign of $\g_o$ along with sending $\eta \rightarrow \frac{1}{\eta}$. We can connect this picture to Glauber's results in Ref. \cite{glauber_timedependent_1963} by taking $\gamma_o=0$. Here we see 
\begin{align}
\begin{aligned}
    \eta_+\eta_- &= \frac{1+\sqrt{1-\gamma_e^2}}{\gamma_e}\frac{1-\sqrt{1-\gamma_e^2}}{\gamma_e}\,, \\
    &= \frac{1-(1-\gamma_e^2)}{\gamma_e^2} = 1\,.
\end{aligned}
\end{align}
So we see that Glauber's solution of $\eta^{|n|}$ is a special case of \C{1pt_eq_fixedNR} because $\eta_-^{|k|} = \eta_+^{k}$ for $k < 0$. With non-reciprocity, the picture is generalized because $\eta_+$ is not generally the inverse of $\eta_-$ but rather, 
\begin{align}
\begin{aligned}
    \eta_+\eta_- &=
    \frac{\gamma_e+\gamma_o}{\gamma_e-\gamma_o} \, ,
\end{aligned}
\end{align}
and $\eta_+\eta_- \neq 1$ except when $\g_o =0$. 
This means the system is asymmetric with $q_k(t_{eq}) \neq q_{-k}(t_{eq})$ and there is a preferential direction for excitations to move: spins at positive and negative lattice points are coupled differently to the fixed spin at zero. 
We can verify that the positive and negative lattices have identical behavior after taking $\gamma_o\rto -\gamma_o$ by noting that 
\begin{align}
\begin{aligned}
    \eta_+(\gamma_o)\eta_-(-\gamma_o) &= \frac{1 + \sqrt{1-(\gamma_e^2-\gamma_o^2)}}{(\gamma_e-\gamma_o)} \frac{1 - \sqrt{1-(\gamma_e^2-\gamma_o^2)}}{(\gamma_e+\gamma_o)}  = 1 \, .
\end{aligned}
\end{align}
So swapping the sign of $\gamma_o$ interchanges the role of $\eta_+$ and $\eta_-$ as expected. 

The complete description of this system is then given by adding the homogeneous counterpart to our one-spin fixed condition; namely, by adding in solutions with $q_0(t) =0$. For these one-spin fixed spin-chains, we note that a spin at $k$ cannot interact with a spin at $-k$. This is a result of the $q_0(t)=1$ condition, which serves as a wall preventing interactions across this point. As an example, if we had $q_{-1}(t)=-1$, this would have no effect on $q_0$ which is fixed, and so $q_{-1}$ has no effect on the spin at $k=1$. As such, we can view a one-spin fixed infinite chain as two semi-infinite chains, with different coupling to $q_0$ as a result of non-reciprocity. 

This gives our solution for $k > 0$,
\begin{align}
    \begin{aligned}
     q_k(t) = \eta_{-}^k + e^{-\alpha t}\sum_{l=1}^\infty (q_l(0) - \eta_-^l) \Bigg[&\sum_{n=-\infty}^\infty I_{k-l+n}(\gamma_e \alpha t)(-1)^{n}J_{n}(\gamma_o \alpha t) \\&-\sum_{n=-\infty}^\infty I_{-k-l+n}(\gamma_e \alpha t)(-1)^{n}J_{n}(\gamma_o \alpha t) \Bigg]\,,
    \end{aligned}
\end{align}
and for $k< 0$, 
\begin{align}
    \begin{aligned}
        q_k(t) = \eta_{+}^k + e^{-\alpha t}\sum_{l=-\infty}^{-1} (q_l(0) - \eta_+^l) \Bigg[&\sum_{n=-\infty}^\infty I_{k+l+n}(\gamma_e \alpha t)(-1)^{n}J_{n}(\gamma_o \alpha t) \\&-\sum_{n=-\infty}^\infty I_{-k+l+n}(\gamma_e \alpha t)(-1)^{n}J_{n}(\gamma_o \alpha t) \Bigg]\,.
    \end{aligned}
\end{align}
In the infinite time limit we reach the equilibrium solution \C{1pt_eq_fixedNR} while also enforcing no communication from one side of the chain to the other.

\subsection{Open spin chains}
\label{Sec-Open-Spin}

To study the open finite chain, we apply periodicity to the semi-infinite case. However, we cannot simply apply periodicity to a system of size $N$ as we did in the periodic chain. We instead define periodicity over a $2(N+1)$ size system and enforce that points $k=0$, $N+1$ on our lattice be zero. To do this, we essentially apply $2N+2$ periodicity to a semi-infinite system with $q_{-k+2N+2}(t) \equiv q_{-k}(t) = -q_k(t)$. This will give our one-point function as
\begin{align}
\label{finite_expval}
\begin{aligned}
    q_k(t) = e^{-\alpha t} \sum_{l=1}^N q_l(0) \Bigg[&\sum_{n,s=-\infty}^\infty I_{k-l+n+s(2N+2)}(\gamma_e \alpha t)(-1)^{n}J_{n}(\gamma_o \alpha t)  \\&-\sum_{n,s=-\infty}^\infty I_{-k-l+n+s(2N+2)}(\gamma_e \alpha t)(-1)^{n}J_{n}(\gamma_o \alpha t) \Bigg] \,.
\end{aligned}
\end{align}
We can verify that this gives the expected behavior for the finite case by noting that both $0,N+1$ are now fixed to be 0. For these two cases, \C{finite_expval} can be re-indexed as follows:
\begin{align}
\label{finite}
\begin{aligned}
    q_{0}(t) = e^{-\alpha t} \sum_{l=1}^N q_l(0) \Bigg[&\sum_{n,s=-\infty}^\infty I_{-l+n+s(2N+2)}(\gamma_e \alpha t)(-1)^{n}J_{n}(\gamma_o \alpha t)  \\&-\sum_{n,s'=-\infty}^\infty I_{-l+n+s'(2N+2)}(\gamma_e \alpha t)(-1)^{n}J_{n}(\gamma_o \alpha t) \Bigg] = 0 \, ,
\end{aligned} \\
\begin{aligned}
    q_{N+1}(t) = e^{-\alpha t} \sum_{l=1}^N q_l(0) \Bigg[&\sum_{n,s=-\infty}^\infty I_{N+1-l+n+s(2N+2)}(\gamma_e \alpha t)(-1)^{n}J_{n}(\gamma_o \alpha t)  \\&-\sum_{n,s'=-\infty}^\infty I_{-N-1-l+n+s'(2N+2)}(\gamma_e \alpha t)(-1)^{n}J_{n}(\gamma_o \alpha t) \Bigg] = 0\,.
\end{aligned}
\end{align}
For $q_0(t)$ the first and second sums cancel with $s'=s$, while for $q_{N+1}(t)$ the sums cancel with $s'=s+1$. As a result, we see that this obeys our desired boundary and periodicity conditions. As a result, the lattice points $1,\dots N$ give a finite chain while $N+2,\dots 2N+1$ give the negative of that finite chain.

\begin{figure}[ht]
    \centering
    \includegraphics[width=0.8\linewidth]{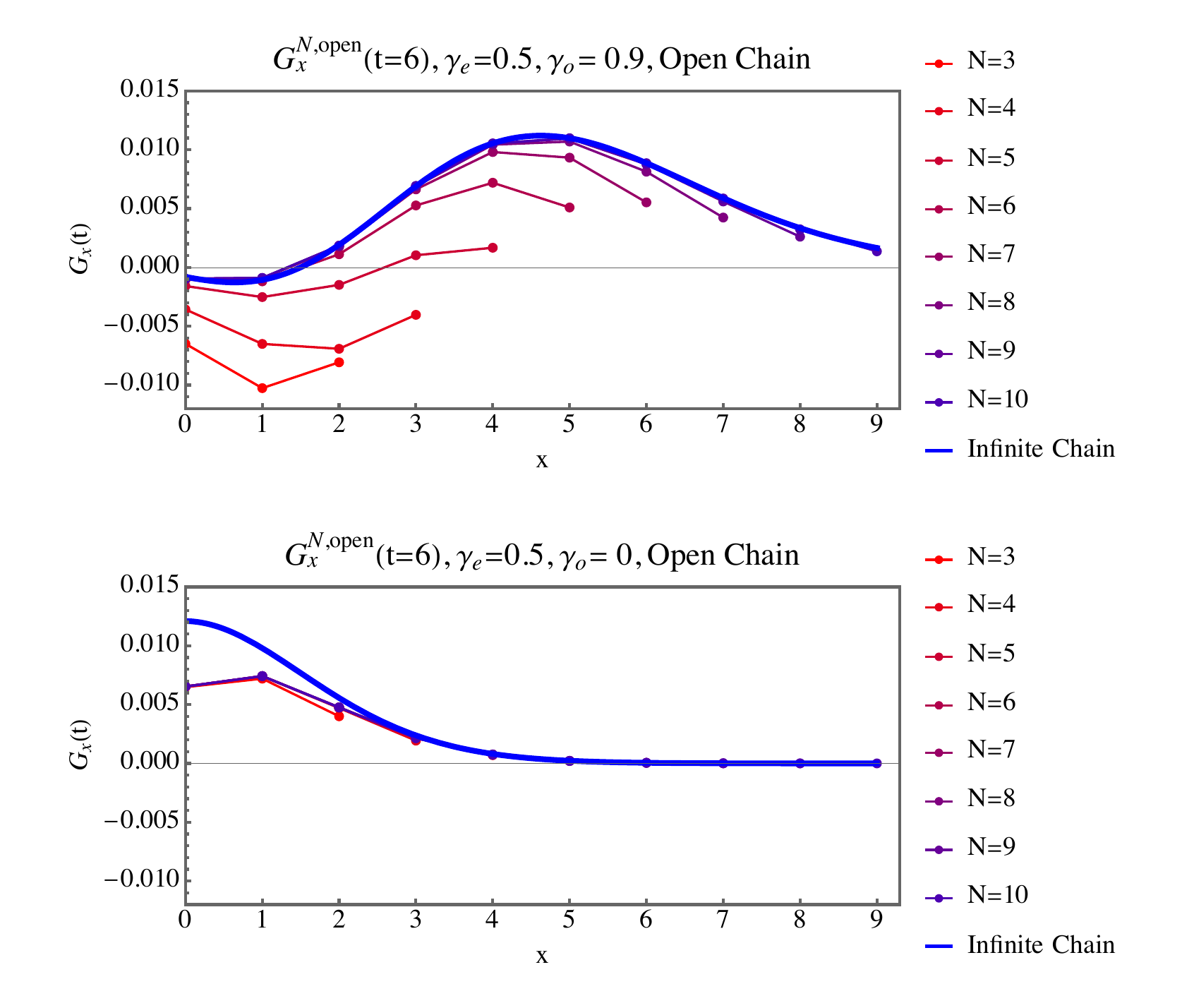}
    \caption{Plotted on top is $G_x(t=6)$ with $\gamma_e=0.5, \gamma_o=0.9$ for open finite chains of size $N$, with $N$ varying from 3 to 10 and  also the infinite chain. The finite chains are plotted with color going from red (small $N$) to blue (large $N$) with the infinite chain in a solid blue. The larger $N$ cases more closely agree with the infinite solution, as we would expect. However, the negative values seen for the $N=3$, $4$, $5$ chains are unexpected because large negative values are not present in the infinite chain or in periodic rings of the same size (see Figure \ref{per_dev}). The lower plot is the reciprocal case for comparison.}
    \label{chain_dev}
\end{figure}

\subsubsection*{\ul{\it Time evolution for open spin chains}}

The time evolution of the open spin chain displays interesting properties related to the boundary conditions and the size of the system, which we denote by $N$. To study this behavior, we take a finite chain of size $N$ and set initial condition $q_1(0) = 1$ (note that $x$ and the subindex of all correlators indicates the distance away from the first site). Taking the behavior of the chain at fixed $t=6$ as in Figure \ref{chain_dev}, we first observe that small $N$ values disagree with the infinite limit far more than larger $N$ chains. While this deviation is certainly expected, the small $N$ values are far more negative than would be expected when compared with the infinite chain correlators or the periodic correlators (seen in Figure \ref{per_dev}). 

\par
\begin{figure}[h]
    \centering
    \includegraphics[width=0.8\linewidth]{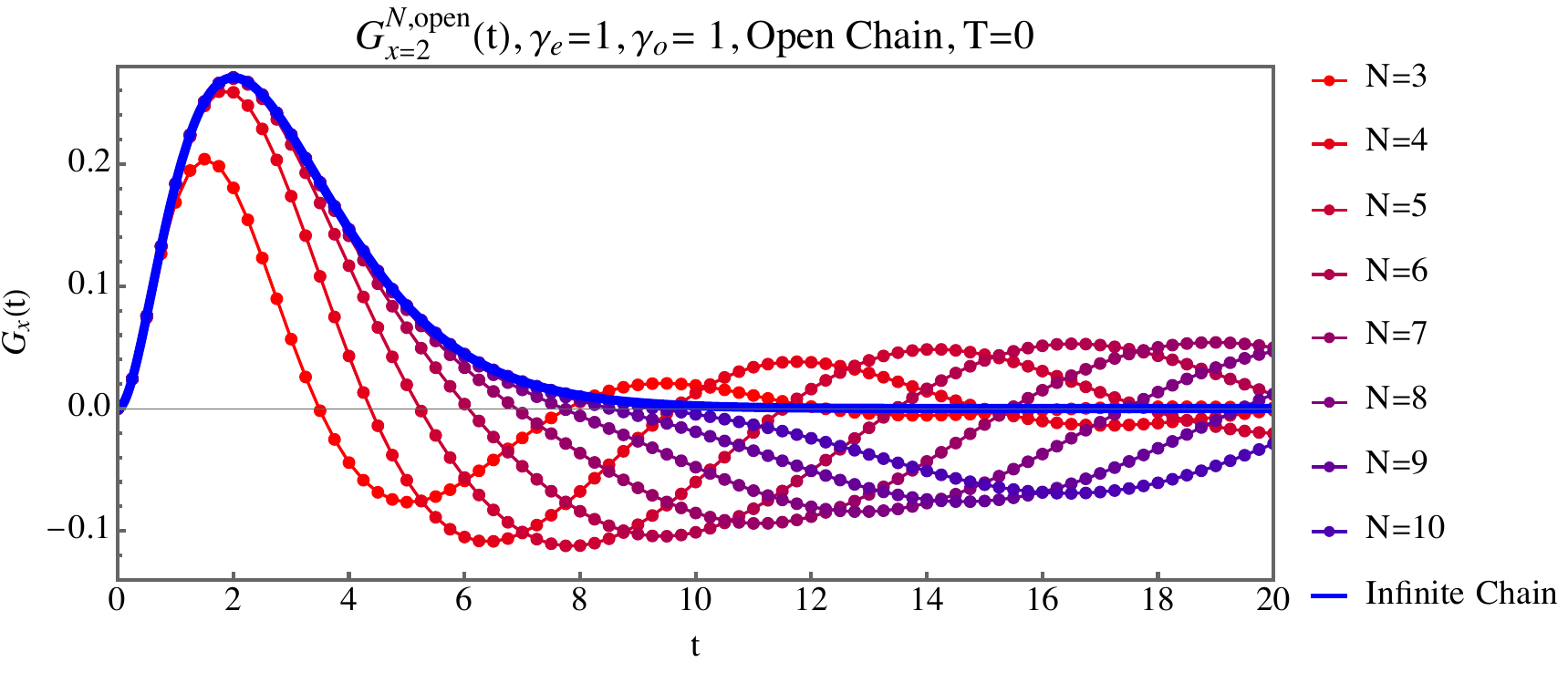}
    \caption{Here the $G_2(t)$ value is plotted for different size chains for $t\in[0,20]$; the dotted plots indicate finite chains with redder colors used for smaller $N$ and bluer colors for larger $N$ and the solid plot indicates the infinite chain. Here, we see how in all finite systems, the point at position 2 reaches some non-negligible negative value and then oscillates to a non-negligible positive value; behavior not observed in any infinite system.}
    \label{chain_t_p2}
\end{figure}

 These negative values are more clear in Figure \ref{chain_t_p2}, where the time development of $G^{N,\text{open}}_2(t)$ is seen. Across all  finite $N$ chains, $G^{N,\text{open}}_2(t)$ becomes negative and oscillates for far longer than we might have expected. Both the infinite chain and the periodic ring stabilized at far earlier times as shown in Figure \ref{per_N}. The open chain instead takes far longer to equilibrate and reaches negative values in each case. These negative values are particularly interesting given $\gamma_e=\gamma_o=1$ implies there are no negative interactions in the system - all coupling strengths $(\gamma_e\pm\gamma_o)/2$ are positive. 

These phenomena are definitely a consequence of non-reciprocity, which we can check by comparing with the reciprocal case, $\g_o=0$, also plotted in Figure \ref{chain_dev}. Even for the reciprocal case, there are noticeable edge effects for the finite $N$ lattices. In this case, we can interpret the smaller positive $G_x(t)$ values visible near $x=0$ as coming from interference with the inverted wave sitting at $-1$ and spreading out over time. The interpretation of what we see in Figure \ref{chain_dev} for the non-reciprocal case is more interesting. In this case because of non-reciprocity, the correlation moves as a wavepacket to the right in Figure \ref{chain_t_p2}. When it reaches the boundary, it is reflected and inverted. At $t=6$ plotted in Figure \ref{chain_dev} we only see a net negative correlation.

\begin{figure}
    \centering
    \includegraphics[width=1.04\linewidth]{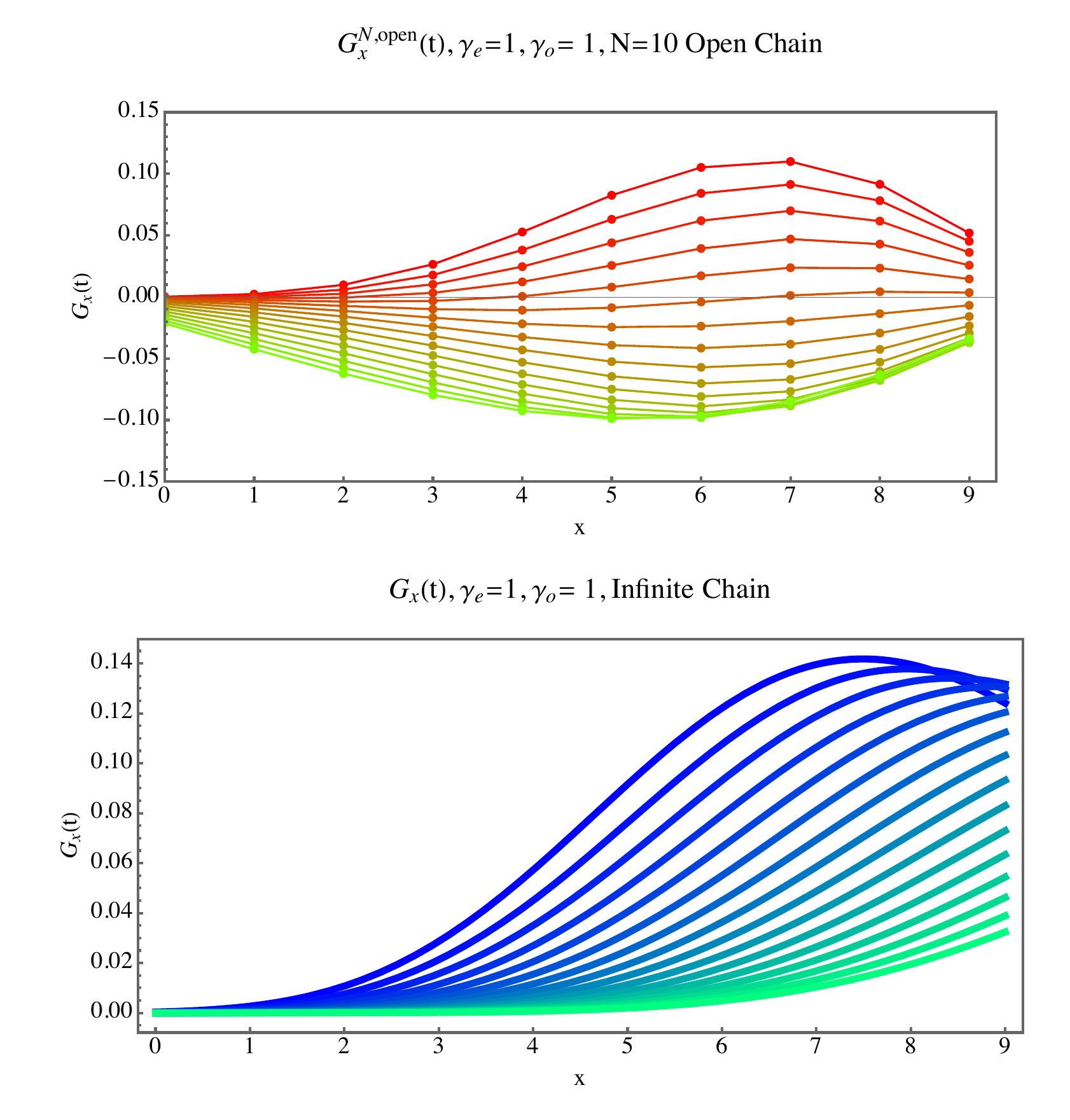}
    \caption{Both plots show $G_x(t)$ for a system with $\gamma_e=\gamma_o=1$. For both plots, each connected line shows $G_x$ at some fixed time $t_i$, with $t_i$ evenly spaced between 8 and 15. The line is colored according to the choice of $t_i$. \textbf{Top:} here we take an $N=10$ finite open spin chain. Earlier lines are indicated by more red colors, with the lines becoming progressively more green for later times. \textbf{Bottom:} here we take the case of an infinite chain with earlier lines now indicated by more blue colors and the lines becoming progressively more green for later times. From this plot, the overall presence of negative $G_x(t)$ correlator values in a finite system can clearly be observed. Using Figure \ref{chain_t_p2}, we observe that around $t\approx 8$ the $N=10$ chain has $G^{10,\text{open}}_x(t)$ roughly similar to the infinite chain $G_x(t)$, but by $t=15$ the entire chain is almost completely negative, in sharp contrast to later times in the bottom plot.}
    \label{chain_tot}
\end{figure}

In Figure \ref{chain_tot}, we see a particularly interesting feature of the finite sized chain. We contrast the case of the $N=10$ finite chain with the first $10$ lattice points of the infinite chain for $\gamma_e=\gamma_o=1$. What is plotted are time snapshots from $t=8$ to $t=15$ of the entire chain. There is a sharp difference in the late time behavior that we ascribe to the wave-like nature of the correlations with non-reciprocity. In the infinite spin chain, correlations travel to the right as a wavepacket while they decay. For the finite spin chain they reflect at the boundary where they are inverted. Subsequently, they move to the left leading to the  negative correlations visible even at late times.

\section{The Two-Point Function and its Scaling Behaviour}

\label{Sec-Two-Point}

Using (\ref{2-point-General}), we can write the general two-point function as
\begin{align} \label{two-point-sol}
    C_{ij}(t,\tau) = \langle s_i(t+\tau)s_j(\tau) \rangle = \sum_{m,n,m'} G_{i-m}(t) G^{(2)}_{m-m',j-n}(\tau) r_{m'n}(0)\,,
\end{align}
where $G_x(t)$ is given by (\ref{1-point-propagator}) and $G^{(2)}_{x_1,x_2}(t)$ by (\ref{2-point-propagator-sol}). Even though this is an exact result, it is not particularly easy to extract physical information from it; this remark is also true for (\ref{equal-two-point-sol}). For example, it is not at all obvious that these expressions are translationally invariant and that the equal-time two-point function (\ref{equal-two-point-sol}) is independent of non-reciprocity. To highlight the main features of these correlators, in this section we solve this system using Laplace transformations and then study their long wavelength and low temperature behavior.

Before doing that, there is already a case where we can extract a lot of information from (\ref{two-point-sol}).
If we set $\tau=0$ and consider uncorrelated initial conditions, $r_{ij}(0) = \delta_{ij}$, we get
\begin{align} \label{2-pt-tau=0-uncor}
    \langle s_{x+i}(t) s_i(0) \rangle = G_{x}(t)\,.
\end{align}
Therefore all the remarks we made about $G_x(t)$ in Sections \ref{Sec-N-Spins} and \ref{Sec-Finite-Spin-Chain} are valid for (\ref{2-pt-tau=0-uncor}). Now we also want to understand the scaling behavior of this system. So we rewrite the Green's  function of \C{1-pt-sol-Fourier} for the periodic ring in momentum space,
\begin{align}
    G_x(t) = \frac{1}{ N}\sum_k \exp\Big[ -\alpha t + \alpha \gamma_e t \cos k a -i \alpha \gamma_o t \sin k a - ik a x \Big]\,.
\end{align}
In this section we set $\alpha=1$ to simplify the expression. Now we let the number of sites go to infinity, the lattice spacing go to zero and we take a low momentum limit $ka < 1$. The momentum sum is replaced by a Gaussian integral which gives, 
\begin{align} \label{Low-E-Gx}
    G_x(t)
    & \approx e^{-(1-\gamma_e) t} \sqrt{\frac{2\pi}{ \gamma_e t}}\exp \left[-\frac{(x+\gamma_o t)^2}{2\gamma_e t} \right]\,.
\end{align}
We can now read off the relaxation timescale for the system, the correlation length and an energy gap:
\begin{align} \label{Gap-etal}
    \tau_{eq} = \frac{1}{1-\gamma_e}, \qquad \xi_{eq} = \frac{1}{2(1-\gamma_e)}, \qquad \mu^2 = 2(1-\gamma_e)\,.
\end{align}
Notice that $\xi_{eq}$ sets the correlation length for the space-time variable $|x+\gamma_o t|$. As we see the energy gap is completely independent of $\gamma_o$; therefore non-reciprocity does not help us close the gap to arrive at a critical phase, which is a feature we discussed in Section \ref{Sec-N-Spins}. However, as we discussed earlier, Figures \hyperref[Gx-Infinite]{\ref*{Gx-Infinite}(A)} and \hyperref[Gx-Infinite]{\ref*{Gx-Infinite}(B)} indicate that non-reciprocity tends to spoil long-time order, but this cannot be seen from the (\ref{Gap-etal}). An important feature that we can already see from (\ref{Low-E-Gx}) is that the central contribution of the non-reciprocity to this correlator is the replacement,
\begin{align} \label{NR-Transformation}
    x \rightarrow x + \gamma_o t\,,
\end{align}
in the well-known result for the $\gamma_o=0$ two-point function \cite{godreche_response_2000}. This transformation makes the correlation between the spin at site $i$ and site $i+x$ behave as a wave packet with velocity $\gamma_o$. This central result was already hinted at in Figure \ref{Gx-Infinite}. We will show it to be true in more general cases, namely those discussed between equations (\ref{2-pt-lim-1}) and (\ref{2-pt-Porod-Equilibrium}).

For $T=0$, $\gamma_e=1$, the gap closes and we are in a critical phase with infinite relaxation time and correlation length. The two-point function is
\begin{align} \label{Gx(t)-Gap-Closed}
    G_x(t) = \sqrt{ \frac{2\pi}{t}} \exp\left[ - \frac{(x+\gamma_o t)^2}{2 t} \right]\,,
\end{align}
where we kept $\gamma_o$ fixed as we took the limit $T\rightarrow 0$. If $\gamma_o=0$ then this function scales homogeneously under the following non-relativistic transformation, 
\begin{align} \label{Scale-Transf-Age}
    x \rightarrow \lambda^{-1} x\, , \qquad t \rightarrow \lambda^{-2} t\,, 
\end{align}
which is part of the Aging group \cite{henkel_aging_2001,henkel_phenomenology_2002,henkel_non-equilibrium_2010}. This group is a generalization of the non-relativistic conformal group for diffusive systems. The non-reciprocity spoils this scaling property. Nonetheless the system is still at criticality and instead of \C{Scale-Transf-Age}, we note that there is still a notion of a good scale transformation for the two-point function but one that looks more like a combination of a boost and a scaling:
\begin{align} \label{Scale-Transf-Age-2}
    x \rightarrow \lambda^{-1}\left(x+ (1-\lambda^{-1}) \gamma_o t \right) \, , \qquad t \rightarrow \lambda^{-2} t\,.
\end{align}

We now proceed to analyze the two-point function for $\tau\neq 0$. The form (\ref{two-point-sol}) is not adequate to study the low-energy behavior of this system. Instead it is better to go back to the equations of motion and analyze the system via Laplace transformation. We start by obtaining an equation of motion for (\ref{equal-two-point-sol}); it is easy to show that
\begin{align} \label{2-pt-fun-eom}
\begin{aligned}
    \frac{d}{dt}  C_{ij} = - C_{ij} + \frac{\gamma_e}{2}(C_{i-1j} + C_{i+1j}) + \frac{\gamma_o}{2}(C_{i-1j} - C_{i+1j})\,.
\end{aligned}
\end{align} 
To solve this equation we use the Laplace transformation for the two times and a Fourier transformation for space assuming an infinite discrete chain, 
\begin{align} \label{Laplace_Fourier}
    &C_{ij}(s_t,s_\tau) = \int dt d\tau\, e^{-s_t t - s_\tau \tau} C_{ij}(t,\tau)\,, \qquad C_{kq}(t,\tau) = \sum_{ij} e^{i(k i + q j)} C_{ij}(t,\tau)\,.
\end{align}
The inverse Fourier transform is,
\begin{align}
    C_{ij}(t,\tau) & = \int\limits_{-\pi}^{\pi} \frac{dk}{2\pi}\frac{dq}{2\pi} e^{-i(ki+qj)} C_{kq}(t,\tau) \,.
\end{align}
The initial condition of (\ref{2-pt-fun-eom}) is $C_{ij}(0,\tau) = r_{ij}(\tau)$. Substituting (\ref{Laplace_Fourier}) into (\ref{2-pt-fun-eom}) gives,
\begin{align}
    s_t C_{kq}(s_t,s_\tau) = -\left(1-\gamma_e \cos k + i \gamma_o \sin k \right) C_{kq}(s_t,s_\tau) + r_{kq}(s_\tau)\,,
\end{align}
whose solution is
\begin{align}
    C_{kq}(s_t,s_\tau) = \frac{r_{kq}(s_\tau)}{\big(s_t+1-\g_e \cos k +i \g_o \sin k \big)}\,.
\end{align}
We now need to find $r_{kq}(s_\tau)$. To do this, it is easier to solve the equations of motion for $r_{ij}(t)$ in Laplace space. The equations of motion for $r_{ij}(t)$ are given by
\begin{align} \label{equal-2-pt-Laplace}
\begin{aligned}
    \frac{d}{dt}{r}_{i j}=&-2r_{i j}+\frac{ \gamma_e}{2}\left(r_{i-1 j}+r_{i +1 j}+r_{i j-1}+r_{i j+1}\right)  \cr & +\frac{\gamma_o}{2}\left(r_{i-1 j}-r_{i+1 j}+r_{i j-1}-r_{i j+1}\right)+ v(t) \delta_{ij}\,, 
\end{aligned}
\end{align}
where $v(t)$ is an auxiliary variable which will help us enforce $r_{jj}=1$. The solution to this equation for $r_{ij}(0)=\delta_{ij}$ is
\begin{align}
    r_{kq}(s_\tau) = \frac{(v(s_\tau)+1)2\pi \delta_{k,-q}}{s_\tau+2-\g_e(\cos k + \cos q ) + i \g_o (\sin k + \sin q) }\,.
\end{align}
The condition $r_{jj}(t)=1$ translates to
\begin{align} \label{equal-2-pt-Condition}
    &\int\limits_{-\pi}^{\pi} \frac{dk}{2\pi} \frac{dq}{2\pi} r_{kq}(s_\tau) = \frac{v(s_\tau) +1 }{\sqrt{(s_\tau+2)^2-4\gamma_e^2 }} = \frac{1}{s_\tau}\,,
\end{align}
which gives $v(s_\tau) =\frac{1}{s_\tau} \sqrt{(s_\tau+2)^2-4\gamma_e^2}-1$, and we get
\begin{align}
    r_{kq}(s_\tau) = \frac{\sqrt{(s_\tau+2)^2-4\g_e^2} \ 2\pi\delta_{k,-q}}{s_\tau[s_\tau+2-\g_e(\cos k + \cos q) + i \g_o (\sin k + \sin q) ]} \,.
\end{align}
Therefore (\ref{equal-two-point-sol}) is also given by the inverse Laplace transform of
\begin{align} \label{equal-2-pt-Laplace-Sol}
    r_{ij}(s_\tau) = \int\limits_{-\pi}^{\pi}\frac{dk}{2\pi} e^{-i(i-j)k}\frac{ \sqrt{(s_\tau+2)^2-4\g_e^2}}{s_\tau[s_\tau+2-2\g_e \cos k ]} \,.
\end{align}
In this way of expressing $r_{ij}(t)$, it is both obvious that this quantity is translationally invariant and independent of the non-reciprocity. Both features are hidden in the expression (\ref{equal-two-point-sol}). Using this result, the two-point function can be written as follows 
\begin{align} \label{2-point-solution-Laplace-Integrand}
    C_{kq}(s_t,s_\tau) = &\frac{1}{\big(s_t+1-\g_e \cos k +i \g_o \sin k\big)} \times \cr &\frac{\sqrt{(s_\tau+2)^2-4\g_e^2}  }{s_\tau\big(s_\tau+2-\g_e (\cos k+\cos q) + i \g_o(\sin k + \sin q)\big)} \,2\pi\delta_{k,-q}\, .
\end{align}
We now inverse Fourier transform this expression and use the delta function to explicitly perform the integration over $q$, 
\begin{align} \label{2-point-solution-Laplace}
    C_{ij}(s_t,s_\tau) = \int\limits_{-\pi}^{\pi} \frac{dk}{2\pi} \frac{e^{-ik(i-j)}}{\big(s_t+1-\g_e \cos k +i \g_o \sin k \big)} \frac{\sqrt{(s_\tau+2)^2-4\g_e^2} }{s_\tau\big(s_\tau+2-2\g_e \cos k \big)}.
\end{align}
Note that the integrand is essentially a product of the one-point function with the integrand of the two-point function at equal time \C{equal-2-pt-Laplace-Sol}. In this form it is again clear that the non-reciprocity disappears from the part of this function that is related to the equal-time two-point function. It is tricky to inverse Laplace transform in both $s_t$ and $s_\tau$ while also integrating over $k$, but doing so leads us to (\ref{two-point-sol}) with $r_{ij}(0) = \delta_{ij}$. Much more interesting than explicitly performing these computations is observing that we can easily extract some very important information about the behavior of this system for large wavelengths and low temperatures, which is what we do next.

Now we take the continuum limit by taking the lattice spacing $a \rightarrow 0$ and take the low momentum limit so that  we can expand the trigonometric functions of \C{2-point-solution-Laplace-Integrand}. If we also Laplace invert the system from $s_t$ back to $t$ we get 
\begin{align}
    C_{x}(t,s_\tau) =\, &\frac{\sqrt{(s_\tau+2)^2-4\gamma_e^2}}{2s_\tau \sqrt{s_\tau+2(1-\gamma_e)}} \frac{1}{\sqrt{2\pi \gamma_e}}  \cr & \times\int\limits_{t(s_\tau/2+1-\gamma_e)}^\infty \frac{dw}{\sqrt{w}} \exp{\left[-w-(s_\tau+2(1-\gamma_e))\frac{(\gamma_o t + x)^2}{4 \gamma_e} \frac{1}{w}\right]}\,,
\end{align}
where $x$ is the distance between the two points in the two-point function. We now use this to understand the general low-energy behavior of the system to see how non-reciprocity changes the low-energy behavior seen in the reciprocal kinetic Ising model, which was studied in detail in \cite{godreche_response_2000}. To do this we analyze some specific limits: first we take the long-time limit, $\tau \gtrsim \tau_{eq}$, by setting $s_\tau^2 \approx 0$ and then we take the low temperature limit which allows us to approximate $2(1-\gamma_e) \approx 1-\gamma_e^2$. We obtain
\begin{align} \label{Cx(t,s)-Low-T}
    C_{x}(t,s_\tau) = \frac{1}{s_\tau\sqrt{2\pi \gamma_e}} \int\limits_{\frac{t}{2}(s_\tau+\mu^2)}^\infty \frac{dw}{\sqrt{w}} \exp{\left[-w- (s_\tau+\mu^2)\frac{(\gamma_o t + x)^2}{4 \gamma_e} \frac{1}{w}\right]}\,.
\end{align}
Now we look at two distinct limits of this system.
\\

\noindent
\textbf{First limit:} here we take the small-$t$ limit such that $t(s_\tau+\mu^2) \approx 0$ and set the lower bound of the integral in (\ref{Cx(t,s)-Low-T}) to zero. We perform the integral and take the inverse Laplace transformation on the variable $s_\tau$ and obtain 
\begin{align} \label{2-pt-lim-1-Int-Form}
    C_x(t,\tau) & = \frac{1}{\sqrt{2\gamma_e}}\frac{\mu|\gamma_o t + x|}{2\sqrt{\pi \gamma_e}} \int\limits_0^{\mu^2 \tau} \frac{du}{u^{3/2}} \exp\left( - u - \frac{\mu^2}{u} \frac{(\gamma_o t + x)^2}{4\gamma_e}  \right)\,.
\end{align}
This can also be rewritten as
\begin{align}
\begin{aligned} \label{2-pt-lim-1}
    C_x(t,\tau) = \frac{1}{2\sqrt{2 \gamma_e}}\bigg[&\exp\left( -\mu \frac{|\gamma_o t + x|}{\sqrt{\gamma_e}}\right)\text{erfc}\left( \frac{|\gamma_ot+x|}{2\sqrt{\gamma_e \tau}} - \mu \sqrt{\tau} \right) \\
    & + \exp\left( \mu\frac{|\gamma_o t + x|}{\sqrt{\gamma_e}}\right)\text{erfc}\left(\frac{|\gamma_ot+x|}{2\sqrt{\gamma_e \tau}} + \mu \sqrt{\tau} \right) \bigg]\,,
\end{aligned}
\end{align}
where we recall that the error function, $\text{erf}(z)$, and complementary error function, $\text{erfc}(z)$, are given by:
\begin{align}
    \text{erf}(z) = \frac{2}{\sqrt{\pi}} \int\limits_0^z dx\, e^{-x^2}, \qquad \text{erfc}(z) = 1 - \text{erf}(z) = \frac{2}{\sqrt{\pi}} \int\limits_z^\infty dx\, e^{-x^2}\,.
\end{align}

\noindent
\textbf{Second limit:} now we let $t$ be any value but we take the equilibrium limit for the variable $\tau$ by setting $\tau \rightarrow \infty$. This can be done on equation (\ref{Cx(t,s)-Low-T}) directly in Laplace space using the final value theorem $C_x(t,eq)=\lim_{s_\tau\rightarrow 0} s_\tau \, C_x(t,s_\tau)$. Doing this transformation and performing the variable substitution $u=1/w$ we obtain 
\begin{align} \label{2-pt-lim2-eq-Int-Form}
    C_x(t,eq) = \frac{1}{\sqrt{2\gamma_e \pi}} \int \limits_0^{2/(\mu^2 t)} \frac{du}{u^{3/2}} \exp\left( -\frac{\mu^2(\gamma_o t + x)^2}{4\gamma_e} u - \frac{1}{u} \right)\,,
\end{align}
which can be written as, 
\begin{align} \label{2-pt-lim2-eq}
\begin{aligned}
    C_x(t,eq) = \frac{1}{2 \sqrt{2 \gamma_e}}\bigg[& \exp\left( -\mu \frac{|\gamma_o t + x|}{\sqrt{\gamma_e}}\right) \text{erfc} \left( \mu \sqrt{\frac{t}{2}}- \frac{|\gamma_o t + x|}{\sqrt{2t \gamma_e}} \right) \\
    & + \exp\left(\mu \frac{|\gamma_o t + x|}{\sqrt{\gamma_e}}\right) \text{erfc} \left( \mu \sqrt{\frac{t}{2}}+ \frac{|\gamma_o t + x|}{\sqrt{2t \gamma_e}} \right) \bigg]\,.
\end{aligned}
\end{align}
After deriving these two expressions, we can study  some interesting cases in detail.
\\

\textbf{(0) \textit{Equilibrium:}} if the system is initially in equilibrium, $\tau \rightarrow \infty$, at $t=0$ both (\ref{2-pt-lim-1}) and (\ref{2-pt-lim2-eq}) give
\begin{align} \label{2-pt-eq-Standard}
    C_x(t=0,eq)&  = \frac{1}{\sqrt{2\gamma_e}} \exp\left(-\mu \frac{|x|}{\sqrt{\gamma_e}} \right)\,,
\end{align}
and there is no information about the non-reciprocity. However, as time passes we can see from (\ref{2-pt-lim-1}) that the non-reciprocity kicks in and changes the two-point function to
\begin{align} \label{2-pt-low-t-tau-eq}
\begin{aligned}
    C_x(t,eq) = \frac{1}{\sqrt{2\gamma_e}} \exp\left(-\mu \frac{|\gamma_ot + x|}{\sqrt{\gamma_e}} \right)\,,
\end{aligned}
\end{align}
for early times $t \ll \tau_{eq}$. At later times when $t \gtrsim \tau_{eq}$ the two-point function settles into (\ref{2-pt-lim2-eq}). Notice that (\ref{2-pt-eq-Standard}) is just the equilibrium two-point function of a $1$-dimensional Ising chain. The fact that the reciprocal ($\gamma_o=0$) kinetic Ising model settles into this state at equilibrium is well known \cite{glauber_timedependent_1963,godreche_response_2000,henkel_non-equilibrium_2010}. 

These results show that at least at early times $t\ll \tau_{eq}$, the main contribution of the non-reciprocal interaction is to deform (\ref{2-pt-eq-Standard}) to (\ref{2-pt-low-t-tau-eq}); namely it performs the transformation (\ref{NR-Transformation}). After this transformation, the correlation length moves with timelike a wavepacket with velocity $\gamma_o$. For longer times $t\gtrsim \tau_{eq}$, this simple result is spoiled and the correlation is given by the expression (\ref{2-pt-lim2-eq}).
\\

\textbf{(I) \textit{Aging or domain growth regime}:} the aging or domain growth regime is the late time behavior of systems undergoing phase ordering kinetics. Phase ordering kinetics are seen in a wide range of physical systems. For example in systems undergoing phase ordering, phase separation, spin-glass relaxation and non-equilibrium critical dynamics \cite{henkel_non-equilibrium_2010,bray_theory_1994}. Take phase ordering dynamics in the Ising model as an example: below $T_c$ there are two distinct vacua with all spins either pointing up or down. After quenching, the system will start forming small patches where the spins inside each patch will be in one of the two vacua. The boundary between these patches and the rest of the system is the domain wall. Initially this boundary is very irregular but as time passes and the patch grows and the boundary becomes smoother. This process is called coarsening and we can associate a length scale $L(\tau)$ to the size of the patch. The scaling hypothesis for aging phenomena states that in the long time limit, when the length scale associated with the domain is $L(\tau) \ll \xi_{eq}$, the correlation function should scale as
\begin{align}
    C_x(\tau) = f \left(\frac{|x|}{L(\tau)}\right) \,.
\end{align}

After preparing our system in some initial state, we take the zero-temperature limit by setting $\mu=0$ and $\gamma_e=1$ but keeping $\gamma_o$ fixed. Since $\xi_{eq}\rightarrow\infty$ this ensures that $L(\tau)\ll \xi_{eq}$ and thus we are in the aging regime of critical dynamics. We also take the early time limit $t\ll\tau_{eq}$ which gives
\begin{align}
    C_x(t=0,\tau) = \frac{1}{\sqrt{2}} \text{erfc}\left( \frac{|x|}{2 \sqrt{\tau}} \right)\,,
\end{align}
for $t=0$. As time passes, the system evolves to
\begin{align} \label{2-pt-ageing}
    C_x(t,\tau) = \frac{1}{\sqrt{2}} \text{erfc}\left( \frac{|\gamma_o t + x|}{2 \sqrt{\tau}} \right)\,.
\end{align}
Therefore we see that
\begin{align}
    f(y) = \frac{\text{erfc}\left(\frac{y}{2}\right)}{\sqrt{2}}\, , \qquad y= \frac{|\gamma_o t + x|}{L(\tau)}\,, \qquad L(\tau) =  \sqrt{\tau} \,.
\end{align}
Recent studies have argued that non-reciprocity can induce a novel kind of frustration which may lead to similar phenomena as geometric frustration, which includes glassiness \cite{hanai_nonreciprocal_2024}. In this context it is reasonable to wonder if the aging regime captured by equation (\ref{2-pt-ageing}) is related to some glassiness induced by non-reciprocity. However the aging behavior of this system seems to be strictly related to the fact that we take the zero-temperature limit, which is also the critical temperature of the one-dimensional Ising model. Therefore this aging behavior seems to be the typical aging behavior seen in critical dynamics and not immediately related to glassiness of the system.

This correlation function is scale invariant by simple an extension of \C{Scale-Transf-Age-2} where we scale both times the same way,
\begin{align} \label{Scale-Transf-Age-3}
    x \rightarrow \lambda^{-1}\left(x+ (1-\lambda^{-1}) \gamma_o t \right) \, , \qquad t \rightarrow \lambda^{-2} t\,, \qquad \tau \rightarrow \lambda^{-2} \tau\, .
\end{align}
In addition to this scaling transformation there is also invariance under the non-trivial scale transformation
\begin{align} \label{Scaling-Non-Trivial}
    x \rightarrow \lambda^{-1} x\,, \qquad t \rightarrow \lambda^{-1} t\,, \qquad \tau \rightarrow \lambda^{-2} \tau\, ,
\end{align}
and the system is self-similar in this stage. Finally at equilibrium
$\tau \rightarrow \infty$, we get
\begin{align}
    C_x(t,eq) = \frac{1}{\sqrt{2}}\,.
\end{align}

\textbf{(II) \textit{Approaching equilibrium}:} this is the limit $\tau \gg \tau_{eq} = 2/\mu^2$ where we can use equation (\ref{Identity-2}) to approximate (\ref{2-pt-lim-1-Int-Form}). We see that at $t=0$ we have 
\begin{align}
    C_x(t=0,\tau) = \frac{1}{\sqrt{2\gamma_e}}\left[\exp\left( -\mu \frac{|x|}{\sqrt{\gamma_e}} \right) - \frac{\mu |x|}{\sqrt{4\pi\gamma_e}} 
    \frac{e^{-\mu^2 \tau}}{(\mu^2\tau)^{3/2}} \right]\,.
\end{align}
As time passes, we settle into the distribution:
\begin{align} \label{2-pt-Approx}
    C_x(t,\tau) = \frac{1}{\sqrt{2\gamma_e}}\left[\exp\left( -\mu \frac{|\gamma_o t + x|}{\sqrt{\gamma_e}} \right) -
    \frac{\mu |\gamma_o t + x|}{\sqrt{4\pi \gamma_e}}\frac{e^{-\mu^2 \tau}}{(\mu^2\tau)^{3/2}} \right]\,.
\end{align}
Further if we are already in equilibrium in the $\tau$ variable, the system approaches equilibrium in the $t$ variable as
\begin{align}
    C_x(t,eq) = \frac{1}{\sqrt{\pi \gamma_e \mu^2 t}} \exp\left( -\frac{\mu^2 t}{2} \right)\,.
\end{align}
This can be seen by expanding (\ref{2-pt-lim2-eq-Int-Form}) for large $t$.
\\

\textbf{(III) \textit{Spatiotemporal Porod}:} this regime is an extension of the standard spatial Porod regime \cite{bray_theory_1994,henkel_non-equilibrium_2010,godreche_response_2000} appropriate for our system. The spatial Porod regime corresponds to observing the system at a scale which is much smaller than the typical size of a domain wall. For the reciprocal kinetic Ising model, $\gamma_o=0$, the spatial Porod regime happens when $1 \ll |x| \ll \xi_{eq} \sim \sqrt{\tau} $, where the correlation function takes the form \cite{godreche_response_2000}
\begin{align}
    C_x(t=0,\tau) \sim 1 - A(\tau) |x|\,.
\end{align}
Non-reciprocity produces the deformation $x\rightarrow x+\gamma_o t$ and therefore we must reconsider what we call the Porod regime. In this case it is natural to study $1\ll |\gamma_o t + x| \ll \xi_{eq} \sim \sqrt{\tau}$ which is a kind of spatiotemporal Porod. To analyze this regime we can use expansion (\ref{Identity-3}) on (\ref{2-pt-lim-1}). At $t=0$ we recover the well-known result 
\begin{align} \label{2-pt-Porod-t=0}
    C_x(t=0,\tau) & = \frac{1}{\sqrt{2 \gamma_e}} \left[1 -\left( \frac{e^{-\mu^2 \tau}}{\sqrt{\pi \tau}} +\mu\ \text{erf}\left(\mu \sqrt{\tau}\right)  \right) \frac{|x|}{\sqrt{\gamma_e}}\right]\,,
\end{align}
which is also valid for reciprocal kinetic Ising models \cite{godreche_response_2000}. However if we turn on non-reciprocity and let time pass, the system is deformed to
\begin{align} \label{2-pt-Porod}
    C_x(t,\tau) = \frac{1}{\sqrt{2 \gamma_e}} \left[1 -\left( \frac{e^{-\mu^2 \tau}}{\sqrt{\pi \tau}} +\mu\ \text{erf}\left(\mu \sqrt{\tau}\right)  \right) \frac{|\gamma_o t + x|}{\sqrt{\gamma_e}}\right] \,.
\end{align}
Then at very low temperatures, $\mu\approx 0$, we have
\begin{align} \label{2-pt-Porod-T=0}
    C_x(t,\tau) = \frac{1}{\sqrt{2}} \left[ 1 - \frac{|\gamma_o t + x|}{\sqrt{\pi \tau}} \right]\,,
\end{align}
and the system is scale invariant under the transformations (\ref{Scaling-Non-Trivial}). Finally if we are in equilibrium and turn on $t$, we can obtain the spatiotemporal Porod regime by applying (\ref{Identity-3}) to (\ref{2-pt-lim2-eq}), 
\begin{align} \label{2-pt-Porod-Equilibrium}
    C_x(t,eq) \approx \frac{1}{\sqrt{2\gamma_e}}\left[ 1- \left( \frac{\exp\left( - \frac{(\gamma_ot+x)^2}{2\gamma_e t} \right)}{\sqrt{\pi}\frac{|\gamma_ot+x|}{\sqrt{2\gamma_e t}}} + \text{erf} \left( \frac{|\gamma_ot+x|}{\sqrt{2\gamma_e t}} \right) \right) \mu \frac{|\gamma_o t+x|}{\sqrt{\gamma_e}} \right] \, ,
\end{align}
and scale invariance is violated.
\\

To summarize some important points: 
the equations above are very similar to what you would find if you were studying a reciprocal kinetic Ising model \cite{godreche_response_2000}. The fundamental difference is that non-reciprocity deforms the two point functions by
\begin{align}
    x \rightarrow x + \gamma_o t\, ,
\end{align}
at least for small enough $t\ll \tau_{eq}$. In some cases, namely, (\ref{2-pt-lim2-eq}) and (\ref{2-pt-Porod-Equilibrium}), this transformation is valid for all times. We also see that the scale transformation (\ref{Scale-Transf-Age}) is generically violated by the two-point function. We also identify two regimes where non-trivial scale invariance, (\ref{Scale-Transf-Age-3}) and (\ref{Scaling-Non-Trivial}), emerges. These are the aging regime (\ref{2-pt-ageing}) and the zero-temperature spatiotemporal Porod regime (\ref{2-pt-Porod-T=0}).

\section{General 1D Systems}

\label{Sec-Random-Couplings}

In this final section, we broaden our gaze and study a more general class of 1D systems. We take a system with $N$ spins. Each spin $s_i$ feels a local magnetic field given by
\begin{align}
    h_i = \beta J_i s_{i_1} +\beta  K_i s_{i_2} \,,
\end{align}
so this magnetic field is determined by at most two distinct spins at two arbitrary points with arbitrary couplings $J_i, K_i \in \R$. Note here that $(s_{i_1}, s_{i_2})$ could be any spins in our system, even $s_i$ itself. We only care that $h_i$ depends on at most $2$ spins. We then take a transition rate generalizing the standard Glauber transition,\footnote{To simplify notation, we have set $\a=1$, where otherwise $w(-s_i \lto s_i) = \frac{\a}{2} \left[1- \tanh(s_ih_i)\right]$.}
\begin{align}
    & w(-s_i \lto s_i) = \frac{1}{2} (1- \tanh(s_ih_i)) \cr
    &=\frac{1}{2}\left(1 - s_i\frac{s_{i_1} + s_{i_2}}{2}\tanh(\beta J_i+\beta K_i) - s_i\frac{s_{i_1} - s_{i_2}}{2}\tanh(\beta J_i-\beta K_i)\right) \,.
\end{align}
If we now study one-point functions, we observe from \C{Eq-t-n-pt DE} that
\begin{align}
    \label{full_gen_q_eqn}
    \del_t q_i(t) &= -2\langle w(-s_i \leftarrow s_i)s_i \rangle \,,\cr
    &= - \left[q_i(t)  - \frac{q_{i_1} + q_{i_2}}{2}\tanh(\beta J_i+\beta K_i) - \frac{q_{i_1} - q_{i_2}}{2}\tanh(\beta J_i-\beta K_i)\right] \,,\cr
    &= - q_i(t)  + C_{ij}q_j \,,
\end{align} 
where $C_{ij}$ has two nonzero elements:
\begin{align} \label{Cij-Initial}
\begin{aligned}
    C_{ij} = & \frac{\tanh(\beta J_i+\beta K_i) + \tanh(\beta J_i-\beta K_i)}{2} \delta_{j i_1} \cr & +  \frac{\tanh(\beta J_i+\beta K_i) - \tanh(\beta J_i-\beta K_i)}{2} \delta_{j i_2}\,.
\end{aligned}
\end{align}
To solve this set of equations explicitly would require careful analysis via either Fourier transform, generating functional or some other method. We also note that with $(i_1, i_2)$ both arbitrary, solving this system in full generality would generally be very difficult. However, we can instead study the system as a matrix: 
\begin{align}
\label{gen-1D-DE}
    \frac{d}{dt}\begin{pmatrix}
        q_1(t) \\
        \vdots \\
        q_N(t)
    \end{pmatrix} &= -\left[ I_{N} - \begin{pmatrix}
         C_{11}&\dots &C_{1N} \\
         \vdots &  \ddots  & \vdots  \\
         C_{N1}&\dots &C_{NN}
    \end{pmatrix}
    \right]
    \begin{pmatrix}
        q_1(t) \\
        \vdots \\
        q_N(t)
    \end{pmatrix}  = M {\vec q} \, .
\end{align}
Note that the matrix $C_{ij}$ is as general as possible with the only constraint that it has at most $2$ nonzero entries in each row. These $C_{ij}$ terms could even be non-vanishing on the main diagonal and we still arrive at the same final inequality. In our following discussion, it will be simpler to first assume $C_{ij}$ has no diagonal terms.

\subsubsection*{\ul{\it Gershgorin's Theorem}}

In what follows we use Gershgorin's theorem~\cite{shivakumar198735} to prove that the eigenvalues of $M$ have real part smaller than zero, $\text{Re}(\lambda_j)<0$, at finite temperature. This proof is different for $N$ finite and for $N= \infty$ and so we discuss these cases separately. We then show $\text{Re}(\lambda_j)<0$ implies no long-time order because the one point function of the system always decays exponentially to zero at finite temperature.  We must pay special attention to cases where the matrix $M$ cannot be diagonalized and must be put in a Jordan normal form \cite{ashida_non-hermitian_2020}.

The summed absolute values of the off-diagonal elements of $C_{ij}$ takes the form,
\begin{align}
    |C_{ii_1}| + |C_{ii_2}|  = & \left|\frac{\tanh(\beta J_i+\beta K_i)+\tanh(\beta J_i-\beta K_i)}{2}\right| \cr +&  \left|\frac{\tanh(\beta J_i+\beta K_i)-\tanh(\beta J_i-\beta K_i)}{2}\right| \,, \cr
    = & \frac{2L_i}{2} = L_i\,,
\end{align}
where $L_i = \max(|\tanh(\beta J_i+\beta K_i)|,|\tanh(\beta J_i-\beta K_i)| )$ and the last line follows by noting that $2 \cdot \max(|x|,|y|)=|x-y|+|x+y|$. Now noting that $\beta  = T\inv$, we observe that the summed absolute values of the off-diagonal terms in each row is always less than $1$ for $T>0$.
For finite $T$, we get a strictly diagonally dominant matrix in~\C{gen-1D-DE} and we can apply Gershgorin's theorem~\cite{shivakumar198735}. 
This states that all of the eigenvalues of this system lie within disks of radius $L_i$ centered at $-1$ in the complex plane, which we denote $D_{L_i}(-1) \subset \CC$. Note, however, that since $|L_i| < 1$  every eigenvalue $\lambda_j $ has $\text{Re}({\lambda_j}) < 0 $.\footnote{If instead we had $C_{ii_1}$ on the main diagonal, Gershgorin's theorem gives $\lambda_j \in D_{C_{ii_2}}(1-C_{ii_1})$ and the same $L_i$ inequality again shows that zero is outside our disk.}

If the system has an infinite number of spins then we must use a generalized version of Gershgorin's theorem~\cite{shivakumar198735} and additionally impose a constraint that $\sup{L_i} < 1$ in our system. This is essentially requiring that $J_i$ and $K_i$ are finite in the $N=\infty$ system. Since $q_k(t)$ is bounded between $\pm 1$, our state vectors $\vec{q}(t)$ always exist in ${l}_\infty$, which is a complete normed vector space consisting of elements with norm
\begin{align}
    ||q||_\infty = \sup\{|q_i|: q_i \in q\} = 1 < \infty\, .
\end{align} 
We can then apply Gershgorin's theorem over ${l}_\infty$ which states that for all $\epsilon > 0$, the eigenvalue $\lambda_i$ satisfies
\begin{align}
    \lambda_i \in \bigcup D_{R_i(1+\epsilon)}(-1)\,,
\end{align}
where $R_i$ is defined as summed absolute values of the off-diagonal terms in each row~\cite{shivakumar198735}.

\par Using our constraint that $\sup L_i = L_{\text{max}} <1$, we can choose $\epsilon > 0$ with $\epsilon < \frac{1-L_{\text{max}}}{2L_{\text{max}}}$ so that $0\notin \bigcup D_{L_i(1+\epsilon)}(-1)$. As a result, all of our eigenvalues will still have negative real component for finite temperature.

\subsubsection*{\ul{\it No long-time order for a finite system at finite temperature}}

Now let us first consider a finite system with $N$ spins. We must be careful about whether the matrix $M$ is diagonalizable or not. To address this we review some well-known results about non-Hermitian matrices with a detailed discussion found in Ref. \cite{ashida_non-hermitian_2020}. The characteristic polynomial for $M$ can generically be written,
\begin{align}
    p_M(\lambda) = \det \big(\lambda I - M \big) = \prod_{j=1}^J(\lambda - \lambda_j)^{m_j^a} \,,
\end{align}
where $\lambda_j$ are the roots of $p_M(\lambda)$ and thus the eigenvalues of the system. Here the matrix $M$ has $J$ distinct eigenvalues and $m_j^a$ is the algebraic multiplicity of the $j^{th}$ eigenvalue. Now a central question is how many linearly independent eigenvectors are associated with a given eigenvalue $\lambda_j$. The number of linearly independent eigenvectors for a given $\lambda_j$ is called the geometric multiplicity and is given by,
\begin{align}
    m_j^g = \dim \text{Ker}\big(M-\lambda_j I \big)\,.
\end{align}
The matrix $M$ is diagonalizable if and only if $m_j^a=m_j^g$ for all $j$. In other words, if eigenvalue $\lambda_j$ is an $m_j^a$ root of the characteristic polynomial, the matrix is diagonalizable if and only if there are also $m_j^a$ linearly independent eigenvectors. Hermiticity is sufficient but not necessary to guarantee that the matrix is diagonalizable. If the matrix is diagonalizable then there is a similarity transformation or change of basis $V \in GL(n, \CC)$ which diagonalizes $M$ with,
\begin{align}
    M = V \left(\begin{array}{cccccc}
\lambda_1 & 0 & 0 & \cdots & 0 & 0 \\
0 & \lambda_2 & 0 & \cdots & 0 & 0 \\
0 & 0 & \lambda_3 & \cdots & 0 & 0 \\
\vdots & \vdots & \vdots & \ddots & \vdots & \vdots \\
0 & 0 & 0 & \cdots & \lambda_{N-1} & 0 \\
0 & 0 & 0 & \cdots & 0 & \lambda_{N}
\end{array}\right)_{N \times N} V^{-1} \,.
\end{align}
The solution of (\ref{gen-1D-DE}) in the new basis $\tilde q = V^{-1} q$ takes the form:
\begin{align}
    \tilde q_j(t) = e^{\lambda_j t} \tilde q_j(0)\,.
\end{align}
Since $\text{Re}(\lambda_j) < 0$ the one-point function always relaxes to equilibrium exponentially and we can never close the gap to have a phase transition. As an aside, this result is also valid for $N\rightarrow \infty$ for diagonalizable $M$.

The matrix $M$, however, can be non-diagonalizable. Cases of this type include systems with parameters which can be tuned to exhibit exceptional points.\footnote{By exceptional point, we mean a locus in parameter space where the structure of the Jordan blocks change if one changes some parameter.} One could hope that tuning the system to some exceptional point could potentially change the situation described above. Perhaps we could close the gap and generate some non-trivial phase transition in this case. 
Let us turn to this case. 

If $m_j^g<m_j^a$ we cannot diagonalize the matrix, but there is always some similarity transformation which allows the matrix $M$ to be written as
\begin{align} \label{Similarity-Transf}
    M=V\left[\bigoplus_{j=1}^J \bigoplus_{\alpha=1}^{m_j^{\mathrm{g}}} \Big( \lambda_j I_{n_{j\alpha}} + J_{n_{j \alpha}} \Big) \right] V^{-1},
\end{align}
where $J_{n}$ is an $n\times n$ upper triangular Jordan matrix given by:
\begin{align}
    J_n \equiv\left(\begin{array}{cccccc}
0 & 1 & 0 & \cdots & 0 \\
0 & 0 & 1 & \cdots & 0 \\
\vdots & \vdots & \ddots & \ddots & \vdots \\
0 & 0 & 0 & \cdots & 1 \\
0 & 0 & 0 & \cdots & 0
\end{array}\right)_{n \times n} \,.
\end{align}
It is easy to see that the Jordan matrices, $J_n$, are nilpotent satisfying $(J_n)^n=0$. The indices $n_{j\alpha}$ which are the size of the Jordan blocks must satisfy the identity,
\begin{align}
    \sum_{\alpha=1}^{m_j^{\mathrm{g}}} n_{j \alpha}=m_j^{\mathrm{a}}, \quad \forall j=1,2, \ldots, J\,.
\end{align}
In this case the solution to (\ref{gen-1D-DE}) in the basis $\vec{\tilde{q}} = V^{-1} \vec{q}$ can be written as
\begin{align}
    \vec{\tilde{q}}(t) = \sum_{j=1}^J e^{\lambda_j t}\left(I_{n_{j\alpha}}+\sum_{\alpha=1}^{m_j^{\mathrm{g}}} \sum_{p=1}^{n_{j \alpha}-1} \frac{t^p}{p!} J_{n_{j\alpha}}^p\right) \vec{\tilde{q}}(0) \,.
\end{align}
Since the sum on $p$ is finite, the polynomial in $t$ will never offset the exponential factor and the system still decays to equilibrium.

\subsubsection*{\ul{\it Special cases with $N=\infty$}}

Models with $N=\infty$ that are not diagonalizable are quite subtle to analyze. As a first case where we can say something, consider the extreme case where we have an $N^{th}$-order exceptional point with $N\rightarrow\infty$. Namely, there is a single eigenvector $\lambda$ with $m^a=N$ and $m^g=1$. The solution to (\ref{gen-1D-DE}) is then
\begin{align} \label{Tilde-q-EP-Finite}
    \vec{\tilde{q}}(t) & = e^{\lambda t}\left(I_{N}+ \sum_{p=1}^{N-1} \frac{t^p}{p!} J_{N}^p\right) \vec{\tilde{q}}(0)\,. 
\end{align}
For finite $N$, we again have a polynomial of bounded order and there is no phase transition. If we allow $N\rightarrow \infty$, there is now an infinite series which could potentially match the exponential.

To address this case, we first show that if equation (\ref{gen-1D-DE}) has an $N^{th}$-order exceptional point then the unique eigenvalue takes the value $\lambda=-1$. From equation (\ref{Similarity-Transf}) we know that (\ref{gen-1D-DE}) can be rewritten in the basis $\vec{\tilde{q}}$ as
\begin{align} \label{Nth-EP-tilde-q}
    \dot{\vec{\tilde{q}}} & = \big(\lambda I_N + J_N \big) \vec{\tilde{q}} \,.
\end{align}
Transforming this back to the $\vec{q}$ basis we get
\begin{align} \label{EP-Proving-lambda}
\begin{aligned}
    \dot{\vec{q}} & = V\big(\lambda I_N + J_N \big) V^{-1}\vec{q} \,, \\
    & = \big(\lambda I_N + V J_N V^{-1} \big) \vec{q} \,.
\end{aligned}
\end{align}
Comparing this with (\ref{gen-1D-DE}) we see that $\lambda = -1$ and that $C = V J_N V^{-1}$. Therefore (\ref{Tilde-q-EP-Finite}) can be written as
\begin{align} \label{vec-q-Cp}
    \vec{q}(t) & = e^{- t}\left(I_N + \sum_{p=1}^{N-1} \frac{t^p}{p!} C^p\right) \vec{q}(0). 
\end{align}
Now let us write the matrix $C$ from (\ref{Cij-Initial}) as
\begin{align}
    C_{ij} = C_{ii_1} \delta_{j i_1} + C_{ii_2} \delta_{j i_2} \,. \label{defC}
\end{align}
Again we emphasize that $|C_{ii_1}|< 1$, $|C_{ii_2}|< 1$ and $|C_{ii_1}|+|C_{ii_2}| = L_i < 1$. We also define
\begin{align}
    \mathcal{L}_i^{(1)} = \sum_j C_{ij} = C_{ii_1} + C_{ii_2} \,,
\end{align}
which satisfies $|\mathcal{L}_i^{(1)}| \leq L_i < 1$. So (\ref{vec-q-Cp}) can be written as
\begin{align} \label{qj-sec6-Cjj1}
    q_j(t) & = e^{- t}\left(q_j(0)+ \sum_{p=1}^{N-1} \frac{t^p}{p!} C_{jj_{1}}C_{j_{1}j_{2}} \ldots C_{j_{p-2}j_{p-1}}C_{j_{p-1}j_{p}} q_{j_p}(0) \right)\,,
\end{align}
where we are summing over repeated indices.  Now $|q_j(0)|\leq 1$ so we generally cannot use the initial conditions to make the sum larger than the exponential since all $|C_{ij}| < 1$. The extreme case is $q_j(0) = m(0)$ for all $j$, where the infinite sum might be able overcome the exponential. In this case, we have
\begin{align} \label{q-j-C-summation}
\begin{aligned}
    q_j(t) & = e^{- t}\left(1+ \sum_{p=1}^{N-1} \frac{t^p}{p!} C_{jj_{1}}C_{j_{1}j_{2}} \ldots\sum_{j_p}C_{j_{p-1}j_{p}} \right)m(0)\,, \\
    & = e^{- t}\left(1+ \sum_{p=1}^{N-1} \frac{t^p}{p!} C_{jj_{1}}C_{j_{1}j_{2}} \ldots C_{j_{p-2}j_{p-1}}\mathcal{L}_{j_{p-1}}^{(1)} \right) m(0)\,.
\end{aligned}
\end{align}
Following the notation of~\C{defC}, we notice that 
\begin{align}
    |C_{j_{p-2}j_{p-2_1}}\mathcal{L}_{j_{p-2_1}}^{(1)} + C_{j_{p-2}j_{p-2_2}}\mathcal{L}_{j_{p-2_2}}^{(1)}| \leq |C_{j_{p-2}j_{p-2_1}} + C_{j_{p-2}j_{p-2_2}}| \leq L_{j_{p-2}} < 1\,.
\end{align}
Therefore to write the summation (\ref{q-j-C-summation}) in a compact form we define
\begin{align}
\begin{aligned}
    \mathcal{L}_{j_{p-q}}^{(q)} & = C_{j_{p-q} j_{p-(q-1)}} \mathcal{L}_{j_{p-(q-1)}}^{(q-1)}\,, \\
    & = C_{j_{p-q}j_{p-q_1}}\mathcal{L}_{j_{p-q_1}}^{(q-1)} + C_{j_{p-q}j_{p-q_2}}\mathcal{L}_{j_{p-q_2}}^{(q-1)}\,,
\end{aligned}
\end{align}
which satisfies
\begin{align}
    |\mathcal{L}_{j_{p-q}}^{(q)}| \leq |\mathcal{L}_{j_{p-q+1}}^{(q-1)}| \leq \ldots \leq |\mathcal{L}_{j_{p-1}}^{(1)}| \leq L_{j_{p-1}} < 1\,.
\end{align}
Therefore we can write (\ref{qj-sec6-Cjj1}) as
\begin{align}
    q_j(t) & = e^{- t}\left(1+ \sum_{p=1}^{N-1} \frac{t^p}{p!} \mathcal{L}_{j}^{(p)} \right)m(0)\,,
\end{align}
which in the $N\rightarrow \infty$ limit gives
\begin{align} \label{Infinite-EP-Sol}
    q_j(t) & = e^{- t}\left(1+ \sum_{p=1}^{\infty} \frac{t^p}{p!} \mathcal{L}_{j}^{(p)} \right)m(0)\,.
\end{align}
Since $|\mathcal{L}_{j}^{(p)}| < 1$ the summation can never overcome the exponential.

Let us discuss two simple examples of infinite order exceptional points to illustrate the mechanism described above. First suppose we have a semi-infinite chain starting at $j=0$ and let the spin at $j$ interact only with the spin on its right so the equations of motion are, 
\begin{align}
    \dot{q} = \left(\begin{array}{ccccccc} -1 & L_1 & 0 & 0 & \cdots & 0 & 0 \\
0 & -1 & L_2 & 0 &\cdots & 0 & 0 \\
0 & 0 & -1 & L_3 & \cdots & 0 & 0 \\
\vdots & \vdots & \vdots & \vdots & \ddots & \vdots & \vdots \\
\end{array}\right) q \, .
\end{align}
Notice that this is a generalization of the critical damping model that we discussed in Section \ref{Sec-N-Spins} with  equations of motion:
\begin{align}
    \dot q_j = - q_j + L_{j+1} q_{j+1}\,.
\end{align}
This system can be put in Jordan normal form using the similarity transformation induced by
\begin{align}
    V &= \left(\begin{array}{ccccccc} 1 & 0 & 0 & 0 & \cdots & 0 & 0 \\
0 & \frac{1}{L_1} & 0 & 0 &\cdots & 0 & 0 \\
0 & 0 & \frac{1}{L_1L_2} & 0 & \cdots & 0 & 0 \\
0 & 0 & 0 & \frac{1}{L_1L_2L_3} & \cdots & 0 & 0 \\
\vdots & \vdots & \vdots & \vdots & \ddots & \vdots & \vdots \\
\end{array}\right) \,.
\end{align}
The equations of motion become
\begin{align}
    \dot{\tilde{q}} =
    \begin{pmatrix} -1 & 1 & 0 & 0 & \cdots & 0 & 0 \\
    0 & -1 & 1 & 0 &\cdots & 0 & 0 \\
    0 & 0 & -1 & 1 & \cdots & 0 & 0 \\
    \vdots & \vdots & \vdots & \vdots & \ddots & \vdots & \vdots \\
    \end{pmatrix} \tilde q \,,
\end{align}
with $\tilde q_j(t) = \left[\prod_{k=1}^j L_k \right] q_j(t)$. The solution to the equations of motion for $j=0$ is
\begin{align} \label{Tilde-q0-Infinite}
    \tilde q_0(t) = e^{- t}\left(\tilde q_0(0) + \sum_{p=1}^\infty\frac{t^p}{p!} \tilde q_p(0) \right).
\end{align}
Therefore in the old basis we find
\begin{align}
    q_0(t) = e^{-t} \left( q_0(0) + \sum_{p=1}^{\infty} \left[\prod_{k=1}^{p} L_k \right] \frac{t^p}{p!} q_p(0) \right)\,.
\end{align}
If we start in a state with $q_j(0) = m(0)$ for all $j$ then we get
\begin{align}
    q_0(t) = e^{-t} \left( 1 + \sum_{p=1}^{\infty} \left[\prod_{k=1}^{p} L_k \right] \frac{t^p}{p!} \right) m(0) \,.
\end{align}
Since $L_k<1$ the summation never outcompetes $e^{-t}$ unless $T=0$ where $L_i=1$. Hence there is no long-time order at finite temperature. 

The second example is the critical damping case from (\ref{1-pt-soul-Green}). For $\gamma_o>0$ we have
\begin{align}
    q_j(t) = \sum_{l=-\infty}^\infty q_{j-l}(0) \theta(|l-j|) e^{-t}\frac{(\gamma t)^{|l-j|}}{|l-j|!} \,.
\end{align}
For the spin at $j=0$ we have
\begin{align}
    q_0(t) = \sum_{l=0}^{\infty} q_{-l}(0) e^{-t}\frac{(\gamma t)^{l}}{l!} \,.
\end{align}
If $q_{-l}(0) = m(0)$ then we get
\begin{align}
    q_0(t) = e^{-(1-\gamma)t} m(0) \,.
\end{align}
Therefore the gap only closes at zero temperature where $\gamma=1$.

\subsubsection*{\ul{\it General cases with $N=\infty$}}

The proof that an infinite order exceptional point does not lead to long-time order at finite temperature  outlined between (\ref{Nth-EP-tilde-q}) and (\ref{Infinite-EP-Sol}) works when there is a single eigenvalue with $m^a=\infty$ for any $m^g$. The proof does not immediately extend to cases with 
two or more distinct eigenvalues with at least one $m^a=\infty$. For example, suppose there are two eigenvalues $\lambda_1$ and $\lambda_2$ with algebraic and geometric multiplicities $(m_1^a,m_1^g)=(N/2,1)$ and $(m_2^{a},m_2^g)=(N/2,1)$. Equation (\ref{gen-1D-DE}) in $\vec{\tilde{q}}$ is
\begin{align}
    \dot{\vec{\tilde{q}}} = \big( \lambda_1 I_{N/2} \oplus \lambda_2 I_{N/2} + J_{N/2} \oplus J_{N/2}  \big) \vec{\tilde{q}}\,,
\end{align}
and thus 
\begin{align}
    V(\lambda_1 I_{N/2} \oplus \lambda_2 I_{N/2})V^{-1} \neq \lambda I_N \,.
\end{align}
So the argument we outlined above does not immediately apply. It would be very interesting to further explore cases with $N=\infty$ and two or more distinct eigenvalues where at least one has $m^a=\infty$. This is the remaining subset of models where spins interact with at most two other spins that might yet exhibit long-time order at finite temperature.

The other natural place to search for systems that might result in a phase transition at $T\neq 0$ with Glauber-like transition rates,
\begin{align}
    w(-s_j \lto s_j) = \frac{1}{2}\big(1-s_j\tanh(h_j)\big) \,,
\end{align}
is to add more than two spin interactions in $h_j$ so  $\tanh(h_j)$ is truly nonlinear in the spin variables. This can be achieved by going to higher dimensions or considering longer range interactions.

\section{Discussion and Conclusions}
\label{conclusions}

In this paper we formulated the non-reciprocal Ising model of Figure \ref{NR-Ising-Lattice} and found an exact solution for its one-point function  \C{1-pt-soul-Green} and \C{1-point-propagator}. We showed that higher point functions of this system factorize and can be expressed in terms of the one-point function in \C{n-point-sol-hom}; therefore \C{1-point-propagator} is a complete solution of the problem. We then proceeded to analyze the behavior of this solution in detail.
At zero non-reciprocity, $\gamma_o=0$, the system is in the overdamped regime and can be found either in a ferromagnetic or antiferromagnetic phase. As we turn on non-reciprocity, the system reaches a critically damped phase at an infinite order exceptional point, and then goes to an underdamped phase. These results are summarized in equations \C{1-point-propagator} and \C{N-Spin-OCU} and are explored in Figure \ref{Gx-Infinite} where we plot the time evolution and spatial profile of the Green’s function of the system in six different regimes: either ferromagnetic or antiferromagnetic; for each choice, either the overdamped, critically damped or underdamped regime. As we can see from this analysis, non-reciprocity leads to an underdamped regime where the system decays more rapidly to the steady-state configuration.

In Section \ref{Sec-Finite-Spin-Chain} we contrast the behavior of an infinite spin chain with the behavior of finite spin chains  with both open and periodic boundary conditions. This contrast can be seen in Figures \ref{Inf_ge-go}--\ref{chain_tot}. An interesting feature is the parity dependence of the time evolution and the stationary state. This dependence is first explored in Figure \ref{per_parity} and summarized in equation \C{FrustrationGammae<0}. When $\gamma_e<0$ this result is an indication of the standard geometric frustration that happens in antiferromagnetic systems, which is illustrated in Figure \ref{Even-Odd}. However in Figure \ref{per_ge} we see that even when $\gamma_e>0$, there is still parity dependence in the system. As explained around equation \C{FrustrationGammae>0} this happens because as we turn on non-reciprocity, we change the overall coupling constant and induce an effective antiferromagnetic behavior leading to frustration despite $\gamma_e>0$. This is an example of a system which was not frustrated since $\gamma_e>0$, but non-reciprocity drives it to a phase where the system becomes effectively antiferromagnetic and thus frustrated in accord with the intuition of Ref. \cite{hanai_nonreciprocal_2024}.

In Section \ref{Sec-Two-Point} we analyzed the low-energy behavior of this system in detail. We started with the equal-time two-point function \C{2-pt-tau=0-uncor} and \C{Low-E-Gx}, which at the critical point $T=0$ is given by \C{Gx(t)-Gap-Closed} and has the symmetry \C{2-pt-tau=0-uncor}. We believe that this symmetry is a general symmetry of the system at the critical point. However proving this and understanding the origin of that symmetry are open questions for future work. We then looked at the two-point function at different times and studied its equilibrium limit \C{2-pt-low-t-tau-eq}, the aging or domain growth regime \C{2-pt-ageing}, the approach to equilibrium \C{2-pt-Approx} and the spatiotemporal Porod regime \C{2-pt-Porod}. We showed that both in the aging regime and the zero-temperature Porod regime \C{2-pt-Porod-T=0} the system is invariant under the extra symmetry \C{2-pt-Porod-T=0}. 

Some recent studies have shown that non-reciprocity can lead to frustration and then to a new notion of glassiness \cite{hanai_nonreciprocal_2024}. However the aging behavior of this system seems related to the fact that the system is at $T=0$ and is therefore undergoing critical dynamics, which is also a regime where aging behavior is seen \cite{bray_theory_1994,henkel_non-equilibrium_2010}. As we found in Section 4 shown in Figure \ref{per_ge} and explained around equation \C{FrustrationGammae>0}, non-reciprocity indeed leads to frustration. It is therefore possible that some of the novel features that we saw in the aging behavior of this system are related to a novel type of glassiness induced by non-reciprocity. Further investigation is needed to explore this possibility.

Finally in Section \ref{Sec-Random-Couplings} we looked at a broader class of models. We allowed each spin to be coupled to at most two other spins and allowed the coupling strengths to be arbitrary. Notice that there is no notion of a geometrical ordering in this system. In this general setup, we showed that there is no long-time order for systems of finite size at finite temperature, regardless of non-reciprocity. For infinite systems the situation is more subtle since infinite order exceptional points might lead to non-trivial long-range order. We showed that if there is a single infinite order exceptional point with a single eigenvalue the system cannot exhibit long-range order at finite temperature. This leaves open the possibility of finite-temperature long-range order for a system at an exceptional point with more than one eigenvalue. The results of Section \ref{Sec-Random-Couplings} on more general kinetic models open up a number of interesting directions for future exploration.

\section*{Acknowledgments}

We would like to thank Michel Fruchart for helpful comments. S.~S. would like to thank the organizers of the 21st Simons Physics Summer Workshop and the ``The Landscape vs the Swampland'' ESI workshop for hospitality during the completion of this work.
M.~S. is supported in part by the Enrico Fermi Institute and in part by NSF Grant No. PHY2014195. G.~W. and S.~S. are supported in part by NSF Grant No. PHY2014195 and by the Australian Research Council (ARC) Discovery Project DP240101409.

\appendix\newpage

\section{Useful Mathematical Identities} \label{Appendix-Bessel}

\textbf{Identity 1:} First we prove the identity,
\begin{align} \label{deffp}
    f_{p,N}(x,y) &= \sum_{m=0}^{N} \frac{1}{N} \exp\left[i\frac{2\pi m}{N}  p + x \cos \frac{2\pi m}{N}  - i y \sin \frac{2\pi m}{N}  \right] \,,\cr &= \sum_{n,s=-\infty}^\infty (-1)^s I_{p+s+n N}(x) J_s(y) \,.
\end{align}
The first step is to use the Jacobi-Anger identities in the appropriate places \cite{Watson-Tretesie-1922}. These identities are
\begin{align}
\begin{aligned}
    \label{JacobiAngerEqns}
    &e^{i z \cos \theta} = \sum_{n=-\infty}^{\infty} i^n J_n(z) e^{i n \theta}\,,   \\
    &e^{i z \sin \theta} = \sum_{n=-\infty}^{\infty} J_n(z) e^{i n \theta}\,.
\end{aligned}
\end{align}
In our specific cases we have
\begin{align}
    \label{JacobiAngerEqns2}
    \quad e^{x \cos \theta} & = \sum_{n=-\infty}^{\infty} i^n J_n(-ix) e^{i n \theta} = \sum_{n=-\infty}^{\infty} I_n(x) e^{in\theta} \,, \\
    \quad e^{-iy \sin \theta} & = \sum_{n=-\infty}^{\infty} J_n(-y) e^{i n \theta} = \sum_{n=-\infty}^{\infty} (-1)^n J_n(y) e^{in\theta}\,,
\end{align}
where we used
\begin{align}
\label{besselIandJreln}
\begin{aligned}
    J_n(-x) & = (-1)^n J_n(x)\,,\\
    I_n(x) & = i^{-n} J_n(ix)\,, \\
    i^nJ_n(-ix) & = (-1)^n i^{2n} i^{-n} J_n(ix) = I_n(x)\,.
\end{aligned}
\end{align}
Now we can evaluate \C{deffp} using $k=2\pi m/Na$ with $a$ the lattice spacing, 
\begin{align}
\begin{aligned}
    f_{p,N}(x,y) & = \sum_{k} \frac{1}{ N} e^{ika p} \sum_{r=-\infty}^\infty I_r(x) e^{i r ka } \sum_{s=-\infty}^\infty (-1)^s J_s(y) e^{i s ka}\,, \\
    & = \sum_{r,s=-\infty}^\infty (-1)^s I_r(x) J_s(y) \sum_{k} \frac{1}{ N} e^{ika(p+r+s)} \,.
\end{aligned}
\end{align}
For $k=2\pi m/Na$ with $m=0,1,\ldots,N-1$ we use
\begin{align}
    &\sum_{k}\frac{1}{N} e^{ika(p+r+s)} = \sum_{m=0}^{ N-1}\frac{1}{N} e^{i\frac{2\pi m}{N}(p+r+s)} = \delta_{0,p+r+s+n N },
\end{align}
to see that the summation is nonzero only if $p+r+s= - n N$ for any integer  $n$. Therefore we get
\begin{align}
\begin{aligned}
    f_{p,N}(x,y)
    & = \sum_{r,s,n=-\infty}^\infty (-1)^s J_s(y) I_r(x) \delta_{r,-p-s-nN} \,, \\
    & = \sum_{s,n=-\infty}^\infty (-1)^s J_s(y) I_{-p-s-n N}(x) \,, \\
    & = \sum_{s,n=-\infty}^\infty (-1)^s J_s(y) I_{p+s+n N}(x)\,,
\end{aligned}
\end{align}
as desired. In the last step we used
\begin{align}
    I_{-n}(x) = i^n J_{-n}(ix) = i^{2n} i^{-n}(-1)^n J_n(ix) = i^{-n} J_n(x) = I_n(x)\,.
\end{align}

\vskip 0.1 in
\noindent
\textbf{Identity 2:} To study the large $\tau$ limit of  equation (\ref{2-pt-lim-1-Int-Form}), which describes the approach to equilibrium, we use
\begin{align} \label{Identity-2}
    \sqrt{\frac{b}{\pi}} \int_0^a \frac{du}{u^{3/2}} e^{-u-\frac{b}{u}} \approx e^{-2 \sqrt{b}} - \sqrt{\frac{b}{\pi }} \frac{e^{-a}}{a^{3/2}}\,,
\end{align}
 which can be checked using \textit{Mathematica}.

\vskip 0.1 in
\noindent
\textbf{Identity 3:}
To get (\ref{2-pt-Porod-t=0}), (\ref{2-pt-Porod}) and (\ref{2-pt-Porod-Equilibrium}) from (\ref{2-pt-lim-1}) and (\ref{2-pt-lim2-eq}) we approximate
\begin{align} \label{Identity-3}
    e^{c} \text{erfc}(a+z) + e^{-c} \text{erfc}(a-z) \approx 2 \left( 1- 2a \frac{e^{-z^2}}{\sqrt{\pi}} - \text{erf}(z) c \right) \,,
\end{align}
for small $c$ and $a$, which can also be checked with \textit{Mathematica}.

\vskip 0.1 in
\noindent
\textbf{Identity 4:}
We now want to prove that
\begin{align}
\label{besselI}
    \exp\left[{\frac{x}{2}\left(t+\frac{1}{t}\right)}\right] = \sum_{n=-\infty}^\infty I_n(x)t^n \,, \\
    \label{besselJ}
    \exp\left[{\frac{x}{2}\left(t-\frac{1}{t}\right)}\right] = \sum_{n=-\infty}^\infty J_n(x)t^n  \,.
\end{align}
Both of these equations follow from typical Bessel function identities. We prove the first equation by noting that it is a direct result of the Jacobi-Anger identity~\cite{Watson-Tretesie-1922}. The second follows by the same steps. As noted above, these identities give \C{JacobiAngerEqns2}
\begin{align}
    e^{x\cos(\theta)} = \sum_{n=-\infty}^\infty I_n(x)e^{in\theta} \,,
\end{align}
which simplifies by substituting $\cos(\theta) = \frac{1}{2}\left(e^{i\theta} + e^{-i\theta} \right)= \frac{1}{2}(t + t\inv)$.
Substituting $t$ into our Jacobi-Anger identity equation gives
\begin{align}
    \exp\left[\frac{x}{2}(t + t\inv)\right] = \sum_{n=-\infty}^\infty I_n(x)t^n \,,
\end{align}
proving our identity. Following the same steps, we can also prove \C{besselJ} given equation \C{besselIandJreln}.

\newpage

%---------------
%---------------
%---------------
%---------------
%---------------
%---------------

\newpage
%\bibliographystyle{amsunsrt-ensp}
% \bibliographystyle{utphys}
% \bibliography{master}

\providecommand{\href}[2]{#2}\begingroup\raggedright\endgroup

\end{document}